\documentclass[12pt]{article}
\usepackage[margin=0.95 in]{geometry}
\usepackage{amsmath} 
\usepackage{amssymb,amsfonts}
\usepackage[all]{xy}
\usepackage{graphicx}
\usepackage[utf8x]{inputenc}
\usepackage{amsmath}
\usepackage{amssymb}
\usepackage{float}
\usepackage{array}
\usepackage{tikz}
\usepackage{mathtools}
\usepackage{mathrsfs} 
\usepackage[hidelinks]{hyperref}
\usepackage{cite}
\numberwithin{equation}{section}
\newcommand{\be}{\begin{equation}}
\newcommand{\ee}{\end{equation}}
\def\bea{\begin{eqnarray}}
\def\eea{\end{eqnarray}}
\newcommand{\qst}{q_{\text{s.t.} }}
\newcommand{\taust}{\tau_{\text{s.t.} }}
\newcommand{\kbos}{k_{\text{bos}}}

\newcommand{\ii}{\mathrm{i}}
\newcommand{\pa}{\partial}

\renewcommand{\b}{\beta}

\newcommand{\qbarst}{\bar{q}_{\text{s.t.} }}
\newcommand{\taubarst}{\bar{\tau}_{\text{s.t.} }}
\newcommand{\zst}{z }
\newcommand{\zbarst}{\bar{z} }
\newcommand{\nust}{\nu }
\newcommand{\nubarst}{\bar{\nu} }
\newcommand{\Zint}{Z_{\text{int}}}
\newcommand{\cint}{c_{\text{int}}}
\newcommand{\Zsgh}{Z_{\text{sgh}}}
\newcommand{\Zmaj}{Z_{\text{Maj}}}
\DeclareMathOperator{\Tr}{Tr}
\newcommand{\nup}{\nu_{1}}
\newcommand{\num}{\nu_{2}}
\newcommand{\nupbar}{\bar \nu_{1}}
\newcommand{\numbar}{\bar\nu_{2}}
\newcommand{\jtilde}{\widetilde{j}}
\newcommand{\ltilde}{\widetilde{l}}
\numberwithin{equation}{section}
\numberwithin{table}{section}

\begin{document}

\hypersetup{pageanchor=false}
\begin{titlepage}
\vbox{
    \halign{#\hfil         \cr
           } 
      }  
\vspace*{15mm}
\begin{center}
{\Large \bf 
Superstrings in Thermal Anti-de Sitter Space
}

\vspace*{15mm}

{\large Sujay K. Ashok$^a$ and Jan Troost$^b$ }
\vspace*{8mm}

$^a$The Institute of Mathematical Sciences, \\
          Homi Bhabha National Institute (HBNI),\\
		 IV Cross Road, C.I.T. Campus, \\
	 Taramani, Chennai, India 600113

\vspace{.8cm}

$^b$ Laboratoire de Physique de l'\'Ecole Normale Sup\'erieure \\ 
 \hskip -.05cm
 CNRS, ENS, Universit\'e PSL,  Sorbonne Universit\'e, \\
 Universit\'e de Paris 
 \hskip -.05cm F-75005 Paris, France	 
\vskip 0.8cm
	{\small
		E-mail:
		\texttt{sashok@imsc.res.in, jan.troost@ens.fr}
	}
\vspace*{0.8cm}
\end{center}

\begin{abstract}

 We revisit the calculation of the thermal free energy for string theory in three-dimensional anti-de Sitter spacetime with Neveu-Schwarz-Neveu-Schwarz flux. The path integral calculation is exploited to confirm the off-shell Hilbert space  and we find that the  Casimir of the discrete representations of the isometry group takes values in a half-open interval. 
 We extend the free energy calculation to the case of superstrings, calculate the boundary toroidal twisted partition function in the Ramond-Ramond sector, and prove lower bounds on the boundary conformal dimension from the bulk perspective. We  classify Ramond-Ramond ground states and construct their second quantized partition function. The partition function exhibits intriguing modular properties.

\end{abstract}

\end{titlepage}
\hypersetup{pageanchor=true}

\tableofcontents

\section{Introduction}
Holography is a conjectured property of quantum gravity. The claim is strongly substantiated in spacetimes with a negative cosmological constant by the anti-de Sitter/conformal field theory correspondence \cite{Maldacena:1997re}. The strongest support for the duality has been gathered in string theory backgrounds with extended supersymmetry. Observables that preserve more supersymmetry are typically under better calculational control.
In this paper we  concentrate on the holographic correspondence in three/two dimensions in a string theoretic framework. The string theory we study is exceptional in that all tree level higher order curvature corrections in the inverse string tension expansion are under exact control.

One of our goals  is to acquire a rigorous understanding of the Ramond-Ramond ground states of the boundary conformal field theory, from the bulk path integral. Since we compute a boundary partition function, the boundary manifold is a torus and we must compute the bulk amplitude in thermal anti-de Sitter space. 

To perform the calculation, we revisit the calculation of the free energy of bosonic string theory in thermal three-dimensional anti-de Sitter space-time  \cite{Maldacena:2000kv}. The spectrum of bosonic string theory in $AdS_3$ was determined in \cite{Maldacena:2000hw} and was confirmed through a path integral calculation on thermal $AdS_3$ \cite{Maldacena:2000kv}. We review the path integral calculation and confirm the off-shell Hilbert space \cite{Maldacena:2000hw} more directly. We find a half-open bound on the allowed values of the quadratic Casimir of the discrete representations of the $AdS_3$ isometry group. Moreover, we extend the free energy calculation to superstring theory and then apply it to the determination of all boundary Ramond-Ramond ground states. From the one-loop amplitude dressed with fugacities corresponding to global charges, we prove positive energy theorems and a BPS bound. Our calculation takes place in a thermally compactified global anti-de Sitter space-time, with supersymmetric boundary conditions along the circle.

The path integral calculation proves that the spectrum of boundary chiral primaries determined in \cite{Argurio:2000tb,Eberhardt:2017pty} is complete. Moreover, we perform the calculation in a manner that is manifestly space-time supersymmetric. We also analyze  the second quantized generating function of boundary chiral primaries. We note that the gapped spectrum of chiral ring elements leads to intriguing modular properties of the generating function. We   find that the second quantized partition function  matches proposals for the dual conformal field theory in the literature and argue that the second quantized partition function is the right quantity to consider in the perturbative Neveu-Schwarz-Neveu-Schwarz string background.

The paper is structured as follows. We revisit the calculation of the one-loop vacuum amplitude in Neveu-Schwarz-Neveu-Schwarz thermal $AdS_3$ \cite{Maldacena:2000kv} in section \ref{BosonicAdS3}. We extend the calculation to the model with world sheet supersymmetry in section \ref{SusyAdS3}.
Then, we apply the knowledge gained to the  $AdS_3 \times S^3 \times T^4$ solution in superstring theory in section \ref{AdS3S3M} and to $AdS_3 \times S^3 \times S^3 \times S^1$ in section \ref{AdS3S3S3S1}. Section \ref{SecondQuantized} is devoted to a thorough discussion of the second quantized ground state partition function, including its modular properties. We summarize our results and draw conclusions in section \ref{Conclusions}.

\section{Thermal Three-dimensional Anti-de Sitter}
\label{BosonicAdS3}
In this section we review the calculation of the thermal $AdS_3$ partition function in bosonic string theory with Neveu-Schwarz-Neveu-Schwarz flux \cite{Maldacena:2000kv}, and its relation to the spectrum of  string theory in $AdS_3$ \cite{Maldacena:2000hw}. See e.g. \cite{Petropoulos:1989fc} for earlier contributions. To the groundbreaking analyses in \cite{Maldacena:2000kv,Maldacena:2000hw}, we add the discussion of a feature that slightly modifies the description of the discrete spectrum.  We moreover draw attention to the fact that the treatment of the descendant states in the continuum remains partial. We also provide an informative off-shell description of the spectrum from the path integral perspective,    reproducing the analysis of \cite{Maldacena:2000hw}. This review with additions prepares the ground for the generalization to the superstring in later sections.

\subsection{The Single String Free Energy} 
We work in a perturbative bosonic string background $AdS_3 \times N$ with NSNS flux.
The $AdS_3$ factor 
is described by a sl$(2,\mathbb{R})$  Wess-Zumino-Witten conformal field theory on the world sheet. The Wess-Zumino term represents the NSNS flux. We write the  $AdS_3$ metric  in the form:
\be
{ds^2} =  {\kbos} \left( d\phi^2 + (dv+ vd\phi)(d\bar v + \bar v d\phi) \right) ~,
\ee
where we shall take the level $\kbos$ to be $\kbos= k+2$.\footnote{The notation is convenient because starting from section \ref{SusyAdS3}, we will almost exclusively use the level $k$ of the supersymmetric generalization.} We are interested in calculating the thermal free energy and as a consequence compactify the global time coordinate in $AdS_3$. In the coordinates at hand, this corresponds to the identification \cite{Maldacena:2000kv}:
\begin{equation}
\phi\sim\phi + \beta~.
\end{equation}
It is possible to supplement the toroidal compactification of the boundary with a simultaneous twist of the phase of the coordinate $v$, which corresponds to introducing a complexified temperature:
\be
v \sim e^{\ii \mu\beta}v~, \qquad \bar v~ e^{-\ii\mu\beta} \bar v~.
\ee
The parameter $\beta$ is the inverse temperature and $\mu$ is the (imaginary) chemical potential for the $AdS_3$ angular momentum or boundary spin. From the point of view of the boundary conformal field theory,  these parameters are naturally encoded in the complex space-time modular parameter $\tau_{\text{s.t.}}$,
\be
\label{taustdef}
\tau_{\text{s.t.}} = \frac{1}{2\pi}(\beta\, \mu + \ii \beta)~.
\ee
Our first goal is to compute the one loop amplitude for strings propagating on thermal $AdS_3$. Our calculation follows the original computation done in \cite{Maldacena:2000kv} and for some of the details we refer to the original treatment. The path integral is weighed by the  world sheet Wess-Zumino-Witten action that includes the NSNS two form flux $B_{(2)}$:
\be
S = \frac{\kbos}{\pi} \int d^2 z\left(\pa\phi\bar\pa\phi+(\pa\bar v + \bar v\pa\phi)(\bar \pa v + v\bar\pa \phi) \right)~. \label{WorldsheetAction}
\ee
We will consider the one-loop free energy in thermal $AdS_3$ such that the world sheet conformal field theory  lives on a torus as well. We denote the world sheet modular parameter by $\tau$. Because the thermal circle is topologically non-trivial, the embedding of the worldsheet into spacetime is characterized by a pair of winding numbers $(n,m)$. The thermal identification implies that the worldsheet fields satisfy the boundary conditions:
\be
\begin{aligned}
\phi(z+2\pi) &= \phi(z) + \beta n~, \qquad \phi(z+2\pi \tau) = \phi(z) + \beta m \, , \\
v(z+2\pi) & = e^{\ii n \mu\beta}v(z)~, \qquad v(z+2\pi \tau) = e^{\ii m \mu\beta}v(z)~.
\end{aligned}
\ee
It is important to note that the thermal circle is topologically non-trivial, while the angular direction is not.
The boundary conditions can be implemented by introducing the  function $\Phi_{n,m}$,
\be
\Phi_{n,m}(z,\tau) = \frac{\ii}{4\pi\tau_2}(z(n\bar \tau -m) -\bar z(n\tau-m))~,
\ee
and defining the worldsheet fields
\be
\phi = \hat\phi + \beta \Phi_{n,m}~, \qquad v = e^{\ii\mu\beta\Phi_{n,m}} \hat v~, \label{TwistedFields}
\ee
where the hatted fields are periodic on the world sheet. After substituting the fields  (\ref{TwistedFields}) into the action (\ref{WorldsheetAction}), we  perform the path integral over the world sheet  fields in the $AdS_3$ factor: 
\be
Z_{\text{sl(2)}}(\beta,\mu, \tau) = \int D\phi Dv D\bar v e^{-S}~.
\ee
The path integration follows the route laid out  in \cite{Ray:1973sb,Gawedzki:1991yu}, and leads to the result \cite{Maldacena:2000kv}
\be
\label{Znmsum}
\begin{aligned}
Z_{\text{sl(2)}}(\beta,\mu, \tau) = 
\frac{\b \sqrt{k}}{2\pi \sqrt{\tau_2}} \sum_{n,m} \frac{e^{-\frac{(k+2)}{4\pi\tau_2}\b^2|m-n\tau|^2 + \frac{2\pi}{\tau_2}(\text{Im} U_{n,m})^2}}
{|\theta_1(U_{n,m},\tau)|^2}  ~.
\end{aligned}
\ee
We used the notation  $\theta_1(\nu, \tau)$ for one of the Jacobi theta functions (see Appendix \ref{apptheta} for details) and the twist $U_{n,m}$ is a function of the thermal circle winding numbers $(n,m)$ as well as the spacetime modular parameter:
\be
\label{Unmdef}
U_{n,m} = \taust(n\bar\tau - m)~.
\ee
The world sheet $AdS_3$ partition function $Z_{\text{sl(2)}}$ is a crucial factor in the one loop vacuum amplitude of 
  bosonic string theory on the background $AdS_3\times N$. 
  We will choose our background to be critical, such that the central charge $c_{int}$ corresponding to the internal compact factor $N$ equals
$
c_{int} = 26-c_{\text{sl(2)}}= 23-\frac{6}{k}~.
$
The one loop vacuum amplitude  is calculated by integrating  the thermal $AdS_3$ partition function contribution, along with the factors from the ghosts and the internal space, over the fundamental domain $F_0$ of the worldsheet modular parameter $\tau$:
\be
Z(\b,\mu) = \int_{F_0}\frac{d^2\tau}{\tau_2^2}~ Z_{\text{sl(2)}}(\beta,\mu, \tau)~Z_{\text{gh}}\Zint~,
\ee
where the contribution $Z_{\text{gh}}$ of the ghost sector and $\Zint$  of the internal conformal field theory are:
\begin{align}
Z_{\text{gh}} &=\tau_2~ |\eta(\tau)|^4~,\\
\Zint &= (q\bar q)^{-\frac{\cint}{24}}\sum_h d(h,\bar h)~q^h {\bar q}^{\bar h}~.
\end{align}
To perform the calculation, we use the unfolding trick of \cite{Polchinski:1985zf}, in which the sum over the winding number $n$ in equation \eqref{Znmsum} is traded for a sum over  copies of the fundamental domain in the upper half plane strip between $\text{Re}(\tau) \in (-\frac{1}{2},\frac{1}{2}]$ \cite{Maldacena:2000kv}:
\be
\label{Zsumoverm}
\begin{aligned}
Z(\b, \mu) = \frac{\b\sqrt{k}}{2\pi} \int_0^{\infty}\frac{d\tau_2}{\tau_2^{\frac{3}{2}}} \int_{-\frac12}^{\frac12}d\tau_1~ \sum_{m=1}^{\infty}~e^{-\frac{km^2\b^2}{4\pi\tau_2}}~ \frac{|\eta(\tau)|^4}{|\theta_1(m\taust, \tau)|^2}~\Zint~.
\end{aligned}
\ee
One of our goals is to confirm the on-shell spectrum of strings propagating on $AdS_3$ \cite{Maldacena:2000kv}. This is made possible by identifying the one loop string partition function with the thermal free energy of the spacetime theory \cite{Polchinski:1985zf}:
\be
\label{ZFf}
\begin{aligned}
Z(\b, \mu) &= -\b\, F(\b, \mu)
~.
\end{aligned}
\ee
We note that the free energy $F$ consists of connected multi-particle contributions and can be written as: 
\begin{equation}
F(\b,\mu) =  \frac{1}{\beta}  \sum_{{\cal H}_1} \log (1-e^{-\beta E+\ii \b \mu J}) = 
-\sum_{{\cal H}_1}\sum_{m=1}^\infty \frac{1}{m \beta} e^{- m \beta E+\ii m\b\mu J}
= -\sum_{m=1} f(m \beta, \mu) \, .
\end{equation}
Here ${\cal H}_1$ is the one-particle Hilbert space, and $-f(\beta,\mu)$ is the single string contribution to the free energy \cite{Maldacena:2000kv}
\begin{equation}
f(\beta, \mu) = \frac{1}{\beta} \sum_{{\cal H}_1} e^{-  \beta E +\ii \b\mu J} \, ,
\end{equation}
where $J$ is the spacetime spin operator.
By comparing the equation \eqref{ZFf} with the string one loop vacuum amplitude \eqref{Zsumoverm}, we obtain an expression for (minus) the single string free energy $f$ \cite{Maldacena:2000kv}:
\be
f(\beta,\mu)=\frac{\sqrt{k}}{2\pi} \int_0^\infty \frac{d \tau_2}{\tau_2^{3/2}} \int_{-1/2}^{1/2} d \tau_1 
~e^{-k \frac{\beta^2}{ 4 \pi \tau_2}} \frac{|\eta(\tau)|^4 } 
{|\theta_1(\tau, \tau_{\text{s.t.}} )|^2}~ \Zint \, . 
\ee
Once the integral over the strip in the $\tau$-plane is performed, one can read off the on-shell spectrum of string theory in $AdS_3\times N$.

Here, we temporarily part ways with the analysis in  \cite{Maldacena:2000kv}. Before we analyze the on-shell content of the one-loop amplitude, we wish to determine an
expression for the one loop amplitude (or rather, the one loop integrand)
in which we can identify the off-shell Hilbert space of the theory. In order to perform this task, we import the technology developed for the cigar sl$(2,\mathbb{R})/$u$(1)$ coset conformal field theory 
\cite{Hanany:2002ev}. To render the off-shell Hilbert space manifest, it is useful to   disentangle the dependence of the one loop amplitude on the boundary modular parameter $\taust$ from its dependence on the world sheet modular parameter $\tau$. To that end, we insert the following expression for the identity:
\be
\label{ident1}
1= \tau_2 \int_0^1 d^2 s \sum_{v,w \in \mathbb{Z}}
\delta^2(\tau_{\text{s.t.}} - (s_1+w) \tau + s_2+v
)~,
\ee
where the holonomies $s_i$ take values in the half-open interval $[0,1)$.
We then use an integral representation of the $\delta$-function in order to isolate the exponential dependence on the spacetime modular parameter:
\be
\label{ident3}
\begin{aligned}
\delta(\tau_{\text{s.t.}}-(s_1\tau - s_2)- (w\tau - v)) &=\int d\lambda_1d\lambda_2 e^{2 \pi i \lambda_1 (\text{Re} (\tau_{\text{s.t.}}) - (s_1+w) \tau_1 + s_2 + v)} e^{2 \pi i \lambda_2 (\text{Im} (\tau_{\text{s.t.}}) - (s_1+w) \tau_2)} \\
&= \int d^2 \lambda e^{2 \pi i \lambda_1 (\frac{\mu \beta}{2 \pi}- (s_1+w) \tau_1 + s_2 + v)} e^{2 \pi i \lambda_2 (\frac{\beta}{2 \pi} - (s_1+w) \tau_2)} \, .
\end{aligned}
\ee
Plugging in the expressions for the identity and the $\delta$-function, and using the elliptic properties of the $\theta_1$-function (see Appendix \ref{apptheta} for details), we obtain another expression for the single string free energy $f(\beta,\mu)$:
\be
\begin{aligned}
f(\beta,\mu) =&   \frac{\sqrt{k}}{2\pi} \int_0^\infty \frac{d \tau_2}{\tau_2^{1/2}} \int_{-1/2}^{1/2} d \tau_1 \int_0^1 d^2 s_i  \sum_{v,w} \int d^2 \lambda
e^{2 \pi i \lambda_1 (\frac{\mu \beta}{2 \pi}- (s_1+w) \tau_1 + s_2 + v)} e^{2 \pi i \lambda_2 (\frac{\beta}{2 \pi} - (s_1+w) \tau_2)} \\
& \hspace{4cm}
e^{-(k+2) (2s_1 w+w^2) \pi \tau_2  } 
e^{ -k \pi \tau_2 s_1^2} ~
\frac{|\eta(\tau)|^4}{|\theta_1( s_1\tau - s_2,\tau )|^2}
~ \Zint\, .
\end{aligned}
\ee
We use the expansion of the inverse $\theta$-function valid when $|q|<|e^{2 \pi i \nu}|<1$ (see e.g. \cite{Pakman:2003kh,Ashok:2013zka}) to write
\be
\begin{aligned}
\frac{1}{\theta_1(\nu, \tau)} &= i \frac{z^{1/2}}{\eta(\tau)^3} \sum_{r \in \mathbb{Z}} z^r S_r(q) \, ,
\end{aligned}
\ee where we introduced the fugacity $z=e^{2\pi\ii\nu}$ as well as the special series $S_r$:
\be
S_r(q) = \sum_{n =0}^{\infty} (-1)^n q^{\frac{n}{2}(n+2r+1)}~.
\ee
The single string free energy $f$ reads: 
\be
\label{bifurcate}
\begin{aligned}
f(\beta,\mu) =&  \int_0^\infty \frac{d \tau_2}{\sqrt{\tau_2}} \int_{-1/2}^{1/2} d \tau_1 \int_0^1 ds_1 \int_0^1 ds_2\sum_{v,w} \int d^2 \lambda e^{2 \pi i \lambda_1 (\frac{\mu \beta}{2 \pi}- (s_1+w) \tau_1 + s_2 + v)} e^{2 \pi i \lambda_2 (\frac{\beta}{2 \pi} - (s_1+w) \tau_2)} \\
&\frac{\sqrt{k}}{2\pi|\eta(\tau)|^2} e^{-(k+2) (2s_1 w+w^2) \pi \tau_2-2\pi\tau_2 s_1  -k \pi \tau_2 s_1^2}\sum_{r,\bar r} e^{2\pi\ii r(s_1\tau-s_2)}~ e^{-2\pi\ii \bar r(s_1\bar\tau-s_2)}~ S_r~ S_{\bar r}~\Zint\, .
\end{aligned}
\ee

\subsection{The Off-shell Hilbert Space}
\label{BosonicOffshell}

At this point there are two ways to proceed. One way is to undo some of the holonomy integrals we  introduced, and that route will rejoin paths with the calculation of the  on-shell single string free energies  \cite{Maldacena:2000kv}. This will be the subject of subsection \ref{bosoniconshell}. However,  from the formula (\ref{bifurcate}) we are also able to 
 derive the off-shell Hilbert space of bosonic string theory on $AdS_3$ obtained in \cite{Maldacena:2000hw}, but now by purely path integral methods. With this in mind we introduce the Gaussian integral \cite{Hanany:2002ev}
\be
e^{ -k \pi \tau_2 s_1^2} = \sqrt{\frac{\tau_2}{k }}\int dc\ e^{-\frac{\pi \tau_2}{k}c^2 -2\pi\ii c\tau_2 s_1}~, 
\ee
to rewrite:
\be
\begin{aligned}
f(\beta,\mu) =& \int_0^\infty  \frac{d \tau_2}{2\pi} \int_{-1/2}^{1/2} d \tau_1\int_0^1 ds_1  \int_0^1 ds_2 \sum_{v,w} \int d^2 \lambda e^{2 \pi i \lambda_1 (\frac{\mu \beta}{2 \pi}- (s_1+w) \tau_1 + s_2 + v)} e^{2 \pi i \lambda_2 (\frac{\beta}{2 \pi} - (s_1+w) \tau_2)} \\
&\hspace{2cm} \int dc\  e^{-\frac{\pi\tau_2}{k}c^2 -2\pi\ii c \tau_2s_1}\  e^{-(k+2) (2s_1 w+w^2) \pi \tau_2-2\pi\tau_2 s_1  }  \\
&\hspace{3cm} \frac{\Zint}{|\eta(\tau)|^2}\sum_{r,\bar r} e^{2\pi\ii r(s_1\tau-s_2)}~ e^{-2\pi\ii \bar r(s_1\bar\tau-s_2)}~ S_r~ S_{\bar r}\, .
\end{aligned}
\ee
We first perform the sum over the integer $v$ to obtain the Dirac comb:
\be
\label{DCL}
\sum_v e^{2\pi\ii v\lambda} = \sum_{n\in\mathbb{Z}}\delta(\lambda_1-n)~.
\ee
The integral over the holonomy $s_2$
 gives the constraint 
$
\lambda_1 = r-\bar r \in \mathbb{Z}~.
$
Combined with the Dirac comb  \eqref{DCL}, this leads to a trivial integral over the multiplier $\lambda_1$. After these three steps we are left with  
\be
\begin{aligned}
f (\beta, \mu)=&  \frac{1}{2\pi} \int_0^\infty d \tau_2 \int_{-\frac{1}{2}}^{\frac{1}{2}} d \tau_1 \int_0^1 d s_1 \sum_{w,r,\bar r} \int d \lambda_2\ e^{2 \pi i (r-\bar r) (\frac{\mu \beta}{2 \pi}- w \tau_1 )} e^{2 \pi i \lambda_2 (\frac{\beta}{2 \pi} - (s_1+w) \tau_2)} \\
& \int dc\ e^{-\frac{\pi\tau_2}{k}c^2 -2\pi\ii c\tau_2 s_1}\  e^{-(k+2) (2s_1 w+w^2) \pi \tau_2  }   ~\frac{ \Zint}{|\eta(\tau)|^2}e^{-2\pi\tau_2 s_1} e^{-2\pi\tau_2s_1(r+\bar r)}~ S_r~ S_{\bar r}\, .
\end{aligned}
\ee
We  perform the $s_1$ holonomy integral:
\be
\begin{aligned}
\int_0^1 ds_1 e^{-2\pi \tau_2 s_1\left(\ii \lambda_2  +\ii c +(k+2)w+(1+r+\bar r) \right)}
= -\frac{1}{2\pi\tau_2}\frac{e^{-2\pi \tau_2 \left(\ii \lambda_2  +\ii c +(k+2)w+(1+r+\bar r) \right)} - 1}{(\ii \lambda_2  +\ii c +(k+2)w+(1+r+\bar r))}~.
\end{aligned}
\ee
We therefore have 
\be
\begin{aligned}
f (\beta, \mu)&=  \frac{1}{2\pi} \int_0^\infty \frac{d \tau_2}{\tau_2} \int_{-\frac{1}{2}}^{\frac{1}{2}} d \tau_1  \sum_{w,r,\bar r} \int d \lambda_2\ e^{2 \pi i (r-\bar r) (\frac{\mu \beta}{2 \pi}- w \tau_1 )} e^{2 \pi i \lambda_2 (\frac{\beta}{2 \pi} - w \tau_2)} \\
& \int \frac{dc}{2\pi}\, e^{-\frac{\pi\tau_2}{k}c^2 }\  e^{-(k+2) w^2 \pi \tau_2  }   ~ \frac{(1-e^{-2\pi \tau_2 \left(\ii \lambda_2  +\ii c +(k+2)w+(1+r+\bar r) \right)}) }{(\ii \lambda_2  +\ii c +(k+2)w+(1+r+\bar r))}~\frac{ \Zint}{|\eta(\tau)|^2} S_r~ S_{\bar r}\, .
\label{TwoTerms}
\end{aligned}
\ee
Next, we relate the  two terms  that arose out of the $s_1$ holonomy integration. Consider the term proportional to the exponential and in particular factors that depend on the integration variable $c$: 
\be
\begin{aligned}
\int dc\ \frac{e^{-\frac{\pi\tau_2}{k}c^2-2\pi \ii \tau_2 c}}{(\ii \lambda_2  +\ii c +(k+2)w+(1+r+\bar r))}
= e^{-\pi k\tau_2}\int dc\ \frac{e^{-\frac{\pi\tau_2}{k}(c+ \ii k)^2}}{(\ii \lambda_2  +\ii c +(k+2)w+(1+r+\bar r))}~.
\end{aligned}
\ee
We  shift the variable $c\rightarrow c-\ii k$ and as a result pick up poles in addition to the new line integral over the  variable $c$. These poles are located at
\be
\label{polelocations}
c = -\lambda_2 + \ii (k+2)w + \ii(1+r+\bar r)~.
\ee
From (analogous yet different) work on the coset theory  \cite{Hanany:2002ev}, we surmise that the resulting residues are the contributions from the discrete representations to the partition function. These will soon be our primary focus.

\subsubsection{The Continuous Representations}

First however, we show that the shifted exponential term combines nicely with the ``1" term in \eqref{TwoTerms} to give the contribution from the continuous spectrum of the theory, modulo  important subtleties. After shifting the integral, the expression for the single string free energy takes the  form: 
\begin{align}
\label{fexp1}
f (&\beta, \mu)=  \frac{1}{2\pi} \int_0^\infty \frac{d \tau_2}{\tau_2} \int_{-\frac{1}{2}}^{\frac{1}{2}} d \tau_1  \sum_{w,r,\bar r} \int d \lambda_2\ e^{2 \pi i (r-\bar r) (\frac{\mu \beta}{2 \pi}- w \tau_1 )} e^{2 \pi i \lambda_2 (\frac{\beta}{2 \pi} - w \tau_2)}~\int\frac{ dc}{2\pi}\ e^{-\frac{\pi\tau_2}{k}c^2 }  \cr
&e^{-(k+2) w^2 \pi \tau_2  }   
\left(\frac{1}{\scriptstyle \ii \lambda_2  +\ii c +(k+2)w+(1+r+\bar r)} -\frac{e^{-\pi k\tau_2}e^{-2\pi \tau_2 \left(\ii \lambda_2   +(k+2)w+(1+r+\bar r) \right)} }{\scriptstyle \ii \lambda_2  +\ii (c -\ii k)+ (k+2)w+(1+r+\bar r)}\right)\frac{ \Zint}{|\eta(\tau)|^2} S_r~ S_{\bar r}~.
\end{align}
In the second term in parenthesis, we   perform a shift in the integer parameters:
\be
w\rightarrow w -1~,\qquad r \rightarrow r +1~, \qquad \bar r \rightarrow \bar r +1~. \label{BosonicFlow}
\ee
This makes the denominator factor in the second term  identical to that of the first. In addition, most of the exponential factors get cancelled. After some algebra, we obtain
\begin{align}
f (\beta, \mu)=&  \frac{1}{2\pi} \int_0^\infty \frac{d \tau_2}{\tau_2} \int_{-\frac{1}{2}}^{\frac{1}{2}} d \tau_1  \sum_{w,r,\bar r} \int d \lambda_2\ e^{2 \pi i (r-\bar r) (\frac{\mu \beta}{2 \pi}- w \tau_1 )} e^{2 \pi i \lambda_2 (\frac{\beta}{2 \pi} - w \tau_2)}  \\
&
\int\frac{ dc}{2\pi}\ \frac{e^{-\frac{\pi\tau_2}{k}c^2 }e^{-(k+2) w^2 \pi \tau_2  }}{\ii \lambda_2  +\ii c +(k+2)w+(1+r+\bar r)} ~   
\frac{ \Zint}{|\eta(\tau)|^2}~ \big(S_r S_{\bar r} - q^{r+1}\bar q^{\bar r+1}S_{r+1}S_{\bar r+1}\big)~. \nonumber
\end{align}
We  use the special series identities
\be
\label{Sids}
q^{r+1}S_{r+1}= S_{-r-1} \quad\text{and}\quad S_r + S_{-r-1} = 1~,
\ee
to write:
\be
\begin{aligned}
f (\beta, \mu)&=  \frac{1}{2\pi} \int_0^\infty \frac{d \tau_2}{\tau_2} \int_{-\frac{1}{2}}^{\frac{1}{2}} d \tau_1  \sum_{w,r,\bar r} \int d \lambda_2\ e^{2 \pi i (r-\bar r) (\frac{\mu \beta}{2 \pi}- w \tau_1 )} e^{2 \pi i \lambda_2 (\frac{\beta}{2 \pi} - w \tau_2)}  \\
&
\int\frac{ dc}{2\pi}\ \frac{e^{-\frac{\pi\tau_2}{k}c^2 }e^{-(k+2) w^2 \pi \tau_2  }}{\ii \lambda_2  +\ii c +(k+2)w+(1+r+\bar r)}~   
\frac{\Zint}{|\eta(\tau)|^2} ~\big(1+(S_r-1) + (S_{\bar r} - 1)\big)~.
\end{aligned}
\label{Continuous}
\ee
In the context of the on-shell states (to be discussed shortly),  it was suggested in \cite{Maldacena:2000kv}  that the two terms we analyzed should recombine into continuous representations, and a proposal was made for the density of states in the continuum sector. See also \cite{Hanany:2002ev} for a related analysis in the cigar coset theory.   We observe that while the two terms combine well for the primary states (corresponding to the ``1" term in the parenthesis in equation (\ref{Continuous})), this is not obvious for the descendant states, whose contributions are encoded in the $(S_r-1)$ terms. A similar observation was made in previous work  on the sl$(2,\mathbb{R})/$u$(1)$ coset theory \cite{Israel:2004ir}, where an analysis of the continuous contributions was performed, and track was kept of the degeneracy of descendant states, as we did here. We would like to  draw attention to the fact that the analysis of a discrete (alternating spin chain) model of the cigar \cite{Ikhlef:2011ay,Bazhanov:2020uju} led to a spectral density of continuous representations that depends on their descent. We see that this may well be relevant for the analytic analysis of the continuum model. These  issues are interesting and non-trivial  -- we refer to \cite{Bazhanov:2020uju} for the integrable state of the art -- and they lie outside our main subject in this paper. We leave them for future research. We concentrate on the pole contributions to the single string free energy, namely on the discrete representations. 
\label{ContinuousSubtleties}
\label{Caveats}

\subsubsection{Contributions from the Discrete Spectrum}

\begin{figure}[ht]
\center{\includegraphics[width=\textwidth] {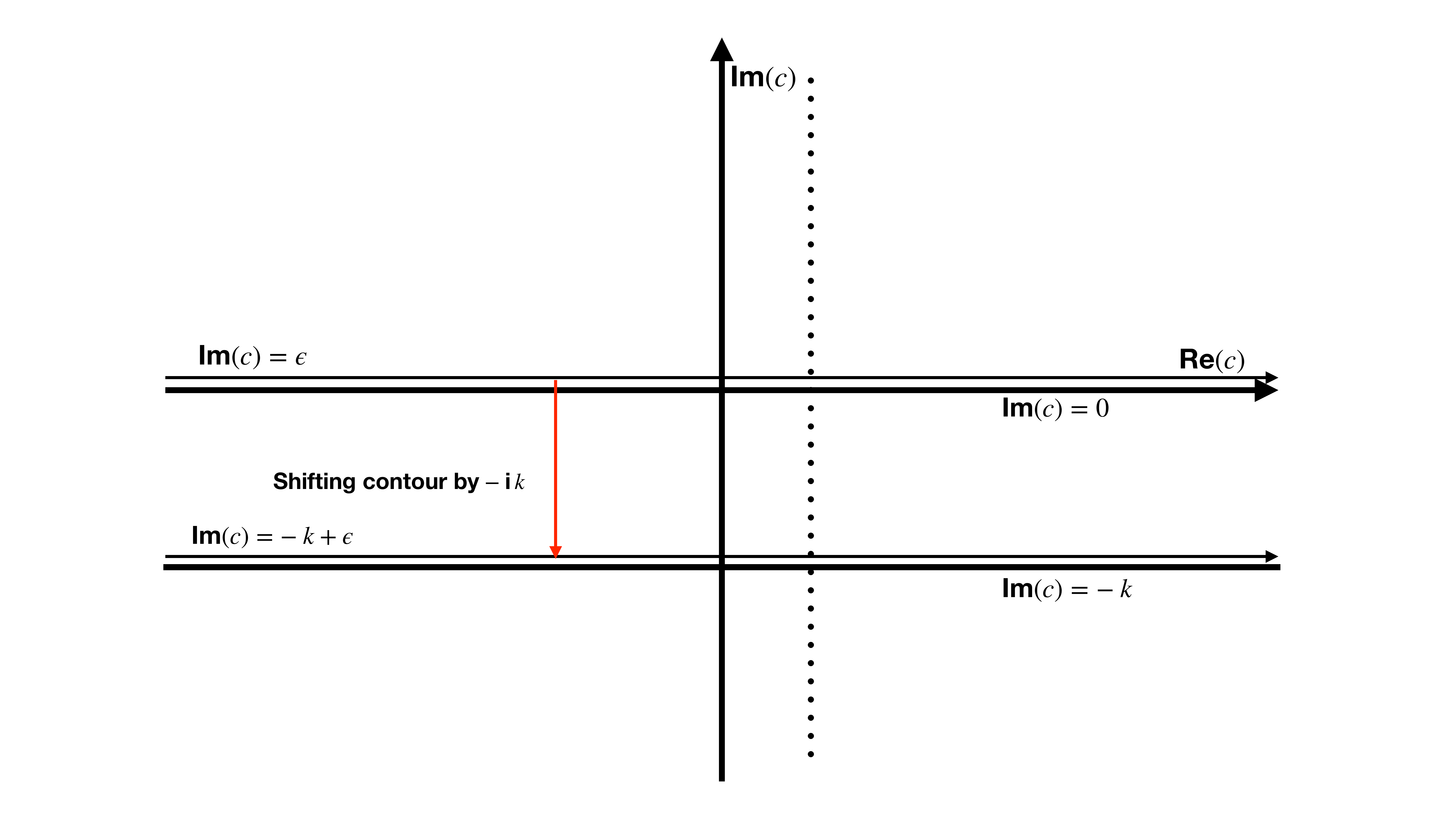}} 
 \caption{\label{fig:my-label} For a fixed value of the Casimir $\lambda_2$ we show the discrete set of poles in the momentum  $c$-plane.  When $\lambda_2=0$, there is a pole along the contour of integration. To avoid the singularity the contour in the $c$-plane is shifted by $\ii\, \epsilon$.
 Upon shifting the $c$-contour by $-\ii\, k$, we see that a finite number of poles is picked up. 
 }  
 \label{shiftbyk}
\end{figure}

We focus on the contributions from the poles that lie in the locations \eqref{polelocations}.  It is important to remember  that the shift in the $c$ integral was by an amount $-\ii k$ such that only the poles in the strip $\text{Im}(c)$ between $0$ and $-\ii k$ will be picked up. This is shown in Figure~\ref{shiftbyk}. 
The residue calculation gives
\be
\begin{aligned}
f_{\text{disc}}(\beta,\mu) =&\frac{1}{2\pi} \int \frac{d\tau_2}{\tau_2}\int_{-\frac{1}{2}}^{\frac{1}{2}}d\tau_1 \sum_{w,r,\bar r} ~
e^{2\pi\ii(r-\bar r)(\frac{\mu\beta}{2\pi} - w\tau_1)}e^{-(k+2)w^2  \pi \tau_2  } 
\\
&\int' \ii\, d\lambda_2 \sum_{\text{poles}(\lambda_2,w,r,\bar r)}e^{\frac{\pi\tau_2}{k}(\ii \lambda_2 + (k+2)w + 1+r+\bar r)^2 }e^{2\pi\ii\lambda_2(\frac{\beta}{2\pi} - w \tau_2)}\frac{ \Zint}{|\eta(\tau)|^2}~ S_r~ S_{\bar r}\, .
\end{aligned}
\ee
We now Wick-rotate the $\lambda_2$ integral onto the imaginary axis by replacing $\lambda_2= \ii \widetilde{\lambda}_2$ and then  shift the $\widetilde{\lambda}_2$ integral so as to get rid of the quadratic dependence on the $(r,\bar r)$ variables.
Moreover, we must pause to discuss an  ambiguity in the discrete contribution.
The variable $c$ is integrated over the real line which can contain a pole. The ambiguity lies in how we treat the pole. 
This ambiguity is intrinsic to the spectrum of the theory. To explain this, we run ahead of ourselves and preview that the variable $c=-i (2j-1)$ after the shift acquires the interpretation of the spin in terms of sl$(2,\mathbb{R})$ representation theory. We have a continuum spectrum labelled by $j=1/2+is$ and a positive momentum $s \in \mathbb{R}^+_0$, touching a spectrum $j > 1/2$ of discrete representations. The integration over the variable $c$ is akin to an (unfolded) integration over the variable $s$, and the integration bumps into the non-normalizable mock discrete representation at $j=1/2$.\footnote{A related subtlety has been understood  in the calculation of the cigar elliptic genus \cite{Troost:2010ud,Eguchi:2010cb,Ashok:2011cy}. In that context,  the partition function is finite, and it becomes entirely manifest that there is a choice in how to separate  continuous from discrete representations. The choice leads to interesting subtleties in the state space interpretation of the spectrum. See e.g. \cite{Giveon:2014hfa} for a discussion of the latter point.} In the following, we choose our integration contour such that we  ensure that the discrete spectrum respects the  bound $j>1/2$. In other words, we ban the possible pole contribution at $j=1/2$ to the continuous sector. To that end, we shift the initial $c$-contour up by $\epsilon$ as in Figure \ref{shiftbyk} such that we do not pick up the pole that lies on the final contour of integration. This choice influences on the continuous contribution.
At the end of these manipulations we find the discrete contribution to the single string partition function:
\be
\label{fdiscrestrictedrange}
\begin{aligned}
f_{\text{disc}}(\beta,\mu) =&\frac{1}{2\pi} \int \frac{d\tau_2}{\tau_2}\int_{-\frac{1}{2}}^{\frac{1}{2}}d\tau_1 \sum_{w,r,\bar r} ~
e^{2\pi\ii(r-\bar r)(\frac{\mu\beta}{2\pi} - w\tau_1)}e^{-(k+2)w^2  \pi \tau_2  } 
\\
&\int_{(0,k]}d\widetilde{\lambda}_2\sum_{\text{poles}(\lambda_2)} e^{\frac{\pi\tau_2}{k}\widetilde{\lambda}_2^2} e^{-2\pi(\widetilde{\lambda}_2+ (k+2)w + 1+r+\bar r) (\frac{\beta}{2\pi} - w \tau_2)}\frac{ \Zint}{|\eta(\tau)|^2}~ S_r~ S_{\bar r}\, .
\end{aligned}
\ee
We have indicated in the half-open integration region of $\widetilde{\lambda}_2$ that we exclude $\widetilde{\lambda}_2$ equal to zero and include $\widetilde{\lambda}_2=k$.

In this discrete sector, we can now render the off-shell Hilbert space of string theory on $AdS_3\times N$ manifest. We first summarize the contributions from the descendants of the $AdS_3$ factor, the internal conformal field theory on $N$ and the ghost sector  as:
\be
\label{compactplus}
\begin{aligned}
 \frac{ \Zint}{|\eta(\tau)|^2}~ S_r~ S_{\bar r} &= (q\bar q)^{-\frac{1}{24}(1+\cint)}\sum_{N, h,\bar N, \bar h} d_{r,h,N}~q^{h+N}~ q^{\bar h+\bar N}\\
 &=(e^{4\pi\tau_2})^{(1-\frac{1}{4k}) }\sum_{N, h,\bar N, \bar h}d_{r,h,N}~ q^{h+N}~ q^{\bar h+\bar N}~.
\end{aligned}
\ee
To declutter the formulae we label the function that counts the degeneracy of states and also the summation variables by only the unbarred variables. In addition we introduce the spin $j$:
\be
\widetilde{\lambda}_2= 2j-1~,
\ee
and separate the $\tau_2$ and $\tau_1$ dependences in the exponent in the expression for the free energy:
\begin{align}
f_{\text{disc}}(\beta,\mu) &=\frac{1}{\pi} \int \frac{d\tau_2}{\tau_2}\int_{-\frac{1}{2}}^{\frac{1}{2}}d\tau_1 \sum_{w,r,h,N} ~d_{r,h,N}~
e^{2\pi\ii\tau_1( (N +h) - w r - (\bar N + \bar h)+w \bar r) }
\cr
&\int_{ (\frac12, \frac{k+1}{2}]}dj(e^{2\pi\tau_2})^{(k+2)\frac{w^2}{2} +\frac{2j(j-1)}{k} + (2j+r+\bar r)w - (N+h+\bar N + \bar h) +2}\,\qst^{j+ (k+2)\frac{w}{2} + r} {\bar q}_{\text{s,t.}}^{j + (k+2)\frac{w}{2} + \bar r} . \nonumber
\end{align} 
We have used the definition (\ref{taustdef}) of the modular parameter of the boundary torus $\tau_{\text{s.t.}}$   and introduced the elliptic nome:
\be
\qst = e^{2\pi\ii\tau_{\text{s.t.}}} = e^{-\beta +\ii\mu \beta}~.
\ee
Finally, we identify the so$(2) \oplus$ so$(2) \subset$ sl$(2,\mathbb{R}) \oplus$ sl$(2,\mathbb{R})$ quantum numbers: 
\be
\label{ads3ids}
m = j+ r ~, \quad \text{and} \quad \bar m = j + \bar r~,
\ee
which allows us to rewrite the discrete contribution to the single string free energy in the expected form:
\be
\label{fdiscL0s}
\begin{aligned}
f_{\text{disc}}(\beta,\mu) =&
 \int \frac{d\tau_2}{\tau_2}\int_{-\frac{1}{2}}^{\frac{1}{2}}d\tau_1  
  \sum_{w,r,h,N} ~d_{r,h,N}~\int_{ (\frac12, \frac{k+1}{2}]}\frac{dj}{\pi}~
 e^{2\pi\ii\tau_1( L_0-\bar{L}_0) }
~(e^{-2\pi\tau_2})^{L_0+\bar{L}_0 - 2}\,\qst^{j_0^3} {\bar q}_{\text{s,t.}}^{\bar{j}_0^3} .
\end{aligned}
\ee
The $\tau_1$-integral imposes the level-matching constraint and we have identified the coefficient of $-2\pi\tau_2$ in the exponent with the zero mode of the worldsheet scaling operator $L_0 + \bar L_0 -2$, which includes the conformal dimensions of the discrete series and their descendants, as well as the contributions from the internal space \cite{Maldacena:2000hw}:
\begin{align}
\label{ads3virasoro}
L_0 &= -\frac{j(j-1)}{k}-w m -(k+2)\frac{w^2}{4} + N+h \, , \\
\bar{L}_0&=  -\frac{j(j-1)}{k}-w\bar m  -(k+2)\frac{w^2}{4} + \bar N+ \bar h ~.
\end{align}
The exponent of the spacetime modular parameter instead is given by the zero mode of the left and right moving so$(2)$ currents $j^3$ and $\bar{j}^3$ of sl$(2,\mathbb{R})$. They measure the energy and momentum in spacetime, or the left- and right-moving spacetime conformal dimensions. Explicitly, the zero modes evaluate as \cite{Maldacena:2000hw}:
\be
j_0^3 = \frac{k+2}{2}w + m~,\qquad \bar{j}_0^3 = \frac{k+2}{2}w + \bar m~.
\ee
A similar story holds in the sector of the continuous representations.
Thus, we demonstrated explicitly  that the path integral of \cite{Maldacena:2000kv}  codes the off-shell spectrum  described in \cite{Maldacena:2000hw}.

Finally, because we kept track of the degeneracy of the descendants states,  we can rewrite the single string contribution to the free energy entirely in terms of the worldsheet characters in the discrete representation.  The discrete characters $\chi_j^w$ in the $w$ spectrally flowed representations are (see Appendix \ref{discchar}):
\begin{equation}
\begin{aligned}
\chi_{j}^w (q,\qst) 
&=\frac{q^{-\frac{j(j-1)}{k} - (k+2) \frac{w^2}{4} -\frac{1}{4k} } \qst^{(k+2)\frac{w}{2}}}{\eta^3(\tau)}
\sum_{r}   q^{-w(j+r)} ~ \qst^{r+j}S_r~.
\end{aligned}
\end{equation}
Thus, the discrete contributions to the single string free energy are transparent:
\begin{equation}
\label{fdiscfinal}
f_{\text{disc}}(\beta,\mu) =   
\int_0^\infty \frac{d \tau_2}{\tau_2^2} \int_{-\frac{1}{2}}^{\frac{1}{2}} d \tau_1  \sum_{w} \int_{ (\frac12, \frac{k+1}{2}]} \frac{ dj}{\pi} \ 
  \chi^w_{j} (q,q_{s.t.})\, \chi^w_{j} (\bar{q},\bar{q}_{s.t.}) ~Z_{\text{gh}}~\Zint \, .
  \end{equation}
We have refined and bridged the results of \cite{Maldacena:2000hw, Maldacena:2000kv}. We stress that we include all spectrally flowed discrete representations $D_j^{+,w}$, with $j$ in the half-open range
\be
\frac12 < j \le\frac{k+1}{2}~.
\ee
Indeed, the spectral flow argument of \cite{Maldacena:2000hw} or equivalently, the joining of the two terms in (\ref{TwoTerms}) through a spectral flow operation (\ref{BosonicFlow}) naturally lead to a half-open interval for the discrete spin. It also meshes well with the calculation of the cigar elliptic genus \cite{Troost:2010ud} as well as the regularized cigar partition function \cite{Bazhanov:2020uju}.
 
\subsection{The Free Energy of the On-shell States}
\label{bosoniconshell}
In this section, we review and  update the derivation of the  on-shell contribution to the single string free energy \cite{Maldacena:2000kv}. Our main task is to efficiently perform the  $\tau_2$-integral in equation \eqref{fdiscL0s}. To that end, we return to our point of on-shell/off-shell bifurcation, equation (\ref{bifurcate}):
\be
\begin{aligned}
f(\beta,\mu) =&  \int_0^\infty \frac{d \tau_2}{\sqrt{\tau_2}} \int_{-\frac{1}{2}}^{\frac{1}{2}} d \tau_1 \int_0^1 ds_1\int_0^1ds_2 \sum_{v,w} \int d^2 \lambda e^{2 \pi i \lambda_1 (\frac{\mu \beta}{2 \pi}- (s_1+w) \tau_1 + s_2 + v)} e^{2 \pi i \lambda_2 (\frac{\beta}{2 \pi} - (s_1+w) \tau_2)} \\
& \frac{\sqrt{k}}{2\pi|\eta(\tau)|^2} e^{-(k+2) (2s_1 w+w^2) \pi \tau_2-2\pi\tau_2 s_1  -k \pi \tau_2 s_1^2}\sum_{r,\bar r} e^{2\pi\ii r(s_1\tau-s_2)}~ e^{-2\pi\ii \bar r(s_1\bar\tau-s_2)}~ S_r~ S_{\bar r}~\Zint~\, .
\end{aligned}
\ee
We perform the $s_1$ holonomy integral  and the $\lambda_2$ integration first. This leads to a delta-function constraint for the variable $s_1$:
\be
\tau_2 s_1 = \frac{\beta}{2\pi} - w\tau_2~.
\label{s1Constraint}
\ee
The $s_1$ variable is  a $U(1)$ holonomy variable on the torus \cite{Hanany:2002ev}. Therefore, it is periodic. It takes values in a half-open interval $[0,1)$. 
We note that the constraint (\ref{s1Constraint}) forces the winding number $w$ to be positive. 
The integration range of $s_1$ 
also leads to a constraint on the range of $\tau_2$ integration:
\be
\frac{\b}{2\pi(w+1)} < \tau_2 \le \frac{\b}{2\pi w}~.
\ee
The range $(\frac{\beta}{2 \pi},\infty)$ is covered by the zero winding sector.
Performing the $s_1$ and $\lambda_2$ integrals, we thus obtain
\begin{multline}
f(\beta,\mu) = \sum_{w \ge 0,r,\bar r} \frac{\sqrt{k}}{2\pi} \int_{\frac{\b}{2\pi(w+1)}}^{ \frac{\b}{2\pi w}} \frac{d \tau_2}{\tau_2^{\frac32}} \int_{-\frac{1}{2}}^{\frac{1}{2}} d \tau_1 \int_0^1ds_2  \int d \lambda_1 \sum_v e^{2 \pi i \lambda_1v} e^{2\pi\ii\lambda_1 ( \frac{\mu \beta}{2 \pi}-w\tau_1) } e^{2\pi\ii s_2(\lambda_1-r+\bar r)}  \\
 e^{2\pi\tau_2 w^2-2w\b} e^{-2\pi(\frac{\b}{2\pi} - w\tau_2)(1+r+\bar r)}~e^{-\frac{k\b^2}{4\pi\tau_2}}
~\frac{\Zint}{|\eta(\tau)|^2}~ S_r~ S_{\bar r}\, .
\end{multline}
We  sum over the integer $v$, leading to a Dirac comb, and integrate over the variable $s_2$. That gives rise to the familiar constraint $\lambda_1 = r-\bar r$, which in term leads to a trivial integration over the multiplier $\lambda_1$. The result of these steps is:
\begin{multline}
f(\beta,\mu) =   \sum_{w,r,\bar r}\frac{\sqrt{k}}{2\pi} \int_{\frac{\b}{2\pi(w+1)}}^{ \frac{\b}{2\pi w}} \frac{d \tau_2}{\tau_2^{\frac32}} \int_{-\frac{1}{2}}^{\frac{1}{2}} d \tau_1   e^{2\pi\ii(r-\bar r) (\frac{ \mu \beta}{2\pi}  -w\tau_1) }  \\
e^{2\pi\tau_2 (w^2+w(1+r+\bar r))} e^{-\b(2w+1+r+\bar r)} ~e^{-\frac{k\b^2}{4\pi\tau_2}}
~\frac{\Zint}{|\eta(\tau)|^2}~ S_r~ S_{\bar r}\, .
\end{multline}
We  use the shorthand \eqref{compactplus} for the other sectors:
\be
\begin{aligned}
 ~\frac{\Zint}{|\eta(\tau)|^2}~ S_r~ S_{\bar r}=(e^{4\pi\tau_2})^{(1-\frac{1}{4k}) }\sum_{N, h,\bar N, \bar h} d_{r,h,N}~q^{h+N} q^{\bar h+\bar N}~,
\end{aligned}
\ee
and obtain the   single particle free energy:
\begin{multline}
f(\beta,\mu) =  \sum_{w,r,h,N} ~d_{r,h,N}~\frac{\sqrt{k}}{2\pi} \int_{\frac{\b}{2\pi(w+1)}}^{ \frac{\b}{2\pi w}} \frac{d \tau_2}{\tau_2^{\frac32}} \int_{-\frac{1}{2}}^{\frac{1}{2}} d \tau_1   e^{2\pi\ii \tau_1(h+N-rw - (\bar h +\bar N -\bar r w)) }  \\
e^{-2\pi\tau_2(h+N + \bar h + \bar N- (w^2+w(1+r+\bar r)) +\frac{1}{2k} -2)  } e^{\ii \mu \b (r-\bar r)-\b(2w+1+r+\bar r)} ~e^{-\frac{k\b^2}{4\pi\tau_2}}~.
\end{multline}
We now parallel  the analysis in \cite{Maldacena:2000kv}.  We introduce the  Gaussian integral:
\be
\label{MOSintegral}
e^{-\frac{k\b^2}{4\pi\tau_2}} = -\frac{8\pi\ii}{\b} \left( \frac{\tau_2}{k} \right)^{\frac{3}{2}}\int~dc~ c~e^{-\frac{4\pi\tau_2}{k}c^2 + 2\ii\beta c}~,
\ee
and obtain 
\be
\begin{aligned}
f(\beta,\mu) =&   \frac{4}{\ii k\b} \sum_{w,r,h,N} ~d_{r,h,N}~\int_{\frac{\b}{2\pi(w+1)}}^{ \frac{\b}{2\pi w}} d \tau_2 \int_{-\frac{1}{2}}^{\frac{1}{2}} d \tau_1   e^{2\pi\ii \tau_1(h+N-rw - (\bar h +\bar N -\bar r w)) }  \\
& 
\int_{-\infty}^{\infty}~dc~ c~e^{-2\pi\tau_2(h+N + \bar h + \bar N- (w^2+w(1+r+\bar r)) +\frac{4c^2+1}{2k} -2)  } e^{2\ii\b c+ \ii \mu \b (r-\bar r)-\b(2w+1+r+\bar r)} ~.
\end{aligned}
\ee
The integral over $\tau_1$  imposes level-matching.
As long as the $\tau_2$ range is finite, we can also perform that integral  \cite{Maldacena:2000kv}. In the winding zero sector, the range of the $\tau_2$ integral is half-infinite. If the coefficient  that multiplies $-2 \pi \tau_2$ in the exponent is positive, the integral is well-defined. If not, we shall define it by analytic continuation. However, we do wish to avoid that the coefficient becomes zero, and therefore perform a Feynman regularization, replacing $c^2 \rightarrow c^2-i \epsilon$.
The $\tau_2$ integration can then be performed, and we
end up with an integral over the  radial momentum $c$:
\be
\begin{aligned}
f(\beta,\mu) =&   \frac{1}{\pi\ii \b } \sum'_{w,r,h,N} ~d_{r,h,N}~
\int_{-\infty}^{\infty}~dc~ c~\frac{ e^{2\ii\b c+ \ii \mu \b (r-\bar r)-\b(2w+1+r+\bar r)}}{(c^2+\frac14-i \epsilon)+k(h+N - r w  - \frac{1}{2} (w^2+w) -1) } \\
&\hspace{.8cm} \left( e^{-\frac{\b}{w+1} (2( h+(N-rw))  - (w^2+w) +\frac{4c^2+1-4 i \epsilon}{2k} -2)}-e^{-\frac{\b}{w} (2( h+(N-rw))  - (w^2+w) +\frac{4c^2+1-4 i \epsilon}{2k} -2)}\right)   ~.
\end{aligned}
\ee
As in the off-shell calculation, we wish to combine the two terms in the parenthesis and associate the result to the continuous part of the spectrum \cite{Maldacena:2000kv}. To that end, we  shift 
the contour in the first term in the parenthesis  upwards along the imaginary axis by $\frac{k(w+1)}{2}$ while the contour in the second term is shifted upwards by $\frac{k w}{2}$. The shifted contours  combine and can be written in terms of the continuous representations, up to the issues discussed in subsection \ref{ContinuousSubtleties}.

We do  need to address a new point. The summation over  the winding $w $ is only over positive numbers $w \ge 0$. Moreover, for the first term that depends on a denominator $w+1$ (in the exponent), the summation is from $w=0$ to infinity, but the second term depending on the denominator $w$ is absent for $w=0$. Recall that we shift the first term by a larger amount than the second term. Thus, all poles that we pick up are cancelled in this contour manipulation, except for the poles of the first term in the range where the imaginary part of $c$ is between $k(w+1)/2$ and $kw/2$. Thus, the discrete pole contributions are present for all $w \ge 0$. However  after recombination and the shift $w+1 \rightarrow w$ in the first term, we only sum over $w \ge 1$ in the continuous sector. Thus, we have established what happens at the boundary of the summation range for $w$.

Let us study in more detail the net set of poles that we pick up when we shift both contours.
The shift of the integrals  lead to additional contributions from possible poles that are  located at 
\be
-c^2 + i \epsilon= \frac{1}{4} + k( N-r w + h -1-\frac{1}{2}(w^2+w)) ~. \label{OnshellPoles}
\ee
The Feynman regularization excludes poles at real $c$. In particular, we note that for zero winding, it excludes a possible pole on the real axis (that will soon turn out to correpond to the representation with spin $j=1/2$).  By the shift of the contours, 
only those poles are included that satisfy the constraint:
\be
\frac{kw}{2} < \text{Im}(c) \le \frac{k(w+1)}{2} ~.
\label{ConstraintOnImaginaryPart}
\ee
Note that the right hand side of equation (\ref{OnshellPoles}) is discrete for a compact manifold $N$ and that therefore the spectrum of on-shell poles we pick up is discrete.\footnote{This contrasts with the off-shell calculation of subsection \ref{BosonicOffshell}.}  Finally, 
the sum over residues gives the contribution of the discrete states in the on-shell free energy on $AdS_3\times M$ \cite{Maldacena:2000hw}: 
\be
\begin{aligned}
 f_{\text{disc}}(\beta,\mu) =\frac{1}{\beta}  \sum'_{w,r,h,N} ~d_{r,h,N}~e^{-\b\big ( -\ii \mu (r-\bar r)+(2w+1+r+\bar r)+\sqrt{1+4k(N+h-r w -1 -\frac{1}{2}(w^2+w))} \big)  }~.
\end{aligned}
\ee
Let us rewrite this in a more insightful manner \cite{Maldacena:2000hw}. The integration over the $\tau_1$ and $\tau_2$ variables imposed both the level matching and the on-shell condition. Using the explicit form of the worldsheet Virasoro generators  \eqref{ads3virasoro}, this implies 
\be
\label{onshellbos}
-\frac{j(j-1)}{k}-w (r+j) -(k+2)\frac{w^2}{4} + N+h = 1~.
\ee
Solving for the spin $j$, we find the relation
\be
2j-1+kw = \sqrt{1+4k(N+h-r w -1 -\frac{1}{2}(w^2+w))}~.
\label{SolutionForSpin}
\ee
Recognizing the square root as the one appearing in the exponents in the free energy, we substitute this in the single string free energy to find
\be
\begin{aligned}
 f_{\text{disc}}(\beta,\mu) &=\frac{1}{\beta}  \sum'_{w,r,h,N} ~d_{r,h,N}~e^{-\b\big ( -\ii \mu (r-\bar r)+ 2j+(k+2)w + r+\bar r\big)}\\
 &=\frac{1}{\beta}  \sum'_{w,r,h,N} ~d_{r,h,N}~\qst^{j+r+\frac{(k+2)w}{2}}
 {\bar q}_{\text{s.t.}}^{j+\bar r+\frac{(k+2)w}{2}} \\
 &=\frac{1}{\beta}  \sum'_{w,r,h,N} ~d_{r,h,N}~\qst^{j_0^3}
 {\bar q}_{\text{s.t.}}^{{\bar j}_0^3}~.
\end{aligned}
\ee
The sum over is over the on-shell states. From the equation (\ref{SolutionForSpin}) for the spin as well as the bound (\ref{ConstraintOnImaginaryPart}) on the imaginary part of the poles, we decide again that we have the bound on spin:
\begin{equation}
\frac{1}{2} < j \le \frac{k+1}{2} \, .
\end{equation}

\subsection{The Positivity of the Spacetime Energy }
As a warm-up exercise for future analyses, we explicitly prove a stability theorem for bosonic string theory on $AdS_3 \times N$. The stability needs to be taken with a grain of salt because of the existence of the closed string tachyon (which generically lies below the Breitenlohner-Freedman bound). We will prove that all excitations except the closed string tachyon are positive energy excitations. Since the energy is the sum of left- and right-moving conformal dimensions, it is stronger to prove that the latter are both positive. We will concentrate on the discrete sector.

 The spacetime left-moving conformal dimension $H$ is given by the exponent of the nome $\qst$:
\begin{equation}
\label{posbound}
H = j+r+ \frac{k+2}{2} w ~.
\end{equation}
Using the on-shell condition $L_0=1$ (see \eqref{onshellbos}) in combination with the above equation, one can eliminate the quantum number $r$  and obtain the following alternative expression for the conformal dimension:
\be
\label{4khw}
4 k w H = 1 -(2j-1)^2 + 4 k h_{int}+k(k+2)w^2 - 4k~.
\ee
Multiplying by $4k$ the equation \eqref{posbound} and adding equation \eqref{4khw}, we obtain:
\be
4k(w+1)H = 4(j-1)(k-j)+2k(k+2)w+k(k+2)w^2+4k(h_{int}+r)~,
\ee
where have denoted $h_{int}=h+N$.
For on-shell states we know that the winding $w$ is positive. Thus, the second and third terms in this expression are always non-negative. For positive $r$ the last term is also non-negative. We now make the point that even for $r<0$, there are no negative contributions to the internal conformal dimension $h_{int}$ arising from the special series descendants $S_r$ because of the identity $S_{r}=q^{-r} S_{-r}$. In fact,  we obtain a positive contribution at least equal to $-r$. Thus $(h_{int}+r) \ge 0$. Therefore the only source of negativity comes from the first term, when $j<1$. These are precisely the tachyonic states that can lead to negative conformal dimensions when $w=0$. 
    
For zero winding, the easiest way to proceed is to revisit the starting point. We see that the on-shell condition only allows solutions in the discrete sector when $h_{int} \ge 1$.
The lowest lying state is when $h_{int}=1$, $j=1$ and we have $r=-1$. Indeed, the state $j^{-}_{-1} |j \rangle$ is the only left-moving state with zero conformal dimension (with a similar story holding for the right-movers) \cite{Maldacena:2000hw}. All other states have strictly positive energy.\footnote{For the continuum sector, assuming $N+h \ge 1$ to exclude the tachyon, one easily proves that there are no states with negative conformal dimension  $\frac{1}{2}(k+2)w+m$ when the constraint  $L_0=1=\frac1k(\frac14+s^2) -wm-\frac14(k+2)w^2+h_{int}$ is satisfied.} This concludes our updated review of the thermal free energy calculation in three-dimensional anti-de Sitter space-time with NSNS flux \cite{Maldacena:2000hw}.

\section{Supersymmetric Thermal Anti-de Sitter}
\label{SusyAdS3} \label{SupersymmetricAdS3}
In this section, we compute the free energy of a supersymmetric world sheet theory on the thermal three-dimensional anti-de Sitter spacetime. This prepares the ground further for the detailed analysis of superstring backgrounds in sections \ref{AdS3S3T4} and \ref{ads3s3s3s1}.

\subsection{World Sheet Supersymmetry and the Global Twist}

 We generalize our analysis of the one loop vacuum energy in thermal $AdS_3$ to the case of the supersymmetric world sheet model. All $AdS_3$ coordinate fields acquire a fermionic superpartner.  We fix the fermionic contribution to the world sheet partition function using the following arguments. 
The $AdS_3$ non-linear sigma-model is a Wess-Zumino-Witten model. World sheet ${\cal N}=(1,1)$ supersymmetry can be attained for any such model by promoting the world sheet quantum fields to ${\cal N}=(1,1)$ superfields. A crucial observation is that the fermions that are thus added to the model can be rendered free through a field redefinition  \cite{DiVecchia:1984nyg}. Thus, we obtain the bosonic $AdS_3$ sigma-model familiar from section \ref{BosonicAdS3} plus a model of three fermions transforming in the adjoint of sl$(2,\mathbb{R}) \oplus \, $sl$(2,\mathbb{R})$. 

In the previous section, we saw that the world sheet bosons  are twisted by a fugacity $U_{n,m}$ \eqref{Unmdef} along the direction generated by the global symmetry $j^3_0$. In the supersymmetric  sigma-model, the relevant global symmetry is the symmetry generated by $J^3_0=j^3_0+j^{F,3}_0$ that consists of both a bosonic $j^3_0$ and a fermionic $j^{F,3}_0$ contribution. Indeed, in superstring theory, the symmetry generator $J^3_0$ must commute with the string theory BRST charge which contains a world sheet supercurrent piece  and consequently coincides with the  $J^3_0$ charge in the supersymmetric Wess-Zumino-Witten model.  The adjoint fermionic modes have charges $(0,+1,-1)$ under this symmetry.  We conclude that the world sheet  fermion partition function  undergoes the twist $U_{n,m}$ on the left and $\bar{U}_{n,m}$ on the right with charges $(0,+1,-1)$ for the three fermions.
Moreover, the twisted world sheet partition function for the two charged fermions carries an exponential factor that renders it modular invariant:
\begin{equation}
Z_{\text{charged fermions}} = \frac{
\theta_i(  U_{n,m},\tau )~ \theta_{\bar{i}}(  \bar{U}_{n,m},\bar{\tau} )}{|\eta|^2}
e^{-\frac{2 \pi (\text{Im}  U_{n,m})^2}{\tau_2}}
\, .
\end{equation}
In the supersymmetric model, this exponential cancels a factor that arises from the anomalous chiral rotation of the bosons.  The indices $i$ and $\bar{i}$ on the $\theta$-functions indicate the boundary condition and world sheet fermion number twist which can take the values $i,\bar{i}\in \{ NS,\widetilde{NS},R, \widetilde{R}\}=\{ 3,4,2,1 \}$ in a standard notation \cite{Polchinski:1998rr}. 

\subsection{The One Loop Amplitude}

The analysis of the one loop vacuum amplitude and the single string free energy proceeds along the lines of the previous section. We unfold the one-loop integral and concentrate on (minus) the single string contribution to the free energy. We thus obtain the starting point of our analysis:
\be
\begin{aligned}
f_{i}^{ \bar i}(\beta,
\mu) &= \frac{1}{2\pi} \int_0^\infty \frac{d \tau_2}{\tau_2^{3/2}} \int_{-\frac{1}{2}}^{\frac{1}{2}} d \tau_1 
 \frac{\sqrt{k} e^{-\frac{k \beta^2} { 4 \pi \tau_2}} }{|\theta_1( \tau_{\text{s.t.}},\tau )|^2}  \frac{ \theta_i (\tau_{\text{s.t.}},\tau)\theta_{\bar i} ( \bar{\tau}_{s.t.},\bar \tau )}{|\eta|^{2}} ~  \Zmaj^i ~
\Zint~ \Zsgh \, .
\end{aligned}
\ee
Again, the internal world sheet conformal field theory  $N$ contributes a partition function factor $\Zint$, the bosonic and fermionic ghost contribution is $\Zsgh$\footnote{We have taken out a factor of $\tau_2$ from the bosonic ghost contribution.}, and $\Zmaj$ is the contribution from the third, uncharged  Majorana fermion along the tangent space of $AdS_3$:
\be
\label{superint}
\Zsgh = \frac{|\eta(\tau)|^6}{|\theta_i(\tau)|^2}~,\qquad \Zmaj^i = \left|\frac{\theta_i}{\eta}\right|~.
\ee
The $i$th fermion sector can equivalently be labelled by a pair of numbers $(a,b)$, each taking one of two values in the index set $\{0,1/2\}$, with the four sectors $ \{ NS,\widetilde{NS},R, \widetilde{R}\}$ assigned the  pairs $(0,0), (0,1/2), (1/2,0)$ and $(1/2,1/2)$ respectively \cite{Polchinski:1998rr}. The corresponding Jacobi theta functions are the contributions to the free energy from the worldsheet fermionic fields. The further calculations proceed much as in section \ref{BosonicAdS3}, and we will therefore provide a more compact description. 

In the first step we introduce the holonomy integrals and integers $v,w$  by means of  the identities  \eqref{ident1} and \eqref{ident3}:
\begin{align}
f_{ab}^{\bar a\bar b}(&\beta,
\mu) =  \int_0^\infty \frac{d \tau_2}{\tau_2^{1/2}} \int_{-\frac{1}{2}}^{\frac{1}{2}} d \tau_1 \int_0^1 ds_1\int_0^1 ds_2 \sum_{v,w} \int d^2 \lambda e^{2 \pi i \lambda_1 (\frac{\mu \beta}{2 \pi} -(s_1+w)\tau_1+s_2+v) + 2 \pi i \lambda_2 (\frac{\beta}{2 \pi} -(s_1+w)\tau_2)} \cr
&\frac{\sqrt{k}}{2\pi}
e^{-2\pi\ii (v (a-\bar a)+ w(b-\bar b))} \frac{ e^{-k \pi (s_1+w)^2 \tau_2} }{|\eta|^{2}} \frac{ \theta^a_b (s_1 \tau-s_2, \tau)\theta^{\bar a}_{\bar b} ( s_1 \bar \tau-s_2, \bar \tau)}{ |\theta_1(  s_1 \tau-s_2, \tau )|^2}~\Zmaj^{a,b} ~\Zint~ \Zsgh  \,.
\end{align}
We  use the $q$-expansions for the $\theta$-functions and the inverse $\theta$-functions to obtain
\begin{multline}
\label{susybifurcate}
f_{ab}^{\bar a\bar b}(\beta,\mu) = \int_0^\infty \frac{d \tau_2}{\tau_2^{1/2}} \int_{-1/2}^{1/2} d \tau_1\int_0^1 ds_1\int_0^1 ds_2\sum_{v,w} \int d^2 \lambda e^{2 \pi i \lambda_1 (\frac{\mu \beta}{2 \pi}- (s_1+w) \tau_1 + s_2 + v)} e^{2 \pi i \lambda_2 (\frac{\beta}{2 \pi} - (s_1+w) \tau_2)} \\
 \frac{\sqrt{k}}{2\pi} \sum_{r,\bar r} e^{2\pi\ii r(s_1\tau-s_2)}~ e^{-2\pi\ii \bar r(s_1\bar\tau-s_2)}e^{ -k \pi \tau_2 s_1^2}  e^{-k(2s_1 w+w^2) \pi \tau_2-2\pi\tau_2 s_1  }  \\
\frac{1}{|\eta(\tau)|^8}\sum_{f,\bar f} e^{2\pi\ii \tau \frac{(f+a)^2}{2}} e^{2\pi\ii(s_1\tau-s_2)(f+a)}e^{-2\pi\ii \bar\tau \frac{(\bar f+\bar a)^2}{2}} e^{-2\pi\ii(s_1\bar \tau-s_2)  (\bar f+\bar a)} \\
e^{-2\pi\ii (v (a-\bar a)+ w(b-\bar b))} e^{2\pi\ii b(a+f)} e^{-2\pi\ii \bar b(\bar a+\bar f)} ~ S_r~ S_{\bar r} ~\Zmaj^{a,b} ~\Zint~ \Zsgh\, .
\end{multline}
The analysis bifurcates at this stage into an off-shell path and an on-shell description.
\subsection{The Off-Shell Hilbert Space}
The off-shell description is obtained with the same initial steps as before. We introduce the Gaussian integral, 
perform the sum over the integer $v$ and the integral over the $s_2$ holonomy. This leads to a delta function that fixes $\lambda_1 =r  + f +a-\bar r  - \bar f-\bar a$, which is the spacetime spin that couples to the fugacity $\mu$. The resulting $\lambda_1$-integral can be done trivially once more, and we then perform the $s_1$ integral to obtain:
\begin{multline}
f_{ab}^{\bar a\bar b} (\beta,\mu)= \int_0^\infty \frac{d \tau_2}{2 \pi \tau_2|\eta(\tau)|^8} \int_{-1/2}^{1/2} d \tau_1 \sum_{w, r,\bar r, f, \bar f} \int d \lambda_2 e^{2 \pi i (r + f+a -\bar r  - \bar f-\bar a)(\frac{\mu \beta}{2 \pi}- w\tau_1)} e^{2 \pi i \lambda_2 (\frac{\beta}{2 \pi} - w \tau_2)} \cr
\int dc\  e^{-\frac{\pi\tau_2}{k}c^2}\  e^{- k\pi \tau_2  w^2 }   \frac{(1-e^{- 2 \pi  \tau_2 (ic+i \lambda_2+kw+1+r+\bar{r}+f+a+\bar{f}+\bar a ) }) }{ic+i \lambda_2+kw+1+r+\bar{r}+f+a+\bar{f}+\bar a } \cr
e^{2\pi\ii \tau \frac{(f+a)^2}{2}}e^{-2\pi\ii\bar \tau \frac{(f+a)^2}{2}} e^{2\pi\ii b(a+f-w)} e^{-2\pi\ii\bar b(\bar f+\bar a-w)}~S_r~ S_{\bar r}~\Zmaj^{a,b} ~\Zint~ \Zsgh~. 
\end{multline}

\subsubsection{Contributions from the Continuum Sector}

The first and second terms in the $c$-integral can be combined and rewritten as a contribution of the continuous representation of sl$(2,\mathbb{R})$. In the second (exponential) term, we perform the shift of  variables:
\be
 (c, w, r, \bar r, f, \bar f) \longrightarrow (c - \ii k, w-1, r+1, \bar r+1, f-1, \bar f-1)~. 
\ee
After a bit of algebra, one can check that, up to contributions from poles that are picked up by the shift of the $c$-contour,  we obtain
\begin{multline}
f_{ab,\text{cont.}}^{\bar a\bar b} (\beta, \mu)=  \int_0^\infty \frac{d \tau_2}{2\pi\tau_2} \int_{-1/2}^{1/2} d \tau_1  \sum_{w,r,\bar r, f, \bar f} \int d \lambda_2\ e^{2 \pi i (r+ f+a -\bar r - \bar f-\bar a) (\frac{\mu \beta}{2 \pi}- w \tau_1 )} e^{2 \pi i \lambda_2 (\frac{\beta}{2 \pi} - w \tau_2)}  \cr
\frac{1}{|\eta(\tau)|^8}\int\frac{ dc}{2\pi}\ \frac{e^{-\frac{\pi\tau_2}{k}c^2 }e^{-k  \pi \tau_2 w^2  }\big(1+(S_r-1)+( S_{\bar r} -1)  \big)}{(ic+i \lambda_2+kw+1+r+\bar{r}+f+a+\bar{f}+\bar a)}\cr
e^{2\pi\ii \tau \frac{(f+a)^2}{2}}e^{-2\pi\ii\bar \tau \frac{(\bar f+\bar a)^2}{2}} e^{2\pi\ii b(a+f-w)} e^{-2\pi\ii\bar b(\bar f+\bar a-w)}~\Zmaj^{a,b} ~\Zint~ \Zsgh~.
\end{multline}
We interpret this as a contribution from the continuous part of the spectrum, with the same caveats  as in subsection \ref{Caveats}.

\subsubsection{Contributions from the Discrete Sector}

The  poles  give rise to the discrete part of the spectrum:
\begin{multline}
\label{susydiscone}
f_{ab,\text{disc}}^{\bar a\bar b}(\beta,\mu) =\frac{1}{2\pi} \int \frac{d\tau_2}{\tau_2}\int_{-1/2}^{1/2}d\tau_1 \sum_{w,r,\bar r,f,\bar f} ~
e^{2\pi\ii(r+f+a-\bar r-\bar f-\bar a)(\frac{\mu\beta}{2\pi} - w\tau_1)}e^{-k  \pi \tau_2 w^2 } 
\\
\frac{ 1}{|\eta(\tau)|^8}\int_{(0,k]}d\lambda_2\sum_{\text{poles}(\lambda_2)} e^{\frac{\pi\tau_2}{k}\lambda_2^2} e^{-2\pi(\lambda_2+ kw + 1+r+\bar r+f+a+\bar f + \bar a) (\frac{\beta}{2\pi} - w \tau_2)}\\
 e^{2\pi\ii \tau \frac{(f+a)^2}{2}}e^{-2\pi\ii\bar \tau \frac{(\bar f+\bar a)^2}{2}} e^{2\pi\ii b(f+a-w)} e^{-2\pi\ii\bar b(\bar f+\bar a-w)} S_r~ S_{\bar r}~\Zmaj^{a,b} ~\Zint~ \Zsgh\, .
\end{multline}
We  write the contributions from the internal sector and the (super-)ghost sectors in the  form:
\be
\label{descendants}
\frac{1}{|\eta(\tau)|^8}S_r~ S_{\bar r}  ~\Zmaj^{a,b} ~\Zint~ \Zsgh = e^{4\pi\tau_2(\frac12 - \frac{1}{4k})} \sum_{h,\bar h, N, \bar N} d_{r,h,N} q^{h+N}{\bar q}^{\bar h+\bar N}~.
\ee
We  collect the exponents of $\tau_1$ and $\tau_2$ in order to identify the eigenvalues of the worldsheet Virasoro generators:
\be
\begin{aligned}
f_{ab,\text{disc}}^{\bar a\bar b}(\beta,\mu) =&\sum_{w,r,f,h,N}d_{r,h,N}~ e^{2\pi\ii b(f+a-w)} e^{-2\pi\ii\bar b(\bar f+\bar a-w)} \int_{(0,k]} \frac{d\lambda_2}{2\pi}\\
&\int_{-1/2}^{1/2}d\tau_1 e^{2\pi\ii\tau_1\big(h+\frac{(f+a-w)^2}{2} + N- wr - \bar h - \frac{(\bar f+\bar a-w)^2}{2} -\bar N + w\bar r\big) }\\
& \int \frac{d\tau_2}{\tau_2} e^{-2\pi\tau_2\big(-\frac{kw^2}{2} - \frac{\lambda_2^2-1}{2k}+ \frac{(f+a-w)^2}{2} +h+ N-w r+ \frac{(\bar f + \bar a-w)^2}{2}+\bar h+ \bar N- w\bar r -w(1+w)-1 \big) }\\
&\hspace{1.2cm} (e^{-\b +\ii\mu\b})^{(\frac{\lambda_2+1}{2}+\frac{kw}{2}+r+f+a)}(e^{-\b -\ii\mu\b})^{(\frac{\lambda_2+1}{2}+\frac{kw}{2}+\bar r+\bar f+\bar a)}~.
\end{aligned}
\ee
We shift the fermionic momentum $(f,\bar f) \rightarrow (f+w, \bar f+w)$, which can be interpreted as spectral flow acting on the fermions. One can then write the single string free energy in the expected form:
\be
\label{fdiscL0susy}
\begin{aligned}
f_{ab,\text{disc}}^{\bar a\bar b}(\beta,\mu) =& \sum_{w,r,f,h,N}d_{r,h,N}~ e^{2\pi\ii b(f+a)} e^{-2\pi\ii\bar b(\bar f+\bar a)}\\
& \int \frac{d\tau_2}{\tau_2}\int_{-1/2}^{1/2}d\tau_1e^{2\pi\ii\tau_1( L_0-\bar{L}_0) }
\int_{(\frac{1}{2},\frac{k+1}{2}]}~\frac{dj}{\pi}~(e^{-2\pi\tau_2})^{L_0+\bar{L}_0 - 1}\,\qst^{J_0} {\bar q}_{\text{s,t.}}^{\bar{J}_0} .
\end{aligned}
\ee
We make the same  identifications of the parameters as in  bosonic $AdS_3$ : 
\be
\lambda_2= 2j-1~, \quad m = j+ r ~, \quad \text{and} \quad \bar m = j + \bar r~. 
\ee
The world sheet scaling generators are given by
\begin{align}
\label{ads3virasorosusy}
L_0 &= -\frac{j(j-1)}{k}-w m -(k+2)\frac{w^2}{4} + N+h + \frac{(f+a)^2}{2}~, \\
\bar{L}_0&=  -\frac{j(j-1)}{k}-w\bar m  -(k+2)\frac{w^2}{4} + \bar N+ \bar h+ \frac{(\bar f+\bar a)^2}{2} ~.
\end{align}
The exponent of the spacetime modular parameter instead is given by the zero mode of the left and right moving currents $J^3$ and $\bar{J}^3$ of sl$(2,\mathbb{R})$, as they are identified with linear combinations of the energy and spin in spacetime. Their eigenvalues are
\be
J_0^3 = \frac{k+2}{2}w + m+f+a~,\qquad \bar{J}_0^3 = \frac{k+2}{2}w + \bar m+\bar f+\bar a~.
\ee
 \subsubsection{Character decomposition}
A little more massaging provides a compact expression for the single string contribution to the free energy:
\begin{align}
\label{fdiscfinalsusy}
f^{\bar a\bar b}_{ab,\text{disc}}(\beta,\mu) =&  \int_0^\infty \frac{d \tau_2}{\tau_2} \int_{-\frac{1}{2}}^{\frac{1}{2}} d \tau_1  \sum_{w} \int_{(\frac{1}{2},\frac{k+1}{2}]} \frac{d j}{ \pi} \ 
  \chi^w_{j} (q,q_{s.t.})\, \chi^w_{j} (\bar{q},\bar{q}_{s.t.})\\
&\hspace{6cm} \times \frac{\theta^a_b(\tau,\taust)\theta^{\bar a}_{\bar b}(\bar \tau,\bar{\tau}_{\text{s.t.}})}{|\eta(\tau)|^2} ~\Zmaj^{a,b} ~\Zint~ \Zsgh\, .  \nonumber
  \end{align}
The final expression clearly exhibits the off-shell Hilbert space for superstrings in $AdS_3\times N$ and also follows immediately from the factorized nature of the one-loop vacuum amplitude combined with the bosonic result of section \ref{BosonicAdS3}. Proving the expression in our pedestrian fashion provides insight into how spectral flow links up the bosons and fermions in the sl$(2,\mathbb{R})$ Wess-Zumino-Witten theory.

\subsection{The Free Energy}

We turn to the calculation of the free energy of the on-shell states.
As before, we start from equation \eqref{susybifurcate} and
perform the $\lambda_2$-integral leading to the $\delta$-function constraint that fixes $s_1$.
We then sum over the integer $v$ and do the $(s_2, \lambda_1)$ integrals. 
The result is an integral over a finite region in the $\tau$-plane:
\be
\begin{aligned}
f^{\bar a\bar b}_{ab}(\beta,\mu) =& \sum_{w,r,\bar r,f,\bar f}\frac{\sqrt{k}}{2\pi}  \int_{\frac{\b}{2\pi(w+1)}}^{ \frac{\b}{2\pi w}} \frac{d \tau_2}{\tau_2^{\frac32}|\eta(\tau)|^8} \int_{-1/2}^{1/2} d \tau_1
e^{2\pi\ii(r+f+a-\bar r-\bar f-\bar a)(\frac{\mu\b}{2\pi} - w\tau_1)} \\
&\hspace{2cm}e^{2\pi\ii \tau \frac{(f+a)^2}{2}} e^{-2\pi\ii \bar\tau \frac{(\bar f+\bar a)^2}{2}} e^{2\pi\ii b(f+a-w)} e^{-2\pi\ii \bar b(\bar f+\bar a-w)}\\
&\hspace{2cm}e^{-2\pi(\frac{\b}{2\pi} - w\tau_2) (r+\bar r+f+\bar f + a+\bar a+1)}~e^{-\frac{k\b^2}{4\pi\tau_2}}~ S_r~ S_{\bar r} ~\Zmaj^{a,b} ~\Zint~ \Zsgh~.
\end{aligned}
\ee
We write down the contribution from the descendant states as in equation \eqref{descendants}. Lastly we introduce the same integral over the radial momentum as in \eqref{MOSintegral} and, after a spectral flow in the fermionic sectors $(f,\bar f) \rightarrow (f+w,\bar f + w)$, we obtain
\be
\begin{aligned}
f^{\bar a\bar b}_{ab}(\beta,\mu) =&   \frac{4}{\ii k\b} \sum_{w,r,f,h,N}d_{r,h,N}~ \int_{-1/2}^{1/2}d \tau_1   e^{2\pi\ii \tau_1(h+\frac{(f+a)^2}{2}+N-rw - \bar h  -\frac{(\bar f+\bar a)^2}{2}- N + \bar rw  ) }  \\
& 
\int_{-\infty}^{\infty}~dc~ c~\int_{\frac{\b}{2\pi(w+1)}}^{ \frac{\b}{2\pi w}} d \tau_2e^{-2\pi\tau_2(h +\frac{(f+a)^2}{2} + N -rw +\bar h +\frac{(\bar f+\bar a)^2}{2}  + \bar N-\bar r w- w(1+w) +\frac{4c^2+1}{2k} -1)  } \\
&~ e^{2\pi\ii b(a+f)} e^{-2\pi\ii\bar b(\bar a+\bar f )}~e^{\ii \mu \b (r+f+a-\bar r-\bar f-\bar a)} e^{-\b(-2\ii c + r+\bar r+f+\bar f + a+\bar a+1+2w)}~.
\end{aligned}
\ee
The $\tau_1$ integral leads to level matching condition:
\be
h+\frac{(f+a)^2}{2}+N-rw  = \bar h  +\frac{(\bar f+\bar a)^2}{2} + N - \bar rw~.
\ee
The $\tau_2$ integral gives rise to
\begin{align}
f^{\bar a\bar b}_{ab}(\beta,\mu) =&   \frac{1}{\pi\ii \b} \sum'_{w,r,f,h,N}d_{r,h,N}~e^{2\pi\ii b(a+f)} e^{-2\pi\ii\bar b(\bar a+\bar f )}~e^{\ii \mu \b (r+f+a-\bar r-\bar f-\bar a)}\\
&\int_{-\infty}^{\infty}~dc~ c~\frac{e^{-\b(-2\ii c + r+\bar r+f+\bar f + a+\bar a+1+2w)}}{h +\frac{(f+a)^2}{2} + N -rw -\frac{1}{2}w(1+w)     +\frac{4c^2+1}{4k} -\frac12} \nonumber \\
&\big(e^{-\frac{2\b}{w+1}(h +\frac{(f+a)^2}{2} + N -rw -\frac{1}{2}w(1+w)     +\frac{4c^2+1}{4k} -\frac12)}
- e^{-\frac{2\b}{w}(h +\frac{(f+a)^2}{2} + N -rw -\frac{1}{2}w(1+w)     +\frac{4c^2+1}{4k} -\frac12) } \big)~.
\nonumber
\end{align}
The prime indicates that level matching is imposed on the summation variables. 
As in the bosonic avatar, one shifts the contour integral in the first term from $\text{Im}(c) = 0$ to $\text{Im}(c) = \frac{k(w+1)}{2}$ and in the second term from $\text{Im}(c) = 0$ to $\text{Im}(c) = \frac{kw}{2}$. We focus on the contribution from the discrete sector that arises from the residues of the poles that are picked up in the region: 
\be
\frac{kw}{2} < \text{Im}(c) \le \frac{k(w+1)}{2}~,
\ee
as a result of the shifts in the contours. The poles are located at
\be
\begin{aligned}
-c^2&= \frac{1}{4}+k\big(h + N  -rw -\frac{1}{2}w(1+w)+\frac{(f+a)^2}{2} -\frac12\big)\\
&=\frac{1}{4}(2j-1+kw)^2~,
\end{aligned}
\ee
where we have used the on-shell condition appropriate for the superstring, $L_0=\frac12$, with the scaling operator $L_0$ given by equation \eqref{ads3virasorosusy}.  
This allows one to solve for the spin $j$:
\be
2j-1+kw = \sqrt{1+4k\left(h +\frac{(f+a)^2}{2} + N -rw -\frac{1}{2}w(1+w) -\frac12\right)}~.
\ee
Substituting for the square root into the expression for the single string free energy we finally obtain
\begin{align}
f^{\bar a\bar b}_{ab,\text{disc}}(\beta, \mu) &=\frac{1}{\beta}\sum_{w,r,f,h,N}d_{r,h,N}~
e^{2\pi\ii b(a+f)} e^{-2\pi\ii\bar b(\bar a+\bar f )} 
e^{-\b\big ( 2j+(k+2)w + r+f+a+\bar r+\bar f+\bar a\big)}e^{ \ii\b \mu (r+f+a-\bar r-\bar f-\bar a)} \nonumber \\
&=\frac{1}{\beta}\sum_{w,r,f,h,N}d_{r,h,N}~e^{2\pi\ii b(a+f)} e^{-2\pi\ii\bar b(\bar a+\bar f )}
\qst^{j+r+f+a+\frac{(k+2)w}{2}}
 {\bar q}_{\text{s.t.}}^{j+\bar r+\bar f+\bar a+\frac{(k+2)w}{2}} \\
 &=\frac{1}{\beta}  \sum_{w,r,f,h,N}d_{r,h,N}~e^{2\pi\ii b(a+f)} e^{-2\pi\ii\bar b(\bar a+\bar f )}~\qst^{J_0^3}
 {\bar q}_{\text{s.t.}}^{{\bar J}_0^3}~. \nonumber 
\end{align}
This is as expected for a partition function that is twisted by fermion number when $b=1/2$.

\subsection{The Stability}

We wish to prove the stability of the theory (up to the instability that arises from the closed string tachyon).
The exponent of the nome $\qst$ which measures the spacetime left-moving conformal dimension of the string states is:
\be
\label{horigin}
H= j+r+f+a+\frac{k+2}{2}w   ~.
\ee
The worldsheet Virasoro generator $L_0$ is given by
\be
L_0 = -\frac{j(j-1)}{k} - w(j+r) - \frac{k+2}{4} w^2 + h_{int} + \frac{(f+a)^2}{2}~.
\ee
We use the on-shell condition $L_0=\frac12$ to eliminate the spin component $r$ and we obtain
\be
4 k w H  =-(2j-1)^2 + 4 k h_{int} + 2k(f+a+w)^2 + k^2w^2+1-2k~.
\ee
We add  $4k$ times the expression  \eqref{horigin} for the dimension $H$  to obtain:
\be
4k(w+1)H = 4j(k+1-j)  + 2k(f+a+w+1)^2+ 4k(h_{int} +r)+ k(kw^2+2 kw - 4)~.
\ee
The first term on the right hand side is positive on account of the bound on the spin $j$ while the second term is manifestly positive. The third term on the right hand side is positive irrespective of the sign of the spin component $r$. For $r \ge 0$ this is obvious while for $r<0$, this follows from the property $q^r S_r = S_{-r}$  of the series $S_r$ that encodes the degeneracies of the sl$(2)$ descendants. 

Thus the only potential source of negativity is from the last term. One can check that that for $j\ge 1$, the dimension $H$ is non-negative for all windings $w$. The only possibility for negative dimension is for $w=0$ and for $\frac12 < j < 1$. These correspond to tachyonic states. In summary, we have extended the calculation of the thermal free energy of string theory in three-dimensional anti-de Sitter spacetime with Neveu-Schwarz-Neveu-Schwarz flux to include world sheet fermions.

\section{Superstrings in Thermal \texorpdfstring{$AdS_3\times S^3\times T^4$}{} }
\label{AdS3S3T4}
\label{AdS3S3M}
In this section, we provide a first application of the general results obtained in sections \ref{BosonicAdS3} and \ref{SusyAdS3} in the context of a supersymmetric compactification of string theory. We calculate the partition function on the thermal $AdS_3 \times S^3 \times T^4$ background. The background arises as the near brane limit of a system of $k$ NS5-branes compactified on $T^4$ with a density of fundamental strings spread on the four-torus.  Since we consider a quotient of global $AdS_3$, we are in the NSNS sector in the boundary theory \cite{Coussaert:1993jp}. We compute the partition function in the NSR formalism for the bulk world sheet string theory. We include fugacities (and their complex conjugates) that keep track of the spacetime conformal dimension and a spacetime $u(1)_R$ charge.  We impose a periodicity in the compactified time direction that is consistent with supersymmetry.
In the course of our calculation, we also make contact with a manifestly spacetime supersymmetric description of the background.

\subsection{The One Loop Vacuum Amplitude with Fugacities}

\label{oneloopvacuumamplitude}

We consider type IIB superstrings propagating on $AdS_3\times S^3\times T^4$ and calculate the one-loop vacuum amplitude. Because of the decoupling of fermions in supersymmetric Wess-Zumino-Witten models, the integrand takes a factorized form, with separate bosonic factors and eight free transverse fermions that are appropriately GSO projected \cite{Gliozzi:1976qd}:
\be
Z_{\text{IIB}} = \frac{1}{2\pi}\int_{F_0} \frac{d^2\tau}{\tau_2} Z_{AdS_3}~Z_{S^3}~ Z^{bos}_{T^4}~Z_{gh}~\frac{1}{4 |\eta|^8}~
\left|\sum_{a,b} e^{2\pi\ii (a+b+2ab)}
(\theta_b^a (\tau))^4
\right|^2  \, .  \label{SusyOneLoop}
\ee
The indices $(a,b)$ take  values in $\{ 0,\frac12 \}$. 
The bosonic $AdS_3$ partition function is described by an sl$(2,\mathbb{R})$ model at bosonic level $\kbos=k+2$; the bosonic level $k-2$ three-sphere partition function is given by  a finite sum over su$(2)$ characters (see Appendix \ref{discchar}): 
\be
Z_{S^3}= \sum_{l=0}^{\frac{k-2}{2}} \left|\chi_l(\tau) \right|^2~.
\ee
The bosonic $T^4$ and ghost partition function are standard.

We  dress the one-loop vacuum amplitude (\ref{SusyOneLoop}) with fugacities for spacetime symmetries, including the spacetime energy, the angular momentum and the spacetime R-charges. 
We  already added the  fugacities $(\taust,\bar{\tau}_{\text{s.t.}})$ which couple to the left/right combinations of the energy and angular momentum in  sections
\ref{BosonicAdS3} and \ref{SusyAdS3}. 
In addition  we introduce the fugacities $(\nust,\nubarst)$ that couple to u$(1)$ R-charges that correspond to left and right rotations on the three-sphere respectively. We thus obtain the  weighted one-loop amplitude:
\be
\begin{aligned}
Z_{\text{IIB}}& = \frac{1}{2\pi}\int_{F_0} \frac{d^2\tau}{\tau_2} Z_{AdS_3}(\taust,\taubarst, \tau)~Z_{S^3}(\nust,\nubarst, \tau)~ Z^{bos}_{T^4}(\tau)~Z_{gh}(\tau) \\
&\hspace{2cm}\frac{1}{4 |\eta|^8}
\left |\sum_{a,b}  e^{2\pi\ii(a+b+2ab)}
\theta^a_b (\taust, \tau)
\theta^a_b (\nust, \tau)
\theta^a_b (0, \tau)
\theta^a_b (0, \tau)\right|^2 \, . 
\label{InitialNSR}
\end{aligned}
\ee
The charged fermion that arose as a partner of the $AdS_3$ bosons is twisted by the fugacity $\taust$ while the charged fermion that arises as a   superpartner of three-sphere bosons is twisted by the fugacity $\nust$.

\subsection{The Single Particle Free Energy}

Given the one loop string amplitude (\ref{InitialNSR}), we can follow the same steps as in section \ref{SusyAdS3}. We unfold and then equate the one loop amplitude with the (twisted) thermal free energy. We  extract the single string contribution to the free energy in the GSO projected type IIB theory:
\be
\begin{aligned}
f(\beta, \mu) &= \frac{1}{2\pi} \int_0^\infty \frac{d \tau_2}{\tau_2^{3/2}} \int_{-\frac{1}{2}}^{\frac{1}{2}} d \tau_1 
 \frac{\sqrt{k} e^{-\frac{k \beta^2} { 4 \pi \tau_2}} }{|\theta_1(\tau,  \tau_{\text{s.t.}} )|^2}
~ |\eta|^{4}  ~ Z_{T^4}^{bos}~ \sum_{l=0}^{(k-2)/2} |\chi_l(\nust, \tau) |^2   \\
&\hspace{2cm}\frac{1}{4 |\eta|^8}\left|\sum_{a,b}  e^{2\pi\ii(a+b+2ab)}
\theta^a_b (\taust, \tau)
\theta^a_b (\nust, \tau)
\theta^a_b (0, \tau)
\theta^a_b (0, \tau) \right|^2~.
\end{aligned}
\ee
To render space-time supersymmetry manifest, we
make use of a Jacobi identity that is rooted in $so(8)$ triality. The abstruse identity reads \cite{WW}:
\begin{equation}
\label{JacobiGS}
\sum_{a,b} e^{2 \pi i (a+b+2 a b)} \prod_{i=1}^4 \theta^a_b(\nu_i,\tau)
= -2 \prod_{i=1}^4 \theta_1 (\mu_i,\tau) \, ,
\end{equation}
where the variables $\nu_i$ map to the variables $\mu_i$ roughly as Cartan torus coordinates under $so(8)$ triality:
\begin{equation}
\begin{aligned}
2 \nu_1 &= -\mu_1+\mu_2+\mu_3+\mu_4 \\
2 \nu_2 &= \phantom{-}\mu_1-\mu_2+\mu_3+\mu_4 \\
2 \nu_3 &= \phantom{-}\mu_1+\mu_2-\mu_3+\mu_4 \\
2 \nu_4 &= \phantom{-}\mu_1+\mu_2+\mu_3-\mu_4 \, .
\end{aligned}
\end{equation}
In string theory, this identity often takes one from a NSR formalism to a Green-Schwarz formalism in which spacetime  supersymmetry becomes manifest.  Applying this formula to our twisted one loop amplitude we obtain
\begin{multline}
\sum_{a,b} e^{2\pi\ii(a+b+2ab)}
\theta^a_b (\taust, \tau)
\theta^a_b (\nust, \tau)
\theta^a_ba (0, \tau)
\theta^a_b (0, \tau) \\
=-2 
\big(\theta_1 (\frac{\tau_{\text{s.t.}}-\nu}{2} ,\tau)\big)^2~\big(\theta_1 (\frac{\tau_{\text{s.t.}}+\nu}{2}, \tau)\big)^2~.
\end{multline}
Furthermore, we spectral flow by  half a unit in the boundary theory (for both the left and right movers) in order to calculate  in the Ramond-Ramond sector of the boundary theory:
\be
\nust \longrightarrow \nust -\taust~,\qquad \nubarst \longrightarrow \nubarst -\taubarst ~.
\ee
Including a standard normalization factor $N=q_{\text{s.t.}}^{\frac{c_{\text{s.t.}}}{24}} z^{-\frac{c_{\text{s.t.}}}{6}}$ depending on the background space-time central charge $c_{\text{s.t.}}$,
we find the free energy
\be
\begin{aligned}
f(\beta, \mu) &= \frac{N}{2\pi} \int_0^\infty \frac{d \tau_2}{\tau_2^{3/2}} \int_{-\frac{1}{2}}^{\frac{1}{2}} d \tau_1 
 \frac{\sqrt{k} e^{-\frac{k \beta^2} { 4 \pi \tau_2}} }{|\theta_1(  \tau_{\text{s.t.}},\tau )|^2}  \sum_{l=0}^{\frac{k-2}{2}} |\chi_l(\nust - \taust, \tau) |^2   \\
&\hspace{4cm}\frac{1}{|\eta|^4}| \theta_1(\tau_{\text{s.t.}} - \frac{\nu }{2}, \tau)|^4
|\theta_1 (\frac{\nu}{2},\tau )|^4~ Z_{T^4}^{bos}(q) .
\end{aligned}
\ee
As before, we  introduce the holonomy integral that represents the $\delta$-function:
\begin{align}
f(\beta,\mu) =& N   \int_0^\infty \frac{d \tau_2}{\tau_2^{1/2}} \int_{-1/2}^{1/2} d \tau_1 \int_0^1 ds_1 \int_0^1 ds_2 \sum_{v,w} \int d^2 \lambda
e^{2 \pi i \lambda_1 (\frac{\mu \beta}{2 \pi}- (s_1+w) \tau_1 + s_2 + v)} e^{2 \pi i \lambda_2 (\frac{\beta}{2 \pi} - (s_1+w) \tau_2)}\nonumber \\ 
&\hspace{2cm}\frac{| \theta_1(s_1\tau -s_2 + w\tau - v -\frac{\nu }{2} , \tau)|^4  }{|\theta_1(s_1\tau - s_2 + w\tau -v,\tau )|^2} \sum_{l=0}^{\frac{k-2}{2}} |\chi_l(\nust - s_1\tau + s_2 - w\tau + v, \tau) |^2   \nonumber \\
&\hspace{4cm}e^{-\pi k \tau_2(s_1+w)^2}
\frac{\sqrt{k}}{2\pi|\eta|^4}
|\theta_1 (\frac{\nu}{2},\tau )|^4~ Z_{T^4}^{bos}(q)  ~.
\end{align}
We simplify the formula using the ellipticity properties of the theta functions as well as the su$(2)$ characters \cite{Feigin:1998sw, Giribet:2007wp}. 
In addition we  use the $q$-expansion for the $\theta_1$-function and its inverse, and the  expansion for the su$(2)$ character (see the Appendices  \ref{apptheta} and \ref{discchar} for details)
\be
\chi_l (\nust - s_1 \tau +s_2,\tau) = q^{\frac{l(l+1)}{k} -\frac{c_{su(2)}}{24}} \sum_{ r'}
C^l_{r'}~e^{2 \pi i r' (s_1 \tau-s_2 - \nust)} \,,
\ee
to write the single string free energy as:
\begin{align}
f(\beta,\mu) =&   \frac{\sqrt{k}N}{2\pi}   \int_0^\infty \frac{d \tau_2}{\tau_2^{1/2}} \int_{-1/2}^{1/2} d \tau_1 \int_0^1 d^2 s_i \sum_{v,w} \int d^2 \lambda
e^{2 \pi i \lambda_1 (\frac{\mu \beta}{2 \pi}- (s_1+w) \tau_1 + s_2 + v)} e^{2 \pi i \lambda_2 (\frac{\beta}{2 \pi} - (s_1+w) \tau_2)}\nonumber \\
& \sum_{l=0}^{\frac{k-2}{2}}\sum_{r,\bar r,f_i, \bar f_i,  r', \bar r'} C^l_{r'} \bar C^l_{ \bar r'}e^{2\pi\ii (r+r'+f_1+f_2+1)(s_1\tau-s_2)}~ e^{-2\pi\ii  (\bar r+\bar r'+\bar f_1+\bar f_2+1)(s_1\bar\tau-s_2)}~S_r~S_{\bar r}\nonumber
\\
&e^{-2\pi\tau_2 s_1} e^{-\pi k \tau_2 s_1^2} e^{2\pi\ii \nust( \frac{ k w}{2} -r' - \frac{f_1+f_2+1}{2})}~e^{-2\pi\ii \nubarst( \frac{ k w}{2}-\bar r' - \frac{\bar f_1+\bar f_2+1}{2}) } e^{\pi\ii(f_1+f_2-\bar f_1-\bar f_2)}\nonumber\\
&\frac{1}{|\eta|^{10}}~q^{\frac12\sum_i (f_i+\frac12)^2+ \frac{l(l+1)}{k} -\frac{c_{su(2)}}{24}} ~ \bar q^{\frac12\sum_i(\bar f_i+\frac12)^2+ \frac{l(l+1)}{k} -\frac{c_{su(2)}}{24}}~
|\theta_1 (\frac{\nu}{2},\tau )|^{4}~ Z_{T^4}^{bos}(q)
 ~. \label{fN=4bif}
\end{align}

\subsection{The Off-shell Hilbert Space}
Once more, we first exhibit the off-shell Hilbert space and the expressions for the Virasoro generators of the worldsheet theory. We repeat the same steps as in the earlier sections. 
The sum over the integer $v$ leads to a Dirac comb for the multiplier $\lambda_1$. The subsequent $s_2$ integration imposes:
\begin{equation}
\lambda_1 = r+ r'+f_1+f_2-\bar{r} - \bar{r}'-\bar{f}_1-\bar{f}_2~.
\end{equation}
We introduce the Gaussian $c$-integral
and perform the $s_1$-integral to obtain: 
\begin{align}
f(\beta,\mu) =& N~  \sum_{l=0}^{\frac{k-2}{2}}\sum_{r,\bar r,f_i, \bar f_i, r', \bar r'} C^l_{ r'} \bar C^l_{ \bar r'}  \int_0^\infty \frac{d \tau_2}{2\pi\tau_2} \int_{-1/2}^{1/2} d \tau_1 \int d\lambda_2 \sum_{w} e^{\pi\ii(f_1+f_2-\bar f_1-\bar f_2)}\nonumber\\
&\int \frac{dc}{2\pi}~e^{-\frac{\pi \tau_2}{k}c^2}\frac{\left(1-e^{-2\pi \tau_2 \left(\ii \lambda_2  +\ii c + r+r'+f_1+f_2+\bar r+\bar r'+\bar f_1+\bar f_2+3\right)}\right) }{\scriptstyle{(\ii \lambda_2  +\ii c + r+r'+f_1+f_2+\bar r+\bar r'+\bar f_1+\bar f_2+3 )}}e^{2 \pi i \lambda_2 (\frac{\beta}{2 \pi} - w\tau_2)}
\nonumber\\
&e^{2 \pi \ii( r+ r'+f_1+f_2-\bar{r} - \bar{r}'-\bar{f}_1-\bar{f}_2) (\frac{\mu \beta}{2 \pi}-w\tau_1)}~e^{2\pi\ii \nust( \frac{ k w}{2} -r' - \frac{f_1+f_2+1}{2}) } ~e^{-2\pi\ii \nubarst( \frac{ k w}{2}-\bar r' - \frac{\bar f_1+\bar f_2+1}{2} ) }
\nonumber\\
& ~q^{\frac12\sum_i( f_i+\frac12)^2+ \frac{l(l+1)}{k}-\frac{c_{su(2)}}{24}} ~ \bar q^{\frac12\sum_i(\bar f_i+\frac12)^2+ \frac{l(l+1)}{k}-\frac{c_{su(2)}}{24}}~
\frac{|\theta_1 (\frac{\nu}{2},\tau )|^{4}}{|\eta|^{10}}~ Z_{T^4}^{bos}(q)~ S_r~S_{\bar r}. \label{fN=4glory}
\end{align}

\subsubsection{Contributions from the continuous spectrum}

We  show that the two terms in the parentheses can be combined as for the bosonic case up to a set of pole contributions. Let us consider the exponential term and make the following redefinitions in the summation and integration variables:
\small
\begin{equation*}
(c, w, r, \bar r, l, r', \bar r', f_i, \bar f_i) \rightarrow (c-\ii k, w-1, r+1, \bar r+1, \frac{k-2}{2} - l,  r' - \frac{k-2}{2},  \bar r'-\frac{k-2}{2}, f_i-1,\bar f_i-1)~.
\end{equation*}
\normalsize
An identity satisfied by the function that captures the degeneracies of the descendant states in the 
affine su$(2)$ representations (see equation \eqref{drnspectral} in Appendix \ref{discchar}) turns out to be  useful: 
 \be
 \label{dSrelations}
 \begin{aligned}
C^{l}_{r'} &= q^{r' -l} C^{\frac{k-2}{2} -l}_{r' - \frac{k-2}{2}}  ~.
 \end{aligned}
 \ee
 After some tedious algebra, the two terms in equation (\ref{fN=4glory}) combine: 
\be
\begin{aligned}
f(\beta,\mu) =&   N~\sum_{l=0}^{\frac{k-2}{2}}\sum_{r,\bar r,f_i, \bar f_i, r', \bar r'}C^l_{r'}\bar C^l_{ \bar r'} ~S_r~S_{\bar r}\int_0^\infty \frac{d \tau_2}{2\pi\tau_2} \int_{-1/2}^{1/2} d \tau_1 \int d\lambda_2 \sum_{w} e^{\pi\ii(f_1+f_2-\bar f_1-\bar f_2)}\\
&\int \frac{dc}{2\pi}~e^{-\frac{\pi \tau_2}{k}c^2}\frac{e^{2 \pi i \lambda_2 (\frac{\beta}{2 \pi} - w\tau_2)} ( 1+(S_r-1) + (S_{\bar r}-1) )}{(\ii \lambda_2  +\ii c + r+r'+f_1+f_2+\bar r+\bar r'+\bar f_1+\bar f_2+3 )}\\
&e^{2 \pi \ii( r+ r'+f_1+f_2-\bar{r} - \bar{r}'-\bar{f}_1-\bar{f}_2) (\frac{\mu \beta}{2 \pi}-w\tau_1)}~e^{2\pi\ii \nust( \frac{ k w}{2} -r' - \frac{f_1+f_2+1}{2}) } ~e^{-2\pi\ii \nubarst( \frac{ k w}{2}-\bar r' - \frac{\bar f_1+\bar f_2+1}{2} ) }\\
&\frac{1 }{|\eta|^{10}}~q^{\frac12\sum_i( f_i+\frac12)^2+ \frac{l(l+1)}{k}-\frac{c_{su(2)}}{24}} ~ \bar q^{\frac12\sum_i(\bar f_i+\frac12)^2+ \frac{l(l+1)}{k}-\frac{c_{su(2)}}{24}}~
|\theta_1 (\frac{\nu}{2},\tau )|^{4}~ Z_{T^4}^{bos}(q)  ~.
\end{aligned}
\ee
The result is similar to the one obtained section \ref{SusyAdS3} and the same caveats apply as in subsection \ref{Caveats}.

\subsubsection{Contributions from the discrete spectrum}

Our main interest is in the contribution from the discrete sector. The shift of the $c$-integral in the second term of (\ref{fN=4glory}) by $-\ii k$ is what allowed us to combine the two terms in the manner shown above. The shift leads to additional contributions that arise from the poles that are encountered in the process. The poles are located at 
\be
 -\ii c  =\ii \lambda_2 + r+r'+f_1+f_2+\bar r+\bar r'+\bar f_1+\bar f_2+3~,
\ee
and the sum of the residues at these poles gives the contribution from the discrete sector. 
We Wick-rotate the $\lambda_2$ integral onto the imaginary axis by replacing $\lambda_2= \ii \widetilde{\lambda}_2$. We further shift the $\widetilde{\lambda}_2$ integral  to get rid of the quadratic dependence on the $(r,r', f_i)$ variables. At the end of these manipulations we find the discrete contribution to the single string free energy:
\be
\label{fdiscinterim}
\begin{aligned}
f_{\text{disc}}(\beta,\mu) =&   N\sum_{l=0}^{\frac{k-2}{2}}\sum_{r,\bar r,f_i, \bar f_i, r', \bar r'}C^l_{r'}\bar C^l_{\bar r'}~S_r~S_{\bar r}  \int_0^\infty \frac{d \tau_2}{2\pi\tau_2} \int_{-1/2}^{1/2} d \tau_1 \sum_{w} e^{\pi\ii(f_1+f_2-\bar f_1-\bar f_2)}\\
&\int_{(0,k]} d\widetilde \lambda_2 ~e^{\frac{\pi \tau_2}{k}\widetilde \lambda_2^2}~ e^{-2 \pi  (\widetilde \lambda_2+( r+r'+f_1+f_2+\bar r+\bar r'+\bar f_1+\bar f_2+3)) (\frac{\beta}{2 \pi} - w\tau_2)} \\
&e^{2 \pi \ii( r+ r'+f_1+f_2-\bar{r} - \bar{r}'-\bar{f}_1-\bar{f}_2) (\frac{\mu \beta}{2 \pi}-w\tau_1)}~e^{2\pi\ii \nust( \frac{ k w}{2} -r' - \frac{f_1+f_2+1}{2}) } ~e^{-2\pi\ii \nubarst( \frac{ k w}{2}-\bar r' - \frac{\bar f_1+\bar f_2+1}{2} ) }\\
&\frac{1 }{|\eta|^{10}}~q^{\frac12\sum_i( f_i+\frac12)^2+ \frac{l(l+1)}{k}-\frac{c_{su(2)}}{24}} ~ \bar q^{\frac12\sum_i(\bar f_i+\frac12)^2+ \frac{l(l+1)}{k}-\frac{c_{su(2)}}{24}}~
|\theta_1 (\frac{\nu}{2},\tau )|^{4}~ Z_{T^4}^{bos}(q)   ~.
\end{aligned}
\ee
The arguments that lead to the finite bound on the $\widetilde \lambda_2$-integral are the same as in the bosonic case.

\subsubsection{Free energy in Terms of Discrete Characters}

Our next goal is to package the expressions into characters of the various sectors. For this purpose we observe that the $f_i$-dependent terms  can be recombined into $\theta_1$-functions:
\be
\sum_{f_i} e^{\pi \ii (f_i+\frac12)}q^{\frac12( f_i+\frac12)^2}q^{-w(f_i +\frac12)}  \big(e^{-\b+ \ii \mu\b}\,  e^{-2\pi\ii\frac{\nu}{2}}\big)^{f_i+\frac12}  = q^{-\frac{w^2}{2}}~\qst^{w}~e^{-2\pi\ii\nust \frac{w}{2}}~\theta_1(\taust-\frac{\nust}{2}, \tau ).
\ee
We also introduce the spin variable $j = \frac12(\widetilde\lambda_2+1)$:
\be
\begin{aligned}
f_{\text{disc}}(\beta,\mu) =&  N \int_0^\infty \frac{d \tau_2}{\tau_2} \int_{-1/2}^{1/2} d \tau_1 \sum_{w}
\int_{(\frac12,\frac{k+1}{2}]} \frac{dj}{\pi}\\
&\sum_{l=0}^{\frac{k-2}{2}}\sum_{r,\bar r,  r', \bar r'}C^l_{r'}\bar C^l_{ \bar r'}~S_r~S_{\bar r} ~
q^{-\frac{j(j-1)}{k}-\frac{1}{4k}-w(j+r+r')} \bar q^{-\frac{j(j-1)}{k}-\frac{1}{4k}-w(j+\bar r+\bar r')} \\
& (e^{-\b + \ii\mu\b})^{ (2w+j+r+r') } (e^{-\b -\ii\mu\b})^{ 2w +j +\bar r+\bar r') } e^{2\pi\ii\nust(\frac{(k-2)w}{2} - r') } e^{-2\pi \ii \nubarst(\frac{(k-2)w}{2} - \bar r')}\\
&\frac{1 }{|\eta|^{10}}~q^{ \frac{l(l+1)}{k}-\frac{c_{su(2)}}{24}-w^2}~\bar q^{\frac{l(l+1)}{k}-\frac{c_{su(2)}}{24}-w^2} |\theta_1(\taust-\frac{\nust}{2}, \tau )|^4
|\theta_1 (\frac{\nu}{2},\tau )|^{4}~ Z_{T^4}^{bos}(q)~.
\end{aligned}
\ee
We recall the definition of the spacetime nome $\qst = (e^{-\b + \ii\mu\b})$ and of the exponentiated fugacity $\zst = e^{2\pi\ii\nust}$ and
separate out the contributions from the $AdS_3$, $S^3$ and $T^4$ factors of spacetime: 
 \be
 \begin{aligned}
 f_{\text{disc}}(\beta,\mu) =&  N \int_0^\infty \frac{d \tau_2}{\tau_2} \int_{-1/2}^{1/2} d \tau_1~
   \int_{(\frac12,\frac{k+1}{2}]} \frac{dj}{\pi}~\sum_{l=0}^{\frac{k-2}{2}}~\sum_{w}\\
 & \frac{1}{|\eta|^3}\sum_r q^{-\frac{j(j-1)}{k}-\frac{1}{4k}-\frac{(k+2)w^2}{4}} \qst^{w+\frac{kw}{2}} (q^{-w}\, \qst)^{ j+r }~S_r\\
& \frac{1}{|\eta|^3}\sum_{\bar r} \bar q^{-\frac{j(j-1)}{k}-\frac{1}{4k}-\frac{(k+2)w^2}{4}} \qbarst^{w+\frac{kw}{2}} (q^{-w}\, \qbarst)^{ j+r } 
 ~S_{\bar r}\\
 &
 q^{ \frac{l(l+1)}{k}-\frac{c_{su(2)}}{24}+\frac{(k-2)w^2}{4}} (\qst \zst^{-1} )^{-\frac{(k-2)w}{2}}\sum_{ r'}C^l_{r'}(q^{-w} \qst \zst^{-1})^{r'} \\
 &\bar q^{\frac{l(l+1)}{k}-\frac{c_{su(2)}}{24}+\frac{(k-2)w^2}{4}}(\qbarst \zbarst^{-1})^{-\frac{(k-2)w}{2}}\sum_{ \bar r'}\bar C^l_{ \bar r'}  (\bar q^{-w} \qbarst \zbarst^{-1})^{\bar r'} \\
 &\frac{1 }{|\eta|^{4}}~ |\theta_1(\taust-\frac{\nust}{2}, \tau )|^4
 |\theta_1 (\frac{\nu}{2},\tau )|^{4}~ Z_{T^4}^{bos}(q) 
 \end{aligned}
 \ee
In the second and third lines, one recognizes the spectral flowed sl$(2,\mathbb{R})$ character. Similarly, in the fourth and fifth lines, one identifies the spectral flowed  flowed su$(2)_{k-2}$ character  \eqref{specflowsu2}:
\be
\label{su2chiw}
\chi_{l}^w(\nu, \tau) = q^{\frac{(k-2)w^2}{4}}  e^{-2\pi\ii \frac{(k-2) w}{2}\nu} \chi_l ( \nu-w\tau, \tau)~.
\ee
As a consequence, the single string free energy can be written in the  compact form:
\be
\begin{aligned}
f_{\text{disc}}(\beta,\mu) =& N  \int_0^\infty \frac{d \tau_2}{\tau_2} \int_{-1/2}^{1/2} d \tau_1 
\int_{(\frac12,\frac{k+1}{2}]} \frac{dj}{\pi} ~\sum_{w}~
 \chi_j^w(\taust, \tau)~ \chi^w_{j} (\bar{\tau}_{\text{s.t.}}, \bar \tau)\\
& \sum_{l=0}^{\frac{k-2}{2}} \chi^w_{l}(\nust -\taust, \tau)~ \chi^w_{l}(\nubarst -\taubarst, \bar\tau)   |\theta_1(\taust-\frac{\nust}{2}, \tau )|^4
\frac{|\theta_1 (\frac{\nu}{2},\tau )|^{4}}{|\eta|^4}~ Z_{T^4}^{bos}(q)~.
\end{aligned}
\ee
We note that both the sl$(2,\mathbb{R})$ and the su$(2)$ characters have been spectrally flowed by $w$ units. 
To identify the world sheet Virasoro generators that capture the off-shell spectrum, we expand the $\theta$-functions in \eqref{fdiscinterim}, rewrite the integrals in terms of the spin $j$ and collect terms in the boundary modular parameter and fugacity  ($\taust$, $\nust)$ and also collect terms in the worldsheet modular parameters $(\tau_1, \tau_2)$. 
%
%
%
In addition we introduce a shorthand for the degeneracies of the worldsheet primaries and descendants
\be
\label{N=4osc}
\frac{(q\bar q)^{-\frac{c_{su(2)}}{24} }}{|\eta|^{10}}C^l_{r'}\bar C^l_{\bar r'}~S_r~S_{\bar r} ~Z_{T^4}^{bos}(q)= (q\bar q)^{\frac{1}{2} + \frac{1}{4k}}\sum_{h, N, \bar h, \bar N}d_{r,r',h,N}^l \bar q^{\bar h + \bar N}~,
\ee
to finally obtain:
\begin{align}
\label{GSform}
f_{\text{disc}}(\beta,\mu) = N  \sum_{l=0}^{\frac{k-2}{2}}&\int_{(\frac12,\frac{k+1}{2}]} \frac{dj}{\pi}
\sum_{r,r',f_i,h, N}~e^{\pi\ii\sum_{i=1}^4(f_i-\bar f_i)}~d^l_{r,r',f_i,h, N}\nonumber \\
& \int_{-1/2}^{1/2} d \tau_1e^{2\pi\ii \tau_1 (L_0-\bar L_0)} \int_0^\infty \frac{d \tau_2}{\tau_2} e^{-2\pi\tau_2 (L_0+\bar L_0-1)}\\
&\qst^{j+r+r'+f_1+f_2+1}~ \qbarst^{j+\bar r+\bar r'+f_1+f_2+1}~ \zst^{ \frac{ k w}{2} -r' + \frac{f_3+f_4 -f_1-f_2}{2}} ~\zbarst^{ \frac{ k w}{2}-\bar r'+ \frac{\bar f_3+ \bar f_4-\bar f_1-\bar f_2}{2}  }.\nonumber
\end{align}
We identify the left-moving worldsheet Virasoro generator:
\be
\label{leftVira}
\begin{aligned}
L_0 &= -\frac{j(j-1))}{k} +\frac{l(l+1)}{k} -w(j+r+r'+f_1+f_2+1) +h+N +\frac12\sum_{i=1}^4(f_i+\frac12)^2 \\
&=\left[ -\frac{j(j-1))}{k} - w(j+r) - \frac{(k+2)w^2}{4} \right] + \left[\frac{l(l+1)}{k}- w r' + \frac{(k-2)w^2}{4}\right]\\
&\hspace{7.4cm} +\frac12\sum_{i=1}^2(f_i-w+\frac12)^2+\frac12\sum_{i=3}^4(f_i+\frac12)^2\\
&= (L_0)_{\text{\text{sl}}(2,\mathbb{R})_{k+2}} + (L_0)_{\text{su}(2)_{k-2}} + (L_0)_{\text{fermions}}~,
\end{aligned}
\ee
with a similar expression for the right-moving Virasoro generator $\bar L_0$ with barred variables.

We  make a few remarks about the form of the  result. 
The form of the Virasoro generator makes contact with the construction of vertex operators in \cite{Giribet:2007wp} in which spectrally flowed operators in the anti-de Sitter space, the three-sphere and  the fermionic factors are paired.
The spacetime fermion number is counted by $F_s=\sum_{i=1}^4 f_i-\bar{f}_i$. We indeed compute $\Tr(-1)^{F_s}$ in the Ramond-Ramond sector of the boundary conformal field theory.
The phase factors $e^{\pi i \sum_{i=1}^4(f_i- \bar f_i)}$ in equation \eqref{GSform} make it clear that the oscillators with odd $f_i$  correspond to spacetime fermions. We thus confirm that we obtained the partition function in a manifestly spacetime supersymmetric (Green-Schwarz) form. 
This is a direct consequence of applying the generalized Jacobi identity and so$(8)$ triality. 
It would be interesting to derive the partition function directly from a  Green-Schwarz \cite{Pesando:1998wm,Rahmfeld:1998zn}, hybrid  \cite{Berkovits:1999im} or integrable supercoset approach  \cite{Sfondrini:2014via}.

\subsection{The Free Energy of On-shell States}

Our second goal is to obtain the on-shell contribution to the single string free energy. We revert to the earlier expression \eqref{fN=4bif}, perform the $\lambda_1$ and $s_2$ integral and expand the $\theta$-functions to end up with: 
\begin{align}
f_{\text{disc}}(\beta,\mu) =&   \frac{\sqrt{k}N}{2\pi}   \sum_{l=0}^{\frac{k-2}{2}}\sum_{r,\bar r,f_i, \bar f_i, r', \bar r'} C^l_{r} C^l_{\bar r'} \int_0^\infty \frac{d \tau_2}{\tau_2^{1/2}} \int_{-1/2}^{1/2} d \tau_1 \int_0^1 ds_1 \sum_{w} e^{\pi\ii\sum_{i=1}^4(f_i-\bar f_i)}\cr
& \int d \lambda_2
e^{2 \pi \ii( r+ r'+f_1+f_2-\bar{r} - \bar{r}'-\bar{f}_1-\bar{f}_2) (\frac{\mu \beta}{2 \pi} - w\tau_1)}e^{2 \pi i \lambda_2 (\frac{\beta}{2 \pi} - (s_1+w) \tau_2)}\cr
&e^{-2\pi\tau_2s_1 (r+r'+f_1+f_2+\bar r+\bar r'+\bar f_1+\bar f_2+3)}
~e^{2\pi\ii \nust( \frac{ k w}{2} -r' + \frac{f_3+f_4 -f_1-f_2}{2}) } ~e^{-2\pi\ii \nubarst( \frac{ k w}{2}-\bar r' + \frac{\bar f_3+\bar f_4-\bar f_1-\bar f_2}{2} ) }
\cr
&\frac{e^{-\pi k \tau_2 s_1^2} }{|\eta|^{10}}~q^{\frac12\sum_{i=1}^4( f_i+\frac12)^2+ \frac{l(l+1)}{k}-\frac{c_{su(2)}}{24}} ~ \bar q^{\frac12\sum_{i=1}^4(\bar f_i+\frac12)^2+ \frac{l(l+1)}{k}-\frac{c_{su(2)}}{24}}~ Z_{T^4}^{bos}(q)~ S_r~S_{\bar r}.
\end{align}
 In familiar fashion, we perform the $\lambda_2$-integral, leading to a $\delta$-function constraint for the $s_1$-variable, which can be solved.
After the combined integration, we obtain a sum over only non-negative winding numbers $w$:
\begin{align}
f_{\text{disc}}(\beta,\mu) =& \frac{\sqrt{k}N}{2\pi}   \sum_{l=0}^{\frac{k-2}{2}}\sum_{r,\bar r,f_i, \bar f_i,  r', \bar r',w\ge0} C^l_{ r'} \bar C^l_{\bar r'} \int_{\frac{\b}{2\pi(w+1)} }^{\frac{\b}{2\pi w}} \frac{d \tau_2}{\tau_2^{3/2}} \int_{-1/2}^{1/2} d \tau_1  e^{\pi\ii\sum_{i=1}^4(f_i-\bar f_i)}\cr
&e^{2 \pi \ii( r+ r'+f_1+f_2-\bar{r} - \bar{r}'-\bar{f}_1-\bar{f}_2) (\frac{\mu \beta}{2 \pi} - w\tau_1)}~e^{2\pi\ii \nust( \frac{ k w}{2} -r' + \frac{f_3+f_4 -f_1-f_2}{2}) } ~e^{-2\pi\ii \nubarst( \frac{ k w}{2}-\bar r' + \frac{\bar f_3+\bar f_4-\bar f_1-\bar f_2}{2} ) }\cr
&e^{-2\pi(\frac{\b}{2\pi} - w\tau_2) (r+r'+f_1+f_2+\bar r+\bar r'+\bar f_1+\bar f_2+3)}~e^{-\frac{\pi k}{\tau_2} (\frac{\b}{2\pi} - w \tau_2 )^2}  \cr
&\frac{1}{|\eta|^{10}}~q^{\frac12\sum_{i=1}^4( f_i+\frac12)^2+ \frac{l(l+1)}{k}-\frac{c_{su(2)}}{24}} ~ \bar q^{\frac12\sum_{i=1}^4(\bar f_i+\frac12)^2+ \frac{l(l+1)}{k}-\frac{c_{su(2)}}{24}}~ Z_{T^4}^{bos}(q)~ S_r~S_{\bar r}
  ~.
\end{align}
We  again code primaries and descendants as in equation \eqref{N=4osc}
and collect the terms in $\tau_1, \tau_2, \b$, and
 $\nu$ in the exponent. 
 The integral over $\tau_1$ imposes the level matching condition:
\be
h+N +\frac12\sum_{i=1}^4(f_i+\frac12)^2 -w( r+ r'+f_1+f_2) = \bar h+ \bar N+ \frac12\sum_{i=1}^4(\bar f_i+\frac12)^2-w(\bar{r} + \bar{r}' + \bar{f}_1 +\bar{f}_2)~,
\ee
%
while (minus) the single string free energy takes the form
\begin{multline}
f_{\text{disc}}(\beta,\mu) = \frac{4N}{\ii\b k} \sum_{l=0}^{\frac{k-2}{2}}\sum'_{r, r', f_i, h, N} d^l_{r,r',f_i,h, N}~ \sum_{w}e^{\pi\ii\sum_{i=1}^4(f_i-\bar f_i)}
z^{ \frac{ k w}{2} -r' + \frac{f_3+f_4 -f_1-f_2}{2} } ~\bar z^{ \frac{ k w}{2}-\bar r' + \frac{\bar f_3+\bar f_4-\bar f_1-\bar f_2}{2} }\\
\int~dc~ c~\int_{\frac{\b}{2\pi(w+1)} }^{\frac{\b}{2\pi w}}d \tau_2
e^{-4\pi\tau_2(h+N+\frac12\sum_{i=1}^4(f_i+\frac12)^2+\frac{kw^2}{4}-w(r+r'+f_1+f_2+\frac{3}{2}) +\frac{c^2}{k}-\frac12+\frac{1}{k}(l+\frac{1}{2})^2) }
\\
e^{ \ii\mu\b( r+ r'+f_1+f_2-\bar{r} - \bar{r}'-\bar{f}_1-\bar{f}_2)} e^{-\b(-2\ii c + r+r'+f_1+f_2+\bar r+\bar r'+\bar f_1+\bar f_2+3+kw)} ~.
\end{multline}
The prime indicates that level matching is imposed on the summation variables. 
The $\tau_2$-integral can  be done:
\be
\begin{aligned}
\frac{4}{\ii\b k}& \int_{\frac{\b}{2\pi(w+1)} }^{\frac{\b}{2\pi w}}d \tau_2
e^{-4\pi\tau_2(h+N+\frac12\sum_{i=1}^4(f_i+\frac12)^2+\frac{kw^2}{4}-w(r+r'+f_1+f_2+\frac{3}{2}) +\frac{c^2}{k}-\frac12+\frac{1}{k}(l+\frac{1}{2})^2) }\\
&=\frac{1}{\pi \ii\b k}\left[ \frac{e^{- \frac{2\b}{w+1}(h+N+\frac12\sum_{i=1}^4(f_i+\frac12)^2-\frac{kw^2}{4}-w(r+r'+f_1+f_2+\frac{3}{2}) +\frac{c^2}{k}-\frac12+\frac{1}{k}(l+\frac{1}{2})^2) }}{\scriptstyle{\frac{c^2}{k}-\frac12+\frac{1}{k}(l+\frac{1}{2})^2 h+N+\frac12\sum_{i=1}^4(f_i+\frac12)^2+\frac{kw^2}{4}-w(r+r'+f_1+f_2+\frac{3}{2}) }}\right.\\
&\hspace{3.5cm} \left. -\frac{e^{- \frac{2\b}{w}(h+N+\frac12\sum_{i=1}^4(f_i+\frac12)^2+\frac{kw^2}{4}-w(r+r'+f_1+f_2+\frac{3}{2}) +\frac{c^2}{k}-\frac12+\frac{1}{k}(l+\frac{1}{2})^2) }}{\scriptstyle{\frac{c^2}{k}-\frac12+\frac{1}{k}(l+\frac{1}{2})^2+h+N+\frac12\sum_{i=1}^4(f_i+\frac12)^2+\frac{kw^2}{4}-w(r+r'+f_1+f_2+\frac{3}{2}) }}\right]~.
\end{aligned}
\ee
Our focus is on the contribution from the discrete sector that arises from the residues of the poles that are picked up in the region: 
\be
\frac{kw}{2} < \text{Im}(c) \le \frac{k(w+1)}{2} ~,
\ee
as a result of the shifts in the contours. The poles are located at
\begin{align}
\label{polelocationsN=4}
-c^2 &= k\left(h+N+\frac12\sum_{i=1}^4(f_i+\frac12)^2+\frac{kw^2}{4}-\frac{w}{2}+\frac{1}{k}(l+\frac{1}{2})^2-w(r+r'+f_1+f_2+1)-\frac12\right) \nonumber \\
&=\frac{1}{4}(2j-1+kw)^2~.  
\end{align}
where we have used the on-shell condition $L_0=\frac12$ to rewrite the right hand side in terms of the spin $j$.  The residues again  represent the contribution of the contribution of the on-shell discrete states to the single string free energy, which can be written in terms of the spacetime fugacities, with the spin $j$  determined by the on-shell condition:
\begin{align}
f_{\text{disc}}(\beta,\mu) 
 =&\frac{N}{\b }\sum_{l=0}^{\frac{k-2}{2}}\sum'_{r, r', f_i, h, N} d^l_{r, r', f_i, h, N} \sum_{w}e^{\pi\ii\sum_{i=1}^4(f_i-\bar f_i)} \\
&\hspace{2cm}\qst^{j+r+ r'+f_1+f_2+1} ~\qbarst^{j+\bar{r} + \bar{r}' + \bar{f}_1+\bar{f}_2+1}~\zst^{ \frac{ k w}{2} -r' + \frac{f_3+f_4 -f_1-f_2}{2} }~  \zbarst^{\frac{ k w}{2}-\bar r' + \frac{\bar f_3+\bar f_4-\bar f_1-\bar f_2}{2}  } ~. \nonumber
\end{align}

\subsection{A BPS Bound}

 We  derive a positivity bound for the left and right-moving conformal dimension of the boundary theory (ignoring for simplicity the ubiquitous constant $c_{\text{s.t.}}/24$ in the Ramond-Ramond sector). For future generalizations, it is instructive to first undo the spectral flow in the su$(2)$ sector. This is easily accomplished by using \eqref{su2chiw} in reverse and redoing the subsequent steps. This essentially amounts to the shifts: 
 \be
 r' \longrightarrow r ' + \frac{k-2}{2}w~,
 \qquad f_2\longrightarrow f_2+w~.
\ee
 The left-moving R-charge and conformal dimension  take the form:
 \begin{align}
 \label{QT4origin}
 Q_{\text{s.t.}}^R &= -r' + \frac{w}{2}+\frac12(f_3+f_4-f_1-f_2)~,\\
 H &= j+r + \frac{k}{2} w + f_1+f_2+1 + r' \,,
 \label{HT4origin}
 \end{align}
 while the associated worldsheet Virasoro generator is given by
 \be
 L_0 = -\frac{j(j-1)}{k} -\frac{k w^2}{4} -w(j+r+f_1+\frac{1}{2})+ \frac{l(l+1)}{k}+\frac12\sum_{i=1}^4( f_i+\frac12)^2+h_{int}~.
 \ee
In what follows we shall write the expression for the dimension $H$ along with the Virasoro generator $L_0$ in tandem so as to keep in mind the on-shell constraint $L_0=\frac12$ that constrains the spins and fermion numbers appearing in $H$. There  is an additive  structure to both $H$ and $L_0$: there is a term from the sl$(2)$ sector and the associated fermions, a contribution from the compact su$(2)$ sector and the associated fermions, and one from the internal manifold and the various oscillators. 

We first work with the sl$(2)$ sector and restrict ourselves to the $r\ge 0$ case. 
We will find it useful to define a shifted spin 
in the sl$(2)$ sector: 
\begin{equation}
\jtilde = j + \frac{kw}{2} - \frac12~. 
\end{equation}
In terms of this spin we have 
\begin{align}
\label{HT4one}
H &= \jtilde+\frac12+(r +f_1+\frac12)+(r' + f_2+\frac12)  \, \\
L_0 &= - \frac{\jtilde^2}{k}   -w(1+r+f_1)+ \frac{1}{k}\big(l+\frac12\big)^2+\frac12\sum_{i =1}^4 ( f_i+\frac12)^2
+h_{int}~.
\label{HL0one}
\end{align}
Let us  consider the contributions to the energy $H$
and the level, namely the contribution to the worldsheet conformal dimension above the conformal dimension of the primary (or above the ground state energy $1/8$ for the fermionic sectors).
It is clear that the compact and fermionic sectors contribute positively to the level. At fixed contribution to the level, we wish to minimize  the contribution to $H$. This leads to the conclusion that $r, r'$ and $f_i$ are minimal at a fixed compact level. Put differently, at fixed compact contribution to the dimension $H$, we must minimize the contribution of a state to the level. Indeed, otherwise this leads to an increase in the non-compact contributions $\widetilde{j}, w, f_1$, which leads to an increase in dimension $H$.  

\begin{figure}[ht]
\center{
\includegraphics[width=\textwidth] {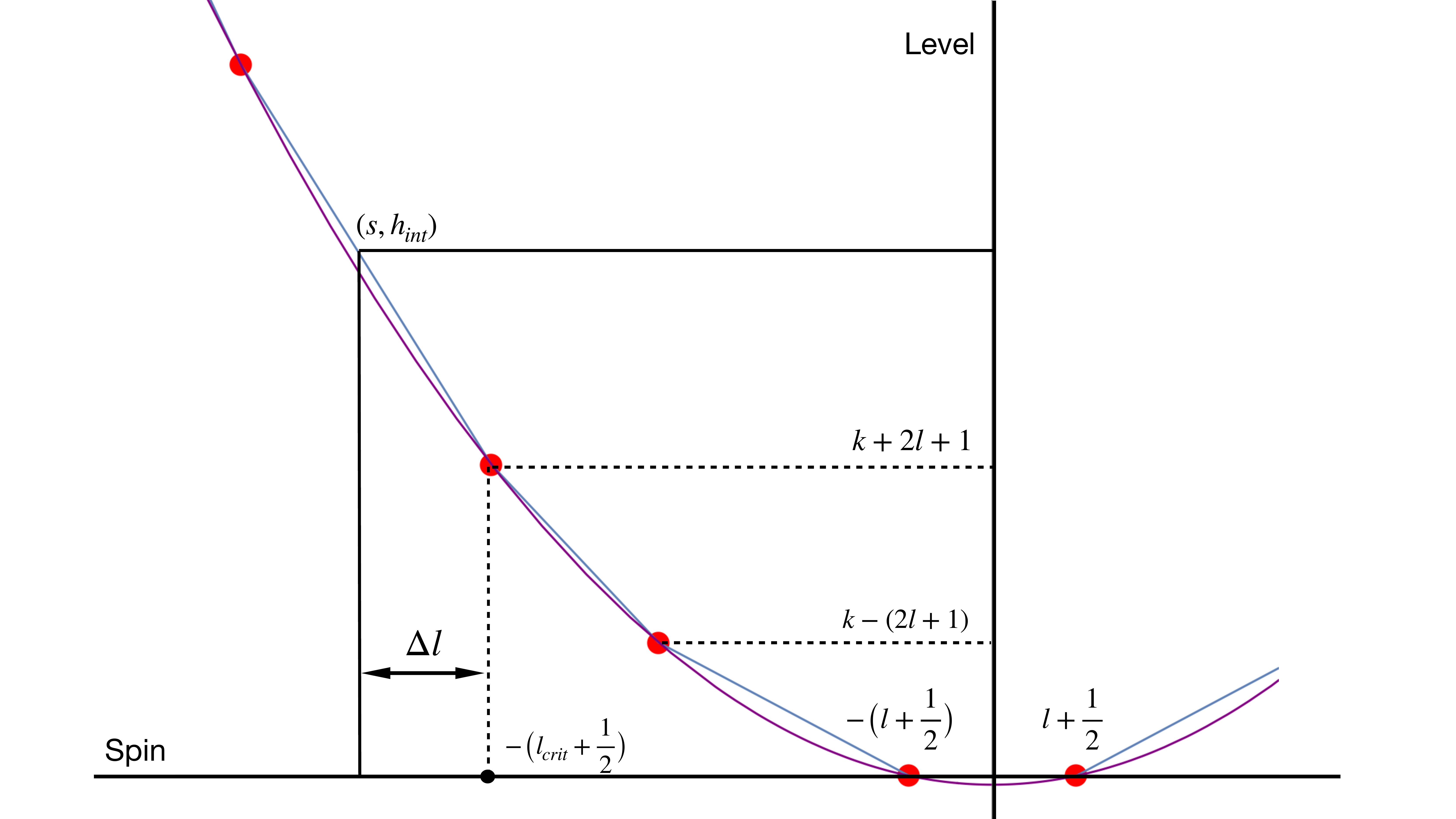}
 \caption{\label{su2susymod} We drew the left quadrant of the weight diagram of an irreducible supersymmetric su$(2)$ current algebra module. The weight diagram is bounded by the blue line segments which are circumscribed by a the purple parabola. Spectral flowed ground states (in red) are the only states that lie on the parabola.
 }}  
\end{figure}

These considerations lead to the following conclusions: first of all we set 
\be
r=0~, \qquad \text{and} \qquad f_{3}~, f_4 \in \{0,-1\}.
\ee
Secondly the combination $r'+f_2$ must be minimized at a given level for both the fermionic $f_2$ contribution and the bosonic contribution $r'$ to the spin. This is precisely the problem that we address in appendix \ref{TheShapeofAffineModules} and we refer the reader to the appendix for details. What we need to proceed is summarized in Figure \ref{su2susymod}. 
The point in the (spin,level) plane marked $(s,h_{int})$ lies on the line segment that is part of the supersymmetric su$(2)$ current algebra module. For a given level we see that the minimal choice of spin is given by:
\be
\begin{aligned}
s = r'+f_2 +\frac12 &= -\big(l_{crit}+\frac12\big) -\Delta l \\
&= -l-\frac{k\widetilde{w}}{2}-\frac12-\Delta l~.
\end{aligned}
\ee
Here we have used the critical value $l_{crit}$ that is derived in Appendix \ref{TheShapeofAffineModules}. It is the spin value of the spectral flowed ground state that lies on the  parabola that circumscribes the weight diagram just below the level $h_{int}$. This is expressed in terms of the spin $l$ and an arbitrary integer $\widetilde w$ that is determined by the segment on which the point $(s,h_{int})$ lies. We have assumed that $\widetilde w$ is even, else we use the mirror value of $l_{crit}$ (see Appendix \ref{TheShapeofAffineModules}). The oscillator contribution to the worldsheet conformal dimension can determined to be  -- see Appendix \ref{TheShapeofAffineModules} -- :
\begin{align}
h_{int} &= \frac{1}{k} \big( (l_{crit}+\frac12)^2-(l+\frac12)^2 \big) + (\widetilde{w}+1) \Delta l  \, .
\end{align}
So far we have parametrized the su$(2)$ spin, fermion number and the oscillator contributions in terms of $(\Delta l, \widetilde w)$. We have also set $r=0$ and  fixed all the fermion numbers except $f_1$ to minimize $H$. By inspection we can see that $f_1=-1$ minimizes $H$.  Plugging all of these values into equations \eqref{HT4one} and \eqref{HL0one} we obtain:
\begin{align}
H_{min} &= \widetilde{j}  -\widetilde{l}-\Delta l \, \\
L_0 &= - \frac{\widetilde{j}^2}{k}  +\frac{\widetilde{l}^2}{k} + (\tilde{w}+1) \Delta l +\frac12
~.
\end{align}
where we have defined
\be
\ltilde = l + \frac{ k \widetilde w}{2} + \frac12~.
\ee

Our final task is to understand how to choose  $(\widetilde{w},\Delta l,l)$ given $\widetilde{j},w$ in order to obtain the strongest bound on the dimension $H$. We eliminate $\widetilde{j}$ using the on-shell condition $L_0=\frac12$ and extremize the resulting expression with respect to $\Delta l$. The ensuing formula for $H$ is precisely the distance of the segment from the parabola and this is minimized when $\Delta l=0$. The state of the su$(2)$ affine module lies {\it on} the parabola. We summarize
\begin{align}
\label{HT4two}
H_{min} &= \widetilde{j} - \widetilde{l}\\
L_0&= -\frac{\widetilde{j}^2}{k} + \frac{\ltilde^2}{k} +\frac{1}{2}~.
\label{L0T42}
\end{align}
The on-shell condition then leads to $\jtilde = \ltilde$. Given the ranges of the spins $j$ and $l$, this fixes $\widetilde w = w$ and we obtain the bound:
\be
H \ge  0~.
\ee
To close a final loophole, we briefly explain why the choice $r=0$ versus $r<0$ is optimal to minimize the energy  in the non-compact sector of a superstring compactification. Let us start with $r=0$, and see how the energy changes when $r<0$. For $r=0$,  we schematically denote the solution of the on-shell condition as $\widetilde{j}=\sqrt{k A}$. This fixes a winding number $w$. We will analyze what happens when we keep all moving parts fixed, except for $r$ and the spin $\widetilde{j}$.  We turn on $r<0$. We have that the energy term $H_{sub}=\widetilde{j}+r$ seemingly goes down (barring what happens to the spin $\widetilde{j}$) and the world sheet scaling dimension $L_0$ is augmented by $(w+1)(-r)$. We re-adjust the spin $\widetilde{j}$ to satisfy the on-shell constraint and find:
\be
\begin{aligned}
\widetilde{j} &=  \sqrt{k(w+1)(-r)+kA}~,\\
H_{sub} &= \widetilde{j}+r = \sqrt{k (w+1)(-r)+k A} + r \, .
\label{ReasoningR}
\end{aligned}
\ee
We want to know the minimum of the energy $H_{sub}$, for $r \le 0$ in a certain range. 
From our original assumption, we have that $k A = \widetilde{j}^2$  lies between $ (kw)^2/4 < kA \le \big(k(w+1)\big)^2/4$.  We should study the  values of $r$ for which the formulas (\ref{ReasoningR}) are valid and analyze the energy $H_{sub}$ in that range. It can be shown that the resulting energy is always larger or equal than the original energy $\sqrt{k A}$, namely that the choice $r=0$ is optimal. Thus our proof is complete.

\subsection{The Ramond-Ramond Sector Ground States}

Finally, let us systematically solve for the states that satisfy the extremal conditions $H=0$ and  $\bar{H}=0$. These are the left and right moving Ramond sector ground states of the boundary theory. A careful look at the proof of the bound demonstrates that ground state quantum numbers appearing in \eqref{HT4origin} must satisfy:
\be
j = l+1~,\quad r' =- l-\frac{k-2}{2} w~,\quad f_1=-1~,\quad f_2 = -w-1  ~.
\ee
They enjoy a four-fold degeneracy captured by the fermion number values
\be
f_3 \in \{-1,0\}\quad \text{and} \quad f_4\in \{-1,0\}~.
\ee
One can check that for all these and only these values do we have $H=0$. It is important to note that the su$(2)$ spin $l$ can take $k-1$ values. Thus there are $4(k-1)$ RR sector ground states. The boundary u$(1)$ R-charge $Q^R_{\text{s.t.}}$ of these states (see \eqref{QT4origin}) is given by: 
\be
\begin{aligned}
Q^R_{\text{s.t.}} 
&=\frac{kw}{2} + j +\frac{f_3+f_4}{2}~,
\end{aligned}
\ee
and similarly for the right-movers. For the four left-moving ground states for a given value of the spin $j$, the R-charges are given in Table \ref{tab:RRgroundstates}. We  defined the handy combination of quantum numbers $n=2j-1+kw$.%
\begin{table}[H]
    \centering
    \begin{tabular}{c|c}
       $(f_3,f_4)$  & $2Q^R_{\text{s.t.}}$  \\
       \hline
       $(-1,-1) $ & $n-1 $\\
       $(-1,\phantom{-}0)$ & $n$\\
       $(\phantom{-}0,-1)$ & $n$\\
       $(\phantom{-}0,\phantom{-}0)$ & $n+1$\\
    \end{tabular}
    \caption{The boundary R-charge of the Ramond ground states for a given value of $n=2j-1+kw$ and  the fermion numbers $f_{3,4}$.}
    \label{tab:RRgroundstates}
\end{table}
\noindent
 The (chiral,chiral) primaries have been described 
 explicitly in terms of vertex operators in the bulk string theory in \cite{Argurio:2000tb, Giribet:2007wp}. They fall into four infinite families, labelled by a positive integer $w \ge 0$ as well as an integer $n=2j-1+kw$,
where $2j-1=1,2,\dots,k-1$. Thus, the floors $n=0,k,2k,\dots$ are missing from the four towers. These vertex operators 
map one-to-one to the Ramond-Ramond sector ground states that we identified.  Through path integral methods, we  have not only confirmed this list, but also proven
that the classification is complete. 

\subsubsection{Summary}

We have focused on the contribution of the discrete states to the single string free energy of superstrings on $AdS_3\times S^3\times T^4$.  From the single string free energy we could read off the left/right conformal dimensions and R-charges of the discrete states in the Ramond-Ramond sector of the boundary theory. We then went on to prove positivity bounds that led to a complete classification of Ramond-Ramond ground states. 

Let us highlight some of the features of our derivation. First of all our approach includes manifest spacetime supersymmetry thanks to the use of the abstruse Jacobi identity. The spectrum  we obtained in the Ramond-Ramond sector is evidently made of supersymmetry multiplets and the proof of the BPS bound proceeds rather straightforwardly on-shell. Lastly, we stress that  our approach is  universal -- it can be applied to any supersymmetric superstring background of the form $AdS_3 \times N$. This includes  string scale compactifications  as well as non-K\"ahler compactification manifolds, as we illustrate in the next section.

\section{\texorpdfstring{Superstrings in Thermal $AdS_3 \times S^3 \times S^3 \times S^1$}{}}
\label{ads3s3s3s1}
\label{AdS3S3S3S1}
In this section, we provide a second application of the calculation of the superstring free energy in thermal $AdS_3$.  We consider a background $AdS_3 \times S^3 \times S^3 \times S^1$ spacetime at supersymmetric levels $(k;k_1,k_2)$ where
\begin{equation}
\frac{1}{k} = \frac{1}{k_1} + \frac{1}{k_2} \, 
\end{equation}
in order to have a critical superstring background. 
We analyze the off-shell and on-shell single string free energy  along the lines of previous sections.  Since the intermediate steps are familiar by now, we exclusively comment on the new features.

\subsection{The Free Energy}

We start out with the twisted single string contribution to the free energy in the GSO projected NSR formalism:
\be
\begin{aligned}
f(\beta, \mu) &= \frac{1}{2\pi} \int_0^\infty \frac{d \tau_2}{\tau_2^{3/2}} \int_{-\frac{1}{2}}^{\frac{1}{2}} d \tau_1 
 \frac{\sqrt{k} e^{-\frac{k \beta^2} { 4 \pi \tau_2}} }{|\theta_1(  \tau_{\text{s.t.}},\tau )|^2}
 |\eta|^{4}     \sum_{l_1=0}^{\frac{k_1-2}{2}} |\chi_{l_1}(\nup, \tau) |^2  ~ \sum_{l_2=0}^{\frac{k_2-2}{2}} |\chi_{l_2}(\num, \tau) |^2~Z_{U(1)}  \\
&\hspace{2cm}\frac{1}{4 |\eta|^8}\sum_{a,b} e^{2\pi\ii(a+b+2ab)}
\theta_a^b (\taust, \tau)
\theta_a^b (\nup, \tau)
\theta_a^b (\num, \tau)
\theta_a^b (0, \tau)\\
&\hspace{3cm}\sum_{\bar{a},\bar{b}} e^{2\pi\ii(\bar a +\bar b + 2 \bar a \bar b) }
\theta_{\bar a}^{\bar b} (\taubarst, \bar\tau)
\theta_{\bar a}^{\bar b} (\nupbar \bar\tau)
\theta_{\bar a}^{\bar b} (\numbar, \bar\tau)
\theta_{\bar a}^{\bar b} (0, \bar\tau)   \, . 
\end{aligned}
\ee
We have introduced twists $\nu_{1,2}$ with respect to the global symmetries of the two three-spheres.
We again use the Jacobi identity: 
\begin{multline}
\sum_{a,b} e^{2\pi\ii(a+b+2ab)}
\theta_a^b (\taust, \tau)
\theta_a^b (\nup, \tau)
\theta_a^b (\num, \tau)
\theta_a^b (0, \tau)\\
=2\theta_1(\frac{\nup+\num-\taust}{2},\tau)\theta_1(\frac{\nup+\num+\taust}{2},\tau)\theta_1(\frac{\nup-\num-\taust}{2},\tau)\theta_1(\frac{\nup-\num+\taust}{2},\tau)~,
\end{multline}
to go to a manifestly supersymmetric description. We identify the boundary R-charge  \cite{Gukov:2004fh, Eberhardt:2017pty}:
\be
Q^R_{\text{s.t.}} =\gamma (K_0^3)_1 + (1-\gamma)(K_0^3)_2~,
\ee 
where the parameter $\gamma$ is determined in terms of the supersymmetric levels: 
\be
\gamma = \frac{k_2}{k_1+k_2}~.
\ee
The charge $(K_0^3)_a$ is the zero mode of the third component of the su$(2)_{k_a}$ super current algebra. We restrict the pair $(\nu_1,\nu_2)$ to the fugacity $\nu$ that couples to R-charge:
\be
\nup =\gamma\, \nu~,\qquad \num=(1-\gamma) \nu ~.
\ee
and similarly for the right-movers. For both the left and right movers we  spectrally flow by half a unit in the boundary theory so that we calculate the single string free energy in the Ramond-Ramond sector of the boundary theory:
\be
\nu \rightarrow \nu - \taust~,\qquad \nubarst \longrightarrow \nubarst -\taubarst~.
\ee
As before this leads to a  normalization constant $N=q_{s.t}^{\frac{c_{\text{s.t.}}}{24}} z^{-\frac{c_{\text{s.t.}}}{6}}$ up front and the single string free energy takes the form:
\begin{align}
f(&\beta, \mu) = \frac{N}{2\pi|\eta|^4} \int_0^\infty \frac{d \tau_2}{\tau_2^{3/2}} \int_{-\frac{1}{2}}^{\frac{1}{2}} d \tau_1 
 \frac{\sqrt{k} e^{-\frac{k \beta^2} { 4 \pi \tau_2}} }{|\theta_1(  \tau_{\text{s.t.}},\tau )|^2} \nonumber \\
 & \sum_{l_1=0}^{\frac{k_1-2}{2}} |\chi_{l_1}(\gamma(  \taust- \nu), \tau) |^2  ~ \sum_{l_2-=0}^{\frac{k_2-2}{2}} |\chi_{l_2}((1-\gamma)( \taust-\nu), \tau) |^2~Z_{U(1)} \\
&| \theta_1(\frac{\nu}{2}, \tau)|^2| \theta_1(\tau_{\text{s.t.}} - \frac{\nu }{2}, \tau)|^2
| \theta_1(\gamma\tau_{\text{s.t.}} -(\gamma-\frac12)\nu), \tau)|^2| \theta_1((1-\gamma)\tau_{\text{s.t.}} +(\gamma-\frac12)\nu), \tau)|^2\nonumber~.
\end{align}
\subsubsection{The Discrete Sector}
We proceed by introducing the integral over the holonomies, expanding the su$(2)$ characters, the $\theta$-functions et cetera and going through the same steps as in the previous sections. 
We skip all of the details and present the contribution of the discrete states to the free energy  as a sum over the off-shell states in the Hilbert space \footnote{In this formula the function $d^{l_a}_{N,h,r,r'_a}$ keeps track of the degeneracies of the descendants that are encoded in the series $S_r$ and $C^{l_a}_{r_a}$:
\be
\label{s3s1oscsum}
\frac{(q\bar q)^{-\frac{1}{24} \sum_{a=1}^2 c^{(a)}_{su(2)}}} {|\eta(\tau)|^{10}}C^{l_1}_{ r'_1} \bar C^{l_1}_{\bar r'_1}C^{l_2}_{ r'_2} \bar C^{l_2}_{\bar r'_2} ~ S_r~S_{\bar r}~Z_{U(1)}=(q\bar q)^{-\frac{1}{2}+\frac{1}{4k}}\sum_{h,\bar h, N, \bar N} d^{l_a}_{N,h,r,r'_a} q^{h+N} \bar q^{\bar h +\bar N }~.
\ee
}:
\be
\label{fAdS3S3S3S1disc2}
\begin{aligned}
f_{\text{disc}}(\beta,\mu) = N   &\sum_{l_1=0}^{\frac{k_1-2}{2}} \sum_{l_2=0}^{\frac{k_2-2}{2}}
\sum_{r,\bar r,f_i, \bar f_i, r'_a, \bar r'_a} e^{\pi\ii(\sum_{i=1}^4( f_i -\bar f_i)} \sum_{h,\bar h, N, \bar N} d^{l_a}_{N,h,r, f_i, r'_a}\\
&\sum_{w}\int_{(\frac12, \frac{k+1}{2}]} \frac{dj}{\pi}
\int_0^\infty \frac{d \tau_2}{\tau_2} \int_{-1/2}^{1/2} d \tau_1~ e^{2\pi\ii\tau_1(L_0-\bar L_0)} ~e^{-2\pi\tau_2(L_0 + \bar L_0 - 1)}\\
&\qst^{j+\frac{(k+2)w}{2} + 1 + r+f_1 + \gamma(r'_1+f_3) + (1-\gamma)(r'_2+f_4)} \qbarst^{j+\frac{(k+2)w}{2} + 1 + \bar r+\bar f_1 + \gamma(\bar r'_1+\bar f_3) + (1-\gamma)(r'_2+\bar f_4)} \\
 &z^{ -\frac12(f_1-f_2)- \gamma r'_1 - (1-\gamma) r'_2 - (\gamma-\frac12) ( f_3-f_4)  } ~\bar z^{ -\frac12( \bar f_1-\bar f_2)- \gamma \bar r'_1 - (1-\gamma) \bar r'_2 - (\gamma-\frac12) (\bar  f_3-\bar f_4) }~.
\end{aligned}
\ee
The worldsheet Virasoro generators are given by
\be
L_0 = -\frac{j(j-1)}{k} -\frac{(k+2)w^2}{4} -w(j+r)+ \sum_{a=1}^2\frac{l_a(l_a+1)}{k_a}+\frac12\sum_{i=1}^4( f_i+\frac12)^2+h+N~, \label{WorldsheetAdS3S3S3S1}
\ee
and similarly for the right-movers, with the barred $(\bar r, \bar f_i, \bar h, \bar N)$ variables. One can clearly identify the contribution to the worldsheet dimension from the spectral flowed spin $j$ representation in the bosonic sl$(2,\mathbb{R})_{k}$ sector, the spin $l_a$ representation in the bosonic su$(2)_{k_a}$, the contribution from the four sets of fermions and lastly the contribution from the internal space and the descendants. 

One can  perform the $(\tau_1, \tau_2)$ integrals to write the free energy as a sum over on-shell states:
\be
\label{S3S1fdisc}
\begin{aligned}
f_{\text{disc}}(\beta,\mu) =&\frac{N}{\b }\sum_{l=0}^{\frac{k-2}{2}}\sum'_{r, r', f_i, h, N} d_{N,r, r'_a, f_i}(h,\bar h) \sum_{w\ge 0}e^{\pi\ii\sum_{i=1}^4(f_i-\bar f_i)}\\
&\qst^{j+\frac{(k+2)w}{2} + 1 + r+f_1 + \gamma(r'_1+f_3) + (1-\gamma)(r'_2+f_4)} \qbarst^{j+\frac{(k+2)w}{2} + 1 + \bar r+\bar f_1 + \gamma(\bar r'_1+\bar f_3) + (1-\gamma)(r'_2+\bar f_4)} \\
&z^{ -\frac12(f_1-f_2)- \gamma r'_1 - (1-\gamma) r'_2 - (\gamma-\frac12) ( f_3-f_4)  } ~\bar z^{ -\frac12( \bar f_1-\bar f_2)- \gamma \bar r'_1 - (1-\gamma) \bar r'_2 - (\gamma-\frac12) (\bar  f_3-\bar f_4) }~.
\end{aligned}
\ee
The prime indicates the level-matching condition $L_0=\bar L_0$ and we note that the sum over winding $w$ is restricted to  the non-negative integers. As before the exponents of the nome $\qst$ and $z$ end up being the same in both the on-shell and off-shell cases, but one has to keep in mind that the on-shell condition $L_0=\frac12$ imposes a  constraint among the various quantum numbers. 

\subsection{A  BPS  Bound} 

From the on shell free energy one can read off the left moving conformal dimension to be 
\begin{align}
\label{Hone}
H &=  j+r+\frac{k+2}{2}w + (\frac{1}{2}+f_1) + 
 \gamma(r_1'+f_3+\frac12)+(1-\gamma) (r_2'+f_4+\frac12)  \, .
\end{align}
The spins are constrained by the on-shell condition $L_0=\frac12$, with the left-moving Virasoro generator given by: 
\be
L_0 = -\frac{j(j-1)}{k} -\frac{(k+2)w^2}{4} -w(j+r)+ \sum_{a=1}^2\frac{l_a(l_a+1)}{k_a}+\frac12\sum_{i=1}^4( f_i+\frac12)^2+h+N~.
\ee
Our goal is to obtain a BPS bound that leads to a minumum value for the dimension $H$. We proceed along the same lines as in the $AdS_3\times S^3\times T^4$ case. We first consider the  sl$(2)$ sector and by the same arguments find that  the minimum value of the dimension $H$ is obtained by setting 
\be
\label{optimalrf}
r=0~, \qquad f_1 = -w-1~.
\ee
As explained in detail previously, the interplay between the dimension $H$ and the worldsheet Virasoro generator ensures that at a fixed compact contribution to $H$, we must minimize the  contribution to the level. As a consequence, the fermion number $f_2$ is set to its ground state values:
\be
\label{optimalf2}
f_2 \in \{ 0,-1 \}~.
\ee
We next consider the supersymmetric su$(2)_a$ affine module and parametrize the spin, fermion number and level in terms of $(\widetilde{w}_a, \Delta l_a)$, where $\widetilde{w}_a$ and $\Delta l_a$ measures its distance from the special values of spin that intersect the parabola shown in figure \ref{su2susymod}. See also Appendix \ref{TheShapeofAffineModules}.
\begin{align}
s_a &= r'_a+f_{a+2} +\frac12 
= -l_a-\frac{k_a\widetilde{w}_a}{2}-\frac12-\Delta l_a\\
h_{int, a} &= \frac{1}{k_a} \big( (l_{a,crit}+\frac12)^2-(l_a+\frac12)^2 \big) + (\widetilde{w}_a+1) \Delta l_a  \, .
\end{align}
Substituting these into the expressions for $H$ and $L_0$, we then extremize $H$ with respect to the free variables. We once again find that $H$ is minimized when $\Delta l_a = 0$. 
At these optimal values, and after a bit of algebra, we obtain the following expressions for $H$ and $L_0$:
\begin{align}
\label{s3s1ineq1}
H_{min} &= \widetilde{j} -\gamma  \ltilde_1- (1-\gamma)\ltilde_2~, \qquad
L_0= -\frac{\widetilde j^2}{k} +\frac{\ltilde_1^2}{k_1} + \frac{\ltilde_2^2}{k_2} +\frac12~.
\end{align}
where we have defined:
\be
\ltilde_1= l_1+\frac{k_1w_1}{2} +\frac12~,\quad \ltilde_2= l_2+\frac{k_2w_2}{2}+\frac12 \quad\text{and}\quad \widetilde{j} = j+\frac{kw}{2}-\frac12~.
\ee
We use the on-shell condition to solve for the spin $\widetilde j$ and substitute into the expression for $H$ to obtain
\be
H_{min} = \sqrt{\gamma \widetilde{l}_1^2 + (1-\gamma)\ltilde_2^2} -\gamma  \ltilde_1- (1-\gamma)\ltilde_2~.
\ee
Here we have used the relation between the levels $k = \gamma k_1 = (1-\gamma) k_2$. Extremizing this expression with respect to the spectral flow parameters $\widetilde {w}_a$ or effectively with respect to the spin $\ltilde_a$, we obtain the equalities 
\be
\widetilde{l}_1 = \ltilde_2  = \widetilde j~, \label{SpinEquality}
\ee
and find that the minimal energy $H_{min}$ equals zero, as expected in the boundary Ramond-Ramond sector, up to the constant shift by $c_{\text{s.t.}}/24\,$. The final result agrees with \cite{Eberhardt:2017pty}.

\subsection{The Ramond-Ramond Sector Ground States}

Let us focus on the ground states of  dimension $H=0$. 
 The boundary u$(1)$ R-charge $Q^R_{\text{s.t.}}$ of these states can be read off from the exponent of the fugacities $\zst$ and $\zbarst$ in \eqref{s3s1oscsum} 
\be
\begin{aligned}
Q^R_{\text{s.t.}} 
&= -\frac12(f_1-f_2)- \gamma r'_1 - (1-\gamma) r'_2 - (\gamma-\frac12) ( f_3-f_4) \\
&= \frac{1}{2}(f_2+f_3+f_4 - f_1) - \gamma(r'_1+f_3)-(1-\gamma)(r'_2+f_4)~.
\end{aligned}
\ee
and similarly for the right-movers. Substituting the  quantum numbers for the ground states that we classified through our proof, we see first of all, that the allowed values of $f_2\in\{0,-1\}$ lead to a twofold degeneracy of states. Explicitly we obtain 
\be
Q^R_{\text{s.t.}} = \frac{f_2}{2} + \widetilde{j}+ \frac{\delta}{2}~.
\ee
where we have defined the shift $\delta = w-w_1-w_2$ which will turn out to take values in the set $\{0,1\}$. Importantly, from the equality  (\ref{SpinEquality})
of spins, we conclude that $2 \widetilde{j}$ takes values in the strictly positive integers, but skips all multiples of both $k_1$ and $k_2$, namely we have $2 \widetilde{j} \in \mathbb{Z}_{>0} \setminus (k_1 \mathbb{Z} \cup k_2 \mathbb{Z})$.

A non-trivial Diophantine task remains:
for each spin $2 \widetilde{j}$ one needs to determine the value of $\delta$. Technically, this coincides with a calculation  carried out in the NS/R formalism in \cite{Eberhardt:2017pty}. In the following reasoning, we crucially use the lemmas we state and prove in Appendix \ref{DiophantineHarmonics}. We will once again suppose that $k \ge 2$. It is not hard to see that the shift $\delta$ equals zero between  $0$ and $k$, since $w=0=w_a$. Then, it becomes one at $2 \widetilde{j}-1=k$ since $w$ jumps to $1$ while $w_a$ remains zero. When $ 2 \tilde{j}-1$ hits a multiple of $k_a$, $w_a$ is augmented by one, and $\delta$ becomes zero again. These steps up and steps down essentially alternate with exceptions proven in Appendix \ref{DiophantineHarmonics}. The net effect is that we create gaps in the spectrum at (the integer part of) the multiples of $k$ while we close the gaps that used to exist at multiples of $k_a$. However, at an interval which corresponds to the case where $k(w+1)$ is a common multiple of $(k_1,k_2)$, we have that $\delta$ remains one throughout.  Thus, this gap in the original spectrum is simply filled.
The net result is that we have gaps in the values of $2 \widetilde{j}-1$ at multiples of $k$, except where these are multiples of the lowest common multiple of $(k_1,k_2)$. See also \cite{Eberhardt:2017pty}.

Thus,  
 the integer combination $m=2j-2+kw+\delta$ takes values in the set \be
 \label{gapss1}
m \in  \mathbb{Z}_{\ge 0} \setminus \big( \left \lfloor{k \mathbb{Z}} \right \rfloor \setminus  \text{l.c.m.}(k_1,k_2)\mathbb{Z} \big) \, .
\ee
We made use of the floor function.
We list below the left-moving ground states along with their R-charges:
\begin{table}[H]
    \centering
    \begin{tabular}{c|c}
       $f_2$  & $2Q^R_{\text{s.t.}}$  \\
       \hline
       $ -1 $ & $m $\\
       $ 0 $ & $m+1$\\
    \end{tabular}
    \caption{The boundary R-charge of the Ramond ground states for a given value of integers $(m, f_2)$. The pair of states form an ${\mathcal N}=4$ multiplet.}
    \label{tab:RRgroundstates:s3s1}
\end{table}
\noindent
It is important to note that our description is manifestly supersymmetric, and proves that the classification performed in \cite{Eberhardt:2017pty} is indeed complete. We note a new phenomenon in this model, which is the  contribution of a  discrete representation at  $j=(k+1)/2$. In section \ref{BosonicAdS3} we showed that a Feynman regularization of the radial momentum integral indeed gives rise to such a state in the spectrum of positive energy states.

As a small check on this result, let us take the infinite level limit $k_2\rightarrow\infty$. In this limit, the parameter $\gamma$ approaches one and the two other level match  $k_1= k\in \mathbb{Z}_{\ge 2}$. The background reduces to the large  radius limit of $AdS_3\times S^3\times \mathbb{T}^4$. We note that the degeneracy of the chiral primaries gets doubled as the fermion number $f_2$ drops out of the expression for the dimension $H$ and it can then take the values $\{-1,0\}$. We take the same limit on the spectrum of Ramond ground states we  obtained above. First of all we see that $\delta=0$ in this case, as the winding ${w}_2$ can be set to zero and ${w}_1=w$. By taking a careful limit of the floor function, we find that  $(m+1) =2j-1+kw \in\mathbb{Z}_{> 0} \setminus  k \mathbb{Z}$. This precisely matches the gaps in the spectrum of ground states in the $T^4$ case.

\section{The Second Quantized Ground States}
\label{SecondQuantized}
In this section we compute the second quantized Ramond-Ramond ground state partition function, and analyze its modular properties. We discuss and compare our results to those in the literature.

\subsection{The Second Quantized Theory}
\label{SecondQuantizedsub}
We wish to study multi-string contributions to the vacuum amplitude. We will concentrate on the multi-string contributions that arise from the single string Ramond-Ramond ground states. Recall  that the  one-loop vacuum amplitude $Z_{T^2}$  in the thermal background is identified with the spacetime free energy $Z_{T^2}=-\beta F$ \cite{Maldacena:2000kv,Polchinski:1998rq}. Moreover, we note that the free energy $F$ consists of connected multi-particle contributions: 
\be
\begin{aligned}
F(\b,\mu) &=  \frac{1}{\beta}  \sum_{{\cal H}_{1,b}} \log (1-e^{-\beta E+\ii \b \mu J}) - \frac{1}{\b}\sum_{{\cal H}_{1,f}}\log(1+e^{-\beta E+\ii \b \mu J})~.
\end{aligned}
\end{equation}
where ${\cal H}_{1,b}$ (${\cal H}_{1,f}$) is the bosonic (respectively fermionic) one-particle Hilbert space.
The generating function for the second quantized theory in which we allow for any number of non-interacting  and disconnected multi-particle loops is obtained by exponentiating the vacuum amplitude:
\begin{equation}
e^{Z_{T^2}} = e^{-\beta F} =\frac {\prod_{{\cal H}_{1,f}} (1+e^{-\beta E+\ii \b \mu J})} {\prod_{{\cal H}_{1,b}} (1-e^{-\beta E+\ii \b \mu J})} \, .
\end{equation}
This is a sum over disconnected vacuum amplitudes of toroidal topology. The thermodynamics that is described here is a grand canonical ensemble of non-interacting particles.
In this section, we will apply this description to the single string Ramond-Ramond ground states.

\subsubsection{The Grand Canonical Order}
\label{GrandCanonical}
Before we exponentiate our single particle free energy for the example of $AdS_3 \times S^3 \times T^4$ on which we concentrate, we wish to enrich it further. 
In our single particle partition sum $f(\beta,\nu,\bar{\nu})$, we kept track
of the left and right angular momenta as well as   the spacetime energy $E$. We propose to refine our single particle sum further by introducing an additional fugacity $\sigma$ that couples to the quantum number $n_i=2j_i-1+kw_i$ of the $i$th single string excitation. In other words, we  track the part of the R-charge quantum number that is universal in the sense that it does not depend on the compactification manifold $M=T^4$. 

The fugacity that couples to this quantum number arises as follows. In the initial NSR frame formula (\ref{InitialNSR}), we introduce  an overall shift of the R-charge fugacity $\nust$ by $2 \sigma$ as well as a fugacity $-\sigma$ in the first two theta-functions corresponding to the $AdS_3 \times S^3$ factors.  

After applying the Jacobi triality as well as spectral flow, we find the single string free energy
\be
\begin{aligned}
f(\beta, \mu, \sigma) &= \frac{N}{2\pi} \int_0^\infty \frac{d \tau_2}{\tau_2^{3/2}} \int_{-\frac{1}{2}}^{\frac{1}{2}} d \tau_1 
 \frac{\sqrt{k} e^{-\frac{k \beta^2} { 4 \pi \tau_2}} }{|\theta_1(  \tau_{\text{s.t.}},\tau )|^2}  Z_{T^4}^{bos}(q)  \sum_{l=0}^{\frac{k-2}{2}} |\chi_l(\nust - \taust, \tau) |^2   \\
&\hspace{4cm}\frac{1}{|\eta|^4}| \theta_1(\tau_{\text{s.t.}} - \frac{\nu  }{2}-\sigma, \tau)|^4
|\theta_1 (\frac{\nu}{2},\tau )|^4~.
\end{aligned}
\ee
Tracing the fugacity $\sigma$ through the calculation performed in section \ref{AdS3S3T4}, and in particular with regard to the Ramond-Ramond ground states, we conclude that  the fugacity $\sigma$ indeed keeps track of the quantum number $n=2j-1+kw$.   We recall that this quantum number takes the values $n=1,2,\dots,k-1,k+1,k+2,\dots$. There is a gap at every integer multiple of the level $k$.

The left and right moving R-charges contain this quantum number and experience an extra shift determined by the Dolbeault cohomology degrees of the complex manifold $M=T^4$. This is clear from the R-charges listed in Table \ref{tab:RRgroundstates} for this example, and it is generically true.  When we allow an arbitrary number of each of these one particle modes, it is convenient to associate a creation oscillator $\alpha^{a,\bar{a}}_{-n}$ to each of them, where we denote the  charge that couples to the fugacity $\sigma$ as a lower index $n$ and the upper $a,\bar{a}$ indices  take values in the Dolbeault cohomology and  keep track of the Dolbeault degree (by abuse of notation). If we denote the exponential of the fugacity $\sigma$ as $p=e^{2 \pi i \sigma}$, we can write down the second quantized partition function $P_{\text{gapped}}$ for the Ramond-Ramond ground states for the case of $M=T^4$:
\be
\label{secondQGF}
P_{\text{gapped}}=\prod_{\stackrel{n \ge 1}{n\ne k\mathbb{Z}}}
\frac{\Big((1-\zst^{-1} p^n)(1-\zst p^n)(1-\zbarst^{-1} p^n)(1-\zbarst p^n)\Big)^{2}}{(1-\zst^{-1}\zbarst^{-1} p^n)(1-\zst \zbarst p^n)(1-\zst \zbarst^{-1} p^n)(1-\zst^{-1} \zbarst p^n) (1-p^n)^4} \, .
\ee
We used the Hodge numbers of the four-torus and excluded the states that fell into the gap.

\subsubsection{The Hodge Polynomial of Hilbert Schemes}
\label{Hilbert}
It is interesting to 
 compare the second quantized partition function (\ref{secondQGF}) to  the generating function of Hodge polynomials of Hilbert schemes of points on $M$  \cite{Dijkgraaf:1996xw,Goettsche}. 
For a smooth projective surface $M$, the set of Hodge polynomials associated to the Hilbert scheme of points is generated by the following function \cite{Dijkgraaf:1996xw,Goettsche}:
\begin{equation}
P=\sum_{N \ge 0} h(S_N,z,\bar{z})t^N =  \prod_{k=1}^\infty
\frac{\prod_{a+\bar{a} \, \text{odd}} (1+z^{a+k-1} \bar{z}^{\bar{a}+k-1} t^k)^{h_{a,\bar{a}}}}{\prod_{a+\bar{a} \, \text{even}} (1-z^{a+k-1} \bar{z}^{\bar{a}+k-1} t^k)^{h_{a,\bar{a}}}} \, ,
\end{equation}
where $h(S_N,z,\bar{z})$ denotes the Hodge polynomial of the Hilbert scheme of $N$ points on the surface $M$, while $h_{a,\bar{a}}$ are the Hodge numbers of the surface $M$. 
The variable $t$ keeps track of the  order of the $S_N$ orbifold group while the variables $(z,\bar{z})$ are fugacities for the left and right  degrees  in the Dolbeault cohomology of the Hilbert scheme. 
When we apply this formula to a smooth projective complex connected surface $M$, we can simplify further since the Hodge numbers satisfy the relations: 
\be
\begin{aligned}
h_{1,0}=h_{0,1}=h_{1,2}=h_{2,1}~,\quad h_{2,0}=h_{0,2}~, \quad
\text{and}\quad h_{0,0}=h_{2,2}=1~.
\end{aligned}
\ee
For instance, for the four-torus $M=T^4$ we have $h_{1,0}=2$, $h_{2,0} =1$ and $h_{1,1}=4$. 
The generating function  simplifies to:
\begin{equation}
\label{GFKahler}
P(\nu,\bar{\nu},\sigma) =  \prod_{n \ge 1}
\frac{\big((1-z^{-1} p^n)(1-z p^n)(1-\bar{z}^{-1} p^n)(1-\bar{z} p^n)\big)^{h_{1,0}}}{(1-z^{-1}\bar{z}^{-1} p^n)(1-z \bar{z} p^n)\big((1-z \bar{z}^{-1} p^n)(1-z^{-1} \bar{z} p^n)\big)^{h_{2,0}} (1-p^n)^{h_{1,1}}} \, .
\end{equation}
We have defined $p=z \bar{z}  t$. From the definition of the generating function $P$, it follows that the second quantized partition function $P_{\text{gapped}}$  \eqref{secondQGF} can be written as a ratio of generating functions:
\be
\label{PgapvsP}
P_{\text{gapped}}(\nu,\bar{\nu},\sigma) = \frac{P(\nust, \nubarst, \sigma)}{P(\nust, \nubarst, k\sigma)}~.
\ee
The division implements the gap in the spectrum. The second quantized partition function is thus a ratio of symmetric orbifold generating functions.

\subsubsection{The $S_W$ Orbifold}
\label{SW}
There is another way in which to present the result for the second quantized partition function for Ramond-Ramond ground states which is illuminating and makes contact with \cite{Argurio:2000tb}. Instead of introducing a fugacity for the quantum number $n=2j-1+kw$, we introduce a fugacity that  keeps track of the winding number $w$ only and write the second  quantized partition function as:
\begin{equation}
P_{\text{gapped}}^W =  \prod_{w=0,1,\dots} 
\prod_{s=1,2,\dots,k-1}
\frac{\prod_{a+\bar{a} \, \text{odd}} (1-z^{a+s+kw-1} \bar{z}^{\bar{a}+s+kw-1} {p'}^{kw})^{h_{a,\bar{a}}}}{\prod_{a+\bar{a} \, \text{even}} (1-z^{a+s+kw-1} \bar{z}^{\bar{a}+s+kw-1} {p'}^{kw})^{h_{a,\bar{a}}}}  \label{WTrack}
\, .
\end{equation}
We will demonstrate that it agrees up to a $p$-independent factor with a  symmetric orbifold partition function of the $w=0$ ground states. 
We  start out with a seed superconformal field theory of central charge $6k$ with a spectrum of R-charges $(n+a,n+\bar{a})$ where $(a,\bar{a})$ is the spectrum of R-charges of a central charge equal to six theory on the manifold $M$ and $n=1,2,\dots,k-1$. The symmetric orbifold generating function for this seed theory reads:
\begin{equation}
P_{{6k}} =  \prod_{m=1}^\infty
\prod_{s=1}^{k-1}
\frac{\prod_{a+\bar{a} \, \text{odd}} (1-z^{a+s+m-1} \bar{z}^{\bar{a}+s+m-1} \tilde{p}^m)^{h_{a+s,\bar{a}+s}}}{\prod_{a+\bar{a} \, \text{even}} (1-z^{a+s+m-1} \bar{z}^{\bar{a}+s+m-1} \tilde{p}^m)^{h_{a+s,\bar{a}+s}}}~.
\end{equation}
We equate the fugacities $z \bar{z} \tilde{p}=q^k$
and use $h_{a+n,\bar{a}+n}=h_{a,\bar{a}}$ to find
 \begin{equation}
P_{{6k}} =  \prod_{m=1}^\infty \prod_{s=1}^{k-1}
\frac{\prod_{a+\bar{a} \, \text{odd}} (1-z^{a+s-1} \bar{z}^{\bar{a}+s-1} q^{km})^{h_{a,\bar{a}}}}{\prod_{a+\bar{a} \, \text{even}} (1-z^{a+s-1} \bar{z}^{\bar{a}+s-1} q^{km})^{h_{a,\bar{a}}}} ~.
\end{equation}
By identifying $q=z \bar{z} p'$, we obtain
\begin{equation}
P_{{6k}} =  \prod_{m=1}^\infty \prod_{s=1}^{k-1}
\frac{\prod_{a+\bar{a} \, \text{odd}} (1-z^{a+s+km-1} \bar{z}^{\bar{a}+s+km-1} {p'}^{km})^{h_{a,\bar{a}}}}{\prod_{a+\bar{a} \, \text{even}} (1-z^{a+s+km-1} \bar{z}^{\bar{a}+s+km-1} {p'}^{km})^{h_{a,\bar{a}}}} ~.\ \label{P6k}
\end{equation}
We see that the symmetric orbifold generating function $P_{6k}$ in (\ref{P6k}) indeed agrees with the second quantized partition function $P^W_{\text{gapped}}$ in (\ref{WTrack}) which tracks the winding number of the Ramond-Ramond sector ground states. We  discuss these formulas in due course.

\subsection{Modular Partition Functions}
\label{Modularity}

First though, we slightly modify our generating functions in order to make it more manifest that they exhibit interesting modular transformation properties. The generating function  $P$ of Hodge polynomials of Hilbert schemes of points given in \eqref{GFKahler}  can be written in terms of theta- and eta-functions up to the following prefactors:
\be
\begin{aligned}
P(\nu, \bar{\nu}, \sigma)&= g(\nu,\bar{\nu})~ p^{\frac{\chi}{24}}~ \widetilde{P}(\nu,\bar{\nu},\sigma) \, ,
\end{aligned}
\ee
where $\chi$ is the Euler character of the K\"ahler manifold $M$ and 
\be
\label{Ptildekahler}
\widetilde{P}(\nu,\bar{\nu},\sigma) =\left( \frac{\theta_1(\nu,\sigma)}{\eta(\sigma)} \frac{\theta_1(\bar{\nu},\sigma)}{\eta(\sigma)} \right)^{h_{1,0}} 
\frac{\eta}{\theta(\nu+\bar{\nu},\sigma)}
\left(\frac{\eta}{\theta(\nu-\bar{\nu},\sigma)}\right)^{h_{2,0}}
\frac{1}{(\eta(\sigma))^{h_{1,1}}}~.
\ee
We now propose to work with this modified generating function $\widetilde{P}$, in which we strip away  the $p$-independent factor $g(\nu,\bar{\nu})$ and a factor of $p^{ \frac{\chi}{24}}$ from the generating function of Hodge polynomials. The  factor $p^{\frac{\chi}{24}}$ may be  familiar from  the generating function of Euler numbers of instanton moduli spaces  that was defined in \cite{Vafa:1994tf}.\footnote{It may well have a similar origin in local curvature dependent terms that arise upon S-dualization. It would also be interesting to analyze the origin of the twist dependent factor $g$. }
The modified generating function $\widetilde{P}$ has good modular and elliptic properties determined by those of its factors. For instance we note that the function behaves under S-modular transformations as:
\begin{equation}
    \widetilde{P}(\frac{\nu}{\sigma},\frac{\bar{\nu}}{\sigma},-\frac{1}{\sigma}) = (-\ii \sigma)^{-\frac{h_{1,1}}{2}}~e^{-\frac{\pi\ii}{\sigma}\big( (\nu^2+\bar\nu^2)(h_{0,0}-h_{1,0}+h_{2,0})+2\nu\bar\nu(h_{2,0}-h_{0,0}) \big)}~ \widetilde{P}(\nu, \bar{\nu}, \sigma)\, .
\end{equation}
It therefore behaves as a multivariable Jacobi form with matrix index determined by the Hodge numbers of $M$. 
Given the relation between the second quantized partition function $P_{\text{gapped}}$ and the generating function of Hodge polynomials $P$  in equation \eqref{PgapvsP}, it is natural to define a modified second quantized partition function of boundary Ramond-Ramond ground states as follows:
\begin{align}
\widetilde{P}_{\text{gapped}}:=p^{\frac{\chi}{24}(k-1)}e^{Z_{T^2}^{RR}} = \frac{\widetilde{P}(\nust, \nubarst, \sigma)}{\widetilde{P}(\nust,\nubarst, k \sigma)}  \, .
\end{align}
 We note that the $p$-independent factor $g(\nust, \nubarst)$ cancels out once we take the ratio of generating functions of the modified Hodge polynomials. Thus, the relation between the previously defined second quantized partition function and the modified one is simply a $p$-dependent factor. The quantity $(1-k)\chi$ plays the role of a central charge, where $k=N_5$ is the number of NS5 branes \cite{Vafa:1994tf}. It is straightforward to check that the gapped partition function $\tilde{P}_{\text{gapped}}$ transforms under modular S-transformations as a multivariable Jacobi form with matrix index:
\begin{equation}
\label{Pgappedkahler}
\widetilde{P}_{\text{gapped}}(\frac{\nust}{\sigma},\frac{\nubarst}{\sigma}-\frac{1}{\sigma}) = N_5^{\frac{h_{1,1}}{2}}~e^{-\frac{k-1}{k}\frac{\pi\ii}{\sigma}\big( (\nu^2+\bar\nu^2)(h_{0,0}-h_{1,0}+h_{2,0})+2\nu\bar\nu(h_{2,0}-h_{0,0}) \big)}~ \widetilde{P}_{\text{gapped}}(\nust,\nubarst,\sigma) \, .
\end{equation}
For the case of $M=T^4$ we note that the index is  zero. The modular transformation property with respect to the variable that keeps track of the orbifold order is intriguing. There may be a hint in the fact that this fugacity mixes with the fugacity corresponding to the spacetime modular parameter $\tau_{\text{s.t.}}$. 

\subsubsection{Other Examples}

When the compact manifold $M$ is a $K3$ manifold we propose the gapped partition function:
\begin{equation}
\widetilde{P}_{\text{gapped}}^{K3} = p^{\frac{\chi}{24}(k-1)} 
\frac{\theta(\nu+\bar{\nu},k\sigma) \theta(\nu-\bar{\nu},k\sigma) \eta^{18}(k \sigma)}{\theta(\nu+\bar{\nu},\sigma) \theta(\nu-\bar{\nu},\sigma) \eta^{18}(\sigma)} \, .
\end{equation}
Here we have used the Hodge numbers of K3 ($h_{1,0}=0$, $h_{1,1}=20$ and $h_{2,0}=1$), in the general formulas in \eqref{Ptildekahler} and \eqref{Pgappedkahler}. 
Its modular properties were already uncovered above.

For the $AdS_3 \times S^3 \times S^3 \times S^1$ models discussed in section \ref{AdS3S3S3S1} there is the intriguing phenomenon of a contribution from the edge of the continuum which indicates that an unambiguous definition of the first and second quantized partition sum may include a contribution from the continuous sector, as in the calculation of (completed mock modular) non-compact elliptic genera \cite{Troost:2010ud}. Still, we can tentatively write down a second quantized partition function based on the spectrum of R-charges we determined. We concentrate on a very simple example in which we have $k_1=2k=k_2$ and a positive integer level $k$. Then the spectrum simplifies to $m \in \mathbb{Z}_{\ge 0} \setminus k (2\mathbb{Z}+1) $. Taking into account the two-fold left/right degeneracy, the second quantized partition sum then reads:
\begin{equation}
\begin{aligned}
P_{\text{gapped}} &= 
\prod_{m \in \mathbb{Z}_{\ge 0} \setminus k (2\mathbb{Z}+1)}^{\infty} \frac{(1-z^{-1} p^{m+1})(1- \bar{z}^{-1} p^{m+1})}{
(1-z^{-1} \bar{z}^{-1} p^{m+1})(1- p^{m+1})}  \, .
\end{aligned}
\end{equation}
More intricate $AdS_3 \times S^3 \times S^3 \times S^1 $ examples with generic levels lead to even more intriguing expressions that we leave for future study.

\subsection{Discussion}
We finish this section with some conceptual remarks on our formulas and their relation to the literature. Indeed,
there remain  open questions associated to the various points of view on the second quantized partition functions.

\subsubsection{Fundamental Strings Are Perturbative}

There is a suggestion, based on an original observation on the cohomology of the moduli spaces of instantons \cite{Vafa:1995bm}, that the boundary dual to $N_1$ D1-branes embedded in $N_5$ D5-branes may correspond to  a point in the moduli space of the symmetric orbifold conformal field theory $Sym_N (M)$ where $N=N_1 N_5$ and the four-manifold $M$ is orthogonal to the D1-branes and parallel to the D5-branes. The central charge of the conformal field theory is $c=6 N_1 N_5$ (in the case of $M=T^4$).  By comparing our second quantized partition function to the symmetric orbifold (or rather Hilbert scheme) generating function in subsection \ref{Hilbert}, we used the conjecture as a point of reference.

Let us  stress though that there important differences between the S-dual Ramond-Ramond background and our NSNS string theory.
 We started out with an NSNS background that consists of $N_5=k$ NS5-branes and $N_1$ fundamental strings. 
 The background number $N_1$ of fundamental strings  is only visible in the NSNS background supergravity solution through the attractor mechanism \cite{Ferrara:1995ih,Maldacena:1998bw}. The latter fixes the string coupling and therefore the three-dimensional Newton coupling as a function of $N_1$ (and $N_5$). Through the Brown-Henneaux central charge formula \cite{Brown:1986nw}, the number $N_1$ thus features in the spacetime central charge $c=6N_1 N_5$. This is the background central charge in the supergravity background around which we choose to do perturbation theory. The background central charge was computed in gravity in \cite{Brown:1986nw} and as a string theory one-point function in  \cite{Troost:2011ud}.  Importantly, 
 the NSNS background has the unique feature of allowing for the addition of perturbative fundamental strings that wind an angular direction in $AdS_3$. These two-dimensional fundamental string world sheets  act as domain walls in the three-dimensional anti-de Sitter spacetime and they separate regions with differing local cosmological constant.
The winding number $w$ of the fundamental strings is a measure for the difference in central charge on one or the other side of the wall:   $\delta c= 6 N_5 \delta N_1 = 6 k w$ as computed in \cite{Giveon:1998ns,Kutasov:1999xu} from the string world sheet perspective. Thus, the central charge of the holographic dual can change as function of the number of perturbative winding string excitations. This is unique (in perturbation theory) to the NSNS background.
 Because we are in a fundamental string picture in which the fundamental strings are light perturbative excitations, there is no stringy exclusion principle \cite{Maldacena:1998bw} at work. This has as a consequence that there is no bound on the central charge and that we therefore automatically obtain a grand canonical partition function.\footnote{Our perspective differs from  the  supergravity analysis of  the elliptic genus,  matched to the boundary elliptic genus at infinite $N$ \cite{deBoer:1998us}.}
 Indeed, this perspective was already taken in the calculation of the partition function of $AdS_3$ string theory at string scale \cite{Eberhardt:2020bgq}, namely at level $N_5=1=k$. The fact that the bulk string theory allows for a change in the central charge of the dual, via the scattering of long winding strings, yet is unitary, creates a puzzle for the proper interpretation of the holographic dual. We refer to  \cite{Saad:2019lba,McNamara:2020uza,Eberhardt:2020bgq} for futher context and discussion of this intriguing aspect of the Neveu-Schwarz-Neveu-Schwarz $AdS_3/CFT_2$ correspondence.

 \subsubsection{Mind the Gap}

The analysis of the perturbative string spectrum \cite{Argurio:2000tb} showed that the bulk string spectrum has fewer chiral primary states than a supersymmetric $Sym_N(M)$ symmetric orbifold conformal field theory. The $AdS_3 \times S^3$ string theory is obtained by descending down the throat of NS5-branes, desingularized by a density of fundamental strings. The linear behaviour of the dilaton down the throat of NS5-branes may no longer be singular, but it leaves its mark on the spectrum of perturbative string excitations: the linear dilaton causes a  gap in world sheet conformal dimensions equal to $h= 1/(4k)$ in perturbative string excitations that is faithfully mirrored by the strings in the continuous representations of the $AdS_3$ isometry group.\footnote{A related phenomenon is the decrease in the number of moduli in non-compact Gepner models compared to local Calabi-Yau manifolds \cite{Ashok:2007ui}.}${}^{,}$\footnote{
One NS5-brane does not generate a throat visible to perturbative fundamental strings. In this case, the bulk spectrum coincides with the symmetric orbifold spectrum \cite{Eberhardt:2018ouy}.} (The slightest perturbation with a Ramond-Ramond flux closes off the throat though and regenerates the symmetric orbifold spectrum \cite{Seiberg:1999xz,Eberhardt:2018vho}.)
Thus, the fundamental strings that would travel up or down the throat of NS5-branes are missing from the spectrum of chiral primaries.  In our description, they correspond to quantum numbers $n$ that are multiples of the number of NS5-branes $k$ (of which the first one lies at $j=(k+1)/2$)
\cite{Eberhardt:2018vho}.
To account for this fact, we worked with a ratio of Hodge polynomial generating functions in subsection \ref{Hilbert}.

It should be remarked that the Hilbert scheme perspective in subsection \ref{Hilbert} interprets  the terms $2j-1$ in the quantum number $n=2j-1+kw$ as representing a change in the boundary spacetime central charge that is a fraction of (six times) the number of NS5-branes $k=N_5$. This attempt at interpretation remains to be substantiated.

Indeed, as we saw previously, changes in the boundary central charge come naturally in units of $6k$. This staircase structure is respected by the counting proposed in subsection \ref{SW}, where we only keep track of the winding numbers $w_i$ of the single string excitations. Their sum
$\sum_i w_i = W$ is the total order of the $S_W$ orbifold.
This coding of the Ramond-Ramond ground states of the boundary conformal field theory agrees with the point of view of \cite{Argurio:2000tb} as well as \cite{Eberhardt:2019qcl} on a conjectured dual orbifold conformal field theory.\footnote{Note  that as in \cite{Argurio:2000tb}, we have a non-trivial bulk operator with trivial qauntum numbers. It plays a crucial role  in the grand canonical partition function since it can trivially augment the order of the orbifold. }

\section{Conclusions}
\label{Conclusions}
In this paper we have revisited the literature on thermal $AdS_3$ partition functions in string theory with NSNS flux \cite{Maldacena:2000hw,Maldacena:2000kv} and obtained a number of improvements. Firstly, we clarified the bound on the spin in the discrete spectrum of the string. It takes values in a half-open interval. This was appreciated in the literature on the cigar sl$(2,\mathbb{R})$/u$(1)$ coset a while back \cite{Troost:2010ud,Eguchi:2010cb,Ashok:2011cy} and agrees with the analysis in integrable systems \cite{Bazhanov:2020uju}.  Moreover, we treated the lower boundary of the winding number range carefully.

Secondly, by introducing the sl$(2,\mathbb{R})$/u$(1)$ coset technology \cite{Hanany:2002ev} into the analysis of the thermal $AdS_3$ partition function, we were able to confirm the proposed off-shell Hilbert space of $AdS_3$ string theory. This makes for a direct path integral bridge between the arguments put forward in \cite{Maldacena:2000hw} and \cite{Maldacena:2000kv}. 

Thirdly, we extended the calculation of the thermal partition function to the case of a supersymmetric world sheet $AdS_3$ string theory, both in the off-shell and the on-shell approach. This allows for the calculation of any thermal partition function on a super string background of the form $AdS_3 \times N$ with NSNS flux.

Fourthly, we applied our technology to compute the one-loop contribution to the boundary Ramond-Ramond twisted index for the $AdS_3\times S^3\times T^4$ and $AdS_3\times S^3\times S^3\times S^1$ backgrounds. The application of a generalized Jacobi identity (or $so(8)$ triality on the world sheet spinor) led us to a Green-Schwarz formulation of the one-loop amplitude. This form of the amplitude is bound to connect well with manifestly supersymmetric formulations of $AdS_3$ string theory.

In all these cases we established positivity bounds. In the case of the supersymmetric string backgrounds, we determined all (boundary) Ramond-Ramond ground states saturating the bound rigorously, thereby providing a complete classification. Our results are in one to one correspondence with the spectrum of boundary chiral primaries proposed  in \cite{Argurio:2000tb,Eberhardt:2017pty}.

We then constructed  a second quantized partition function for the Ramond-Ramond ground state excitations. Due to the gap in the spectrum generated by the NS5-branes, the partition function takes an original form. We found that it can be written as a ratio of generating functions of Hodge polynomials of Hilbert schemes, or as fitting the mold of a winding string orbifold. We also made the intriguing observation that the second quantized partition function has  good modular properties with respect to the fugacity associated to the universal part of the boundary R-charge. 

There are a large number of avenues open for further investigation. It would be important to obtain the  density of states for the continuous part of the spectrum including descendants from the path integral. We would like to understand better the origin of the modular properties of the second quantized partition function of the Ramond-Ramond ground states. Relatedly, it would be useful to develop an S-dual, gauge theory picture for the gapped spectrum (at strong coupling). Finally, understanding how the gapped states fit into the topological AdS/CFT correspondence \cite{BenettiGenolini:2017zmu,Eberhardt:2019ywk,Costello:2020jbh,Li:2020nei,Dei:2020zui} will be worthwhile.

\appendix

\section*{Acknowledgments}
We would like to thank our colleagues for creating  stimulating research environments, both virtual and real. We thank Lorenz Eberhardt and Matthias Gaberdiel for patient explanations.

\section{Theta and Eta Functions}
\label{apptheta}
In this appendix, we stipulate our conventions for the Dedekind eta-function and the Jacobi theta-functions, and review their elliptic and modular properties. 
The Jacobi theta function with characteristic $(a,b)$ has the  power series expansion:
\be
\theta^a_b (\nu,\tau) = \sum_f q^{\frac{(f+a)^2}{2}} z^{f+a} e^{2 \pi\ii b (f+a)} \, .
\ee
We introduced the notations $q=e^{2\pi\ii\tau}$ and $z=e^{2\pi\ii\nu}$. The power series expansion can be used to derive the ellipticity properties of the $\theta$-functions. For integer $v, w \in\mathbb{Z}$,  we have
\be
\begin{aligned}
\theta^a_b (\nu + w \tau - v,\tau) &=\sum_f  q^{\frac{(f+a)^2}{2}} (zq^w e^{-2\pi\ii v})^{f+a} e^{2 \pi \ii b (f+a)}\\
&= e^{-2\pi\ii v a} \sum_f q^{\frac{(f+a)^2}{2} + w (f+a)} z^{f+a} e^{2\pi\ii b(f+a)}\\
&= q^{-\frac{w^2}{2} } z^{-w}e^{-2\pi\ii (v a+ wb)} \sum_f q^{\frac{(f+a+w)^2}{2} + w f} z^{f+a+w} e^{2\pi\ii b(f+a+w)}\\
&= q^{-\frac{w^2}{2} } z^{-w}e^{-2\pi\ii (v a+ wb)} \theta^a_b (\nu,\tau)~.
\end{aligned}
\ee
For the twist $\nu = s_1\tau-s_2$ we obtain
\be
\theta_{b}^a(s_1\tau - s_2 + w\tau -v,\tau) = q^{-\frac{w^2}{2} } e^{-2\pi \ii w (s_1\tau - s_2)} e^{-2\pi \ii (va+wb)} \theta_{1}(s_1\tau - s_2, \tau )~.
\ee
We  denote the Jacobi theta function with characteristic $a=\frac12=b$ as $\theta_1(\nu,\tau)$. We  recall its infinite product representation 
\be
\theta_1(\nu,\tau) = -2\sin\pi\nu~q^{\frac18}\prod_{n=1}^{\infty}(1-q^n)(1-zq^n)(1-z^{-1}q^n)~.
\ee
The Dedekind $\eta$-function equals 
\be
\eta(\tau) = q^{\frac{1}{24}}\prod_{n=1}^{\infty}(1-q^n)~.
\ee
The modular $S$-transformation of these functions is:
\begin{align}
    \theta_1(\frac{\nu}{\tau}, -\frac{1}{\tau}) &= (-\ii\tau)^{\frac12}~e^{\frac{\pi\ii\nu^2}{\tau}}~\theta_1(\nu, \tau)~,\\
    \eta(-\frac{1}{\tau}) &= (-\ii\tau)^{\frac12}~\eta(\tau)~.
\end{align}

\section{Affine Characters}
\label{discchar}

We briefly summarize a few properties of affine algebra characters.
\subsection{Discrete Affine \texorpdfstring{sl$(2,\mathbb{R})$}{} Characters}
We define the discrete characters of the sl$(2,\mathbb{R})$ current algebra at level $k+2$:
\be
\chi_j(q,z) = \Tr_{D_j^+} \left( q^{L_0 - \frac{c}{24}} \, z^{j_0^3}\right)~.
\ee
The central charge $c$ of the sl$(2,\mathbb{R})$ Sugawara Virasoro algebra is 
$c = 3 + \frac{6}{k}$. The character is well studied, see e.g. \cite{Pakman:2003kh} for details. It is explicitly given by:
\be
\begin{aligned}
\chi^+_j(q,z) &= \frac{q^{-\frac{j(j-1)}{k} -\frac{1}{4k} - \frac{1}{8}}z^j}{\prod_{n=1}^\infty (1-q^n)(1-z q^{n-1})(1-z^{-1} q^n)} \\
&= \frac{q^{-\frac{j(j-1)}{k} -\frac{1}{4k} - \frac{1}{8}}z^j}{\prod_{m=1}^\infty (1-q^m)^3}
\sum_{r,n}  z^r (-1)^n q^{n/2(n+2r+1)} \\
&= \frac{q^{-\frac{j(j-1)}{k} -\frac{1}{4k}} z^j}{\eta^3(\tau)}
\sum_{r \in \mathbb{Z}} z^r S_r~,
\end{aligned}
\ee
where we have defined the special series $S_r$ labeled by an integer $r$:
\be
S_r  = \sum_{n=0}^{\infty}(-1)^n q^{\frac{n}{2}(n+2r+1)}~.
\ee
The character for the spectrally flowed representation $D_j^{+,w}$ is similarly given by
\be
\begin{aligned}
\chi^{w}_j (q,z) &= \frac{q^{-\frac{j(j-1)}{k}-\frac{1}{4k} - \frac{1}{8}} z^{j+(k+2)\frac{w}{2}}}{\prod_{m=1}^\infty (1-q^m)^3}
\sum_{r,n} q^{\frac{(j+r)^2}{(k+2)}-\frac{(j+r+(k+2)w/2)^2}{(k+2)}} z^r (-1)^n q^{\frac{n}{2}(n+2r+1)}\\
&= \frac{q^{-\frac{j(j-1)}{k}-\frac{1}{4k} - \frac{1}{8}} z^{j+(k+2)\frac{w}{2}}}{\prod_{m=1}^\infty (1-q^m)^3}
\sum_{r} q^{\frac{(j+r)^2}{(k+2)}-\frac{(j+r+(k+2)w/2)^2}{(k+2)}}z^r S_r \\
&= \frac{q^{-\frac{j(j-1)}{k} - (k+2) \frac{w^2}{4}-\frac{1}{4k} - \frac{1}{8}} z^{(k+2)\frac{w}{2}}}{\prod_{m=1}^\infty (1-q^m)^3}
\sum_{r,n}  (z q^{-w})^{r+j} (-1)^n q^{\frac{n}{2}(n+2r+1)}\\
&= \frac{q^{-\frac{j(j-1)}{k} - (k+2) \frac{w^2}{4} -\frac{1}{4k} } z^{(k+2)\frac{w}{2}}}{\eta^3(\tau)}
\sum_{r}  (z q^{-w})^{r+j} S_r~.
\label{sl2rcharspfl}
\end{aligned}
\ee
We list properties of the series $S_r$. It satisfies the equations:
\begin{equation}
q^{r} S_r = S_{-r} \, , \qquad S_r + S_{-r-1} = 1 \, .
\end{equation}
We moreover have that the minimal surviving power in $S_r$ for $r \ge 0$ equals zero and, despite appearances, it is $q^{-r}$ when $r \le 0$, by the first property recorded above. The series $S_r$ also appears in the power series expansion of the inverse $\theta_1$-function:
\be
\begin{aligned}
\frac{1}{\theta_1(\nu, \tau)} &= i \frac{z^{1/2}}{\eta(\tau)^3} \sum_r z^r S_r(q) \, ,
\end{aligned}
\ee 
where $z=e^{2\pi\ii\nu}$. A proof of these formulae can be found in \cite{Pakman:2003kh, Ashok:2013zka}.

\subsection{Affine \texorpdfstring{su$(2)$}{} Characters}
\label{affine}
We discuss the su$(2)$ affine characters and  their properties. The level $k-2$, spin $l$ character is:
\be
\chi_l(\nu,\tau) 
= \frac{\Theta_{2l+1,k}(\nu,\tau) - \Theta_{-2l-1,k}(\nu,\tau)}{\Theta_{1,2}(\nu,\tau) - \Theta_{-1,2}(\nu,\tau)}~,
\ee
where the spin $l$ takes the values $l= 0,\frac12, \ldots \frac{k-2}{2}$. The level $k$ $\Theta$-functions are defined as:
\be
\Theta_{m,k}(\nu, \tau) = \sum_{n\in \mathbb{Z} + \frac{m}{2k}} q^{km^2} z^{-kn}~,
\ee
with $q=e^{2\pi\ii\tau}$ and $z=e^{2\pi\ii\nu}$ as before.
Spectral flow by $w$ units in the su$(2)$ theory  maps to another spin $l'$ representation \cite{Feigin:1998sw}:
\be
\label{specflowsu2}
q^{\frac{(k-2) w^2}{4}} e^{-2\pi\ii\nu\frac{(k-2)w}{2}}~\chi_{l}(\nu-w\tau,\tau) = 
\begin{cases}
\chi_l(\nu,\tau) & \text{for} \quad w\in 2\mathbb{Z}\\
\chi_{\frac{k-2}{2}-l}(\nu,\tau) & \text{for} \quad w\in 2\mathbb{Z}+1 ~.
\end{cases}
\ee
For even spectral flow the spin is invariant, while for odd spectral flow the spin $l$ is mapped to its mirror, $(k-2)/2-l$.
We use the following expansion for the su$(2)$ character:
\be
\chi_{l}(\nu,\tau) = q^{\frac{l(l+1)}{k} -\frac{c_{su(2)}}{24}} \sum_{r'} C^l_{r'}e^{2\pi\ii \nu r'}~,
\ee
where we have defined the functions
\be
C^l_{r'} = \sum_{n\ge 0} C^l_{r' n} q^n~.
\ee
Spectral flow by a single unit is equivalent to shifting the fugacity $\nu$ by the modular parameter $\tau$, and we obtain
\be
q^{\frac{(k-2)}{4}} e^{-2\pi\ii\nu\frac{(k-2)}{2}}\chi_l(\nu-\tau, \tau) = q^{\frac{(k-2)}{4}} ~ q^{\frac{l(l+1)}{k} -\frac{c_{su(2)}}{24}} \sum_{ r'}  q^{-r'} C^l_{r'} e^{2\pi\ii \nu (r' - \frac{k-2}{2})}~.
\ee
Let us  consider the right hand side equation \eqref{specflowsu2} for flow $w=1$. We obtain
\be
\begin{aligned}
\chi_{\frac{k-2}{2}-l}(\nu,\tau)= q^{\frac{l(l+1)}{k} + \frac{k-2}{4} - l -\frac{c_{su(2)}}{24}} \sum_{r''} C^{\frac{k-2}{2} -l}_{r'',n} e^{2\pi\ii \nu r''} ~.
\end{aligned}
\ee
By equating the two expressions and comparing the respective coefficients with $r'' = r'- \frac{k-2}{2}$, we find the useful relation:
\be
\label{drnspectral}
q^{r' -l} C^{\frac{k-2}{2} -l}_{r' - \frac{k-2}{2}} = C^{l}_{r'}~.
\ee

\subsection{The Shape of Affine Modules}
\label{TheShapeofAffineModules}
\begin{figure}[ht]
\center{\includegraphics[width=\textwidth] {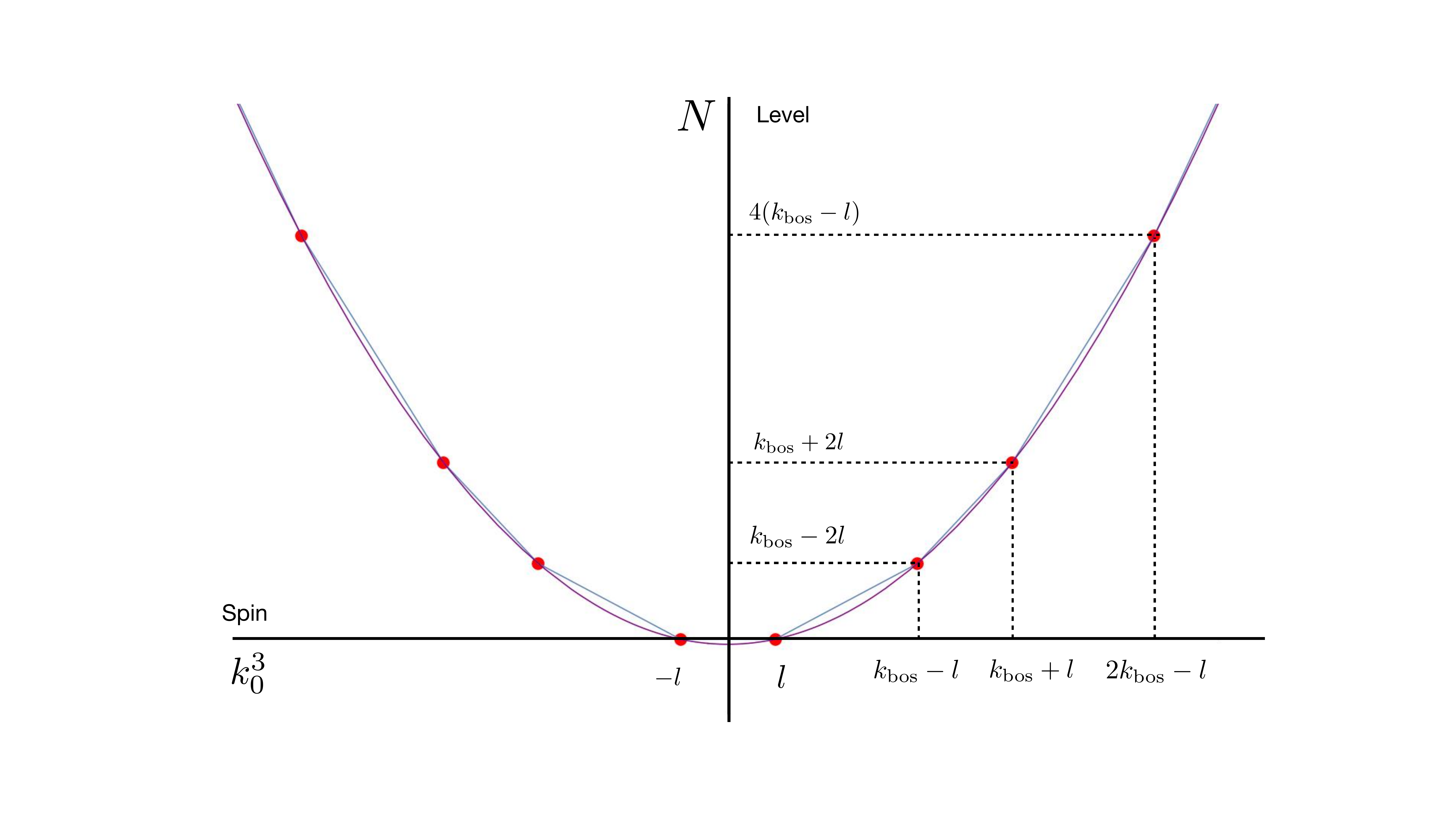}} 
 \caption{\label{fig:my-label2} The weight diagram of an irreducible su$(2)$ current algebra module. The border of the weight diagram is given by the (blue) straight lines. The circumscribing (purple) parabola touches the weight diagram if and only if we are at a spectral flowed ground state, indicated by (red) dots.
 }  
 \label{CurrentAlgebraModule}
\end{figure}

In this section, we wish to understand better the shape of the weight diagram of an irreducible su$(2)$ affine module.
The weight diagram is an approximation with line segments of a parabola in the (spin,level) plane, with the level growing quadratically with the maximal spin. We denote the level of the bosonic current algebra by $\kbos$. At level zero, we have states of spin between $-l$ and $l$ for a current algebra of spin $l$. For a given su$(2)$ spin that arises in the irreducible module, we wish to put a lower bound on the level. Thus, for a spin component below $l$, the bound on the level is zero. We can increase the spin at a minimal cost in level by acting with the generator $k_{-1}^+$. We can act $\kbos-2l$ times before annihilating the state. Thus, we have a straight line in the (spin, level) plane that goes from $(l,0)$ to $(\kbos-l,\kbos-2l)$, as well as its charge conjugate mirror. (See figure \ref{CurrentAlgebraModule}.) 
The exploration of the boundaries of the weight space continues by acting with the operator $k^+_{-2}$ a maximal number of $2l$ times. Along the straight line with a larger slope, we reach the point $(\kbos+l,\kbos+2l)$. We continue the production of the boundary line segments of the weight space of the current algebra module by alternating the action of  $k^+_{-n}$ for odd $n$, $\kbos-2l$ times and for even $n$, $2l$ times. This provides us with the precise shape of the boundary of the weight diagram. The points that lie at the end of a line segment satisfy the equation:
\begin{equation}
(\text{spin})^2-l^2 = \kbos N
\end{equation}
and lie on a parabola. The spectral flowed ground states correspond -- one to one -- to the ends of the line segments. This follows from the formulas:
\begin{eqnarray}
k^3_0 (w) &=& k^3_0 + \frac{\kbos}{2} w
\nonumber \\
L_0 &=& \frac{l(l+1)}{\kbos+2} + w k^3_0 + \frac{\kbos w^2}{4} \, ,
\end{eqnarray}
as well as the initial extremal value $|k_0^3|=l$ at $N=0=L_0-\frac{l(l+1)}{\kbos+2}$. Thus, at each integer $w$, we have a spectrally flowed ground state that lies on the parabolic approximation to the weight diagram. We will call the associated spins $l_{crit}$ critical.  Crucially, at levels other than those parameterized by the integer spectral flow number $w$, the bound in the (spin, level) plane is stricter.

\begin{figure}[ht]
\center{\includegraphics[width=\textwidth] {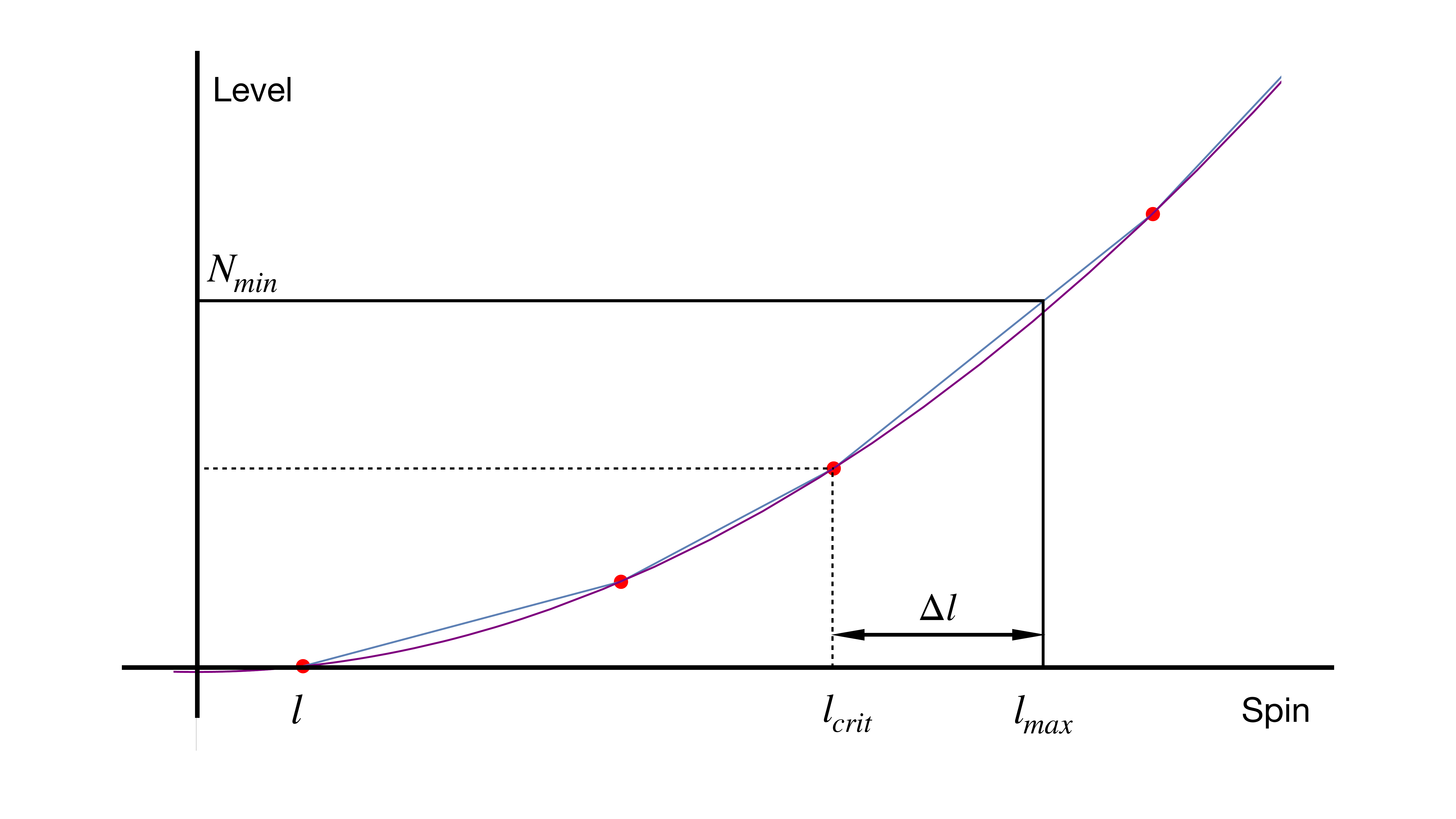}} 
 \caption{\label{CurrentAlgebraModulev4} We show the right quadrant of the weight diagram of an irreducible su$(2)$ current algebra module and indicate all the variables that appear in the various formulae.
 }  
\end{figure}

The optimal bound is the one that interpolates linearly between  the critical points. Thus, suppose that we have a maximal spin of the form $l_{max}$ at a given level $N$. This is shown in Figure \ref{CurrentAlgebraModulev4}. 
We associate an integer $w$ to $l_{max}$ that is provided by the critical spin $l_{crit}$ below or equal to $l_{max}$. We parameterize:
\begin{equation}
l_{max} = l_{crit} + \Delta l
\end{equation}
 where the positive quantity $\Delta l$ is the parameter that serves as the linear interpolation parameter between critical spins.
We then have the bound:
\begin{equation}
\label{boundonlevel}
N \ge N_{min}=\frac{l_{crit}^2-l^2}{\kbos} + (w+1) \Delta l \, ,
\end{equation}
where we have shown the level that saturates the bound in figure \ref{CurrentAlgebraModulev4}. 
For reference, we provide the explicit parameterization that depends on the parity of $w$. For even $w$, we have: 
\begin{equation}
l_{max} = l_{crit}+\Delta l = l + \frac{\kbos w}{2} + \Delta l
\end{equation}
while for  odd $w$ the parameterization is:
\begin{equation}
l_{max}=l_{crit}+\Delta l =-l+\frac{\kbos (w+1)}{2}+ \Delta l\, .
\end{equation}

\subsubsection{The Inclusion of Fermions}
Finally, we describe how the analysis of the bosonic current algebra module changes when we include chiral Ramond fermion excitations transforming in the adjoint of su$(2)$. Each linear segment generated by the action of the bosonic generator $k^+_{-n}$ is lengthened by the action of one fermionic generator $\psi^+_{-n}$. The result is that the action of $2l$ generators is replaced by $2l+1$ generators, leading to the replacement $l \rightarrow l + \frac12$. Moreover, the action of $\kbos-2l$ generators is substituted for by the action of $\kbos-2l+1 = (\kbos+2) -2 (l+ \frac12)$ generators, making for a replacement $\kbos \rightarrow k$ where $k=\kbos+2$ is the supersymmetric level. Thus, the previous bounds are valid, taking into account the fact that the maximal spin component is accounted for by the sum of bosonic and fermionic spin components, and implementing the substitution $(l,\kbos) \rightarrow (l+\frac12,k)$. Importantly, the final picture of the combined bosonic-fermionic current algebra weight space remains of the shape of figure \ref{CurrentAlgebraModule}, namely line segments inscribed in a parabola, touching the parabola at points determined by simultaneously spectral flowed bosons and fermions.

\section{Diophantine Harmonics}
\label{DiophantineHarmonics}
We consider $k_1, k_2 \in \mathbb{Z}_{>0}$ and let
\be
\label{harmonicmean}
\frac{1}{k}=\frac{1}{k_1}+\frac{1}{k_2}~.
\ee
We list three lemmas that prove statements  made in Appendix E of \cite{Eberhardt:2017pty}. 

\paragraph{Lemma 1} (the big jump):
No two multiples of $k_1$ or $k_2$ can fall in the interval $\big(k w,k(w+1)\big]$. \\

\noindent
Proof: the equation \eqref{harmonicmean} for positive integers $k_i$ implies that $k<k_a$. This proves the lemma.

\paragraph{Lemma 2} (two birds with one stone): 
When a multiple of $k_1$ and a multiple of $k_2$ both lie in the interval  $(kw,k(w+1)]$ then they are equal, as well as equal to $k(w+1)$. As a consequence $k(w+1)$ is then a common multiple of $(k_1,k_2)$ and a multiple of the lowest common multiple of $(k_1,k_2)$. \\

\noindent
Proof: $w k < m_1 k_1 \le (w+1)k$ implies $0< m_1 k_1 + (m_1-n) k_2 \le k_2$. We also have the equation with $(1 \leftrightarrow 2)$ exchanged. Adding the two proves that $w=m_1+m_2-1$. Plugging this into the original equation as well as its $(1 \leftrightarrow 2)$ exchanged counterpart implies the result.

\paragraph{Lemma 3} (horror vacui):
When an interval $(k w,k(w+1)]$ contains neither a multiple of $k_1$ nor a multiple of $k_2$, then $kw$ is a multiple of the lowest common multiple of $(k_1,k_2)$.  \\

\noindent
Proof: $
n_1 k_1 \le kw $ and $ k(w+1)< (n_1+1) k_1$. We conclude that
$(k_1+k_2)n_1 \le k_2 w$ and $k_2 (w+1) < (k_1+k_2) (n_1+1)$ as well as its $(1 \leftrightarrow 2)$ counterpart.
Combining these four inequalities we prove that $w=n_1+n_2$. From that we conclude, by using the original inequalities that $k_1 n_1=k_2 n_2$. And from that we prove that $kw$ is a multiple of both $k_1$ and $k_2$, equal to $k_a n_a$. 

These three lemmas show also that in  intervals $(kw, k(w+1)]$ that have no common multiple of $(k_1,k_2)$ at the boundary, there lies precisely one multiple of $k_1$  or one multiple of $k_2$.

\bibliographystyle{JHEP}

\end{document}